\documentclass[aps,prc,superscriptaddress,showpacs,nofootinbib,floatfix,twocolumn]{revtex4}

\usepackage{graphicx}   
\usepackage{rotating}
\usepackage{dcolumn}
\usepackage{bm}
\usepackage{amssymb}
\usepackage{multirow}

\newcommand{\pt}{\mbox{$p_\mathrm{T}\,$}}

\begin{document}

\title{Dihadron azimuthal correlations in Au+Au collisions at $\sqrt{s_{\rm{NN}}}$ = 200 GeV}

\newcommand{\abilene}{Abilene Christian University, Abilene, TX 79699, USA}
\newcommand{\banaras}{Department of Physics, Banaras Hindu University, Varanasi 221005, India}
\newcommand{\bnl}{Brookhaven National Laboratory, Upton, NY 11973-5000, USA}
\newcommand{\caucr}{University of California - Riverside, Riverside, CA 92521, USA}
\newcommand{\charlesczech}{Charles University, Ovocn\'{y} trh 5, Praha 1, 116 36, Prague, Czech Republic}
\newcommand{\ciae}{China Institute of Atomic Energy (CIAE), Beijing, People's Republic of China}
\newcommand{\cns}{Center for Nuclear Study, Graduate School of Science, University of Tokyo, 7-3-1 Hongo, Bunkyo, Tokyo 113-0033, Japan}
\newcommand{\colorado}{University of Colorado, Boulder, CO 80309, USA}
\newcommand{\columbia}{Columbia University, New York, NY 10027 and Nevis Laboratories, Irvington, NY 10533, USA}
\newcommand{\czechtech}{Czech Technical University, Zikova 4, 166 36 Prague 6, Czech Republic}
\newcommand{\dapnia}{Dapnia, CEA Saclay, F-91191, Gif-sur-Yvette, France}
\newcommand{\debrecen}{Debrecen University, H-4010 Debrecen, Egyetem t{\'e}r 1, Hungary}
\newcommand{\elte}{ELTE, E{\"o}tv{\"o}s Lor{\'a}nd University, H - 1117 Budapest, P{\'a}zm{\'a}ny P. s. 1/A, Hungary}
\newcommand{\fit}{Florida Institute of Technology, Melbourne, FL 32901, USA}
\newcommand{\fsu}{Florida State University, Tallahassee, FL 32306, USA}
\newcommand{\gsu}{Georgia State University, Atlanta, GA 30303, USA}
\newcommand{\hiroshima}{Hiroshima University, Kagamiyama, Higashi-Hiroshima 739-8526, Japan}
\newcommand{\ihepprot}{IHEP Protvino, State Research Center of Russian Federation, Institute for High Energy Physics, Protvino, 142281, Russia}
\newcommand{\illuiuc}{University of Illinois at Urbana-Champaign, Urbana, IL 61801, USA}
\newcommand{\instpasczech}{Institute of Physics, Academy of Sciences of the Czech Republic, Na Slovance 2, 182 21 Prague 8, Czech Republic}
\newcommand{\isu}{Iowa State University, Ames, IA 50011, USA}
\newcommand{\jinrdubna}{Joint Institute for Nuclear Research, 141980 Dubna, Moscow Region, Russia}
\newcommand{\kaeri}{KAERI, Cyclotron Application Laboratory, Seoul, Korea}
\newcommand{\kek}{KEK, High Energy Accelerator Research Organization, Tsukuba, Ibaraki 305-0801, Japan}
\newcommand{\kfki}{KFKI Research Institute for Particle and Nuclear Physics of the Hungarian Academy of Sciences (MTA KFKI RMKI), H-1525 Budapest 114, POBox 49, Budapest, Hungary}
\newcommand{\korea}{Korea University, Seoul, 136-701, Korea}
\newcommand{\kurchatov}{Russian Research Center ``Kurchatov Institute", Moscow, Russia}
\newcommand{\kyoto}{Kyoto University, Kyoto 606-8502, Japan}
\newcommand{\labllr}{Laboratoire Leprince-Ringuet, Ecole Polytechnique, CNRS-IN2P3, Route de Saclay, F-91128, Palaiseau, France}
\newcommand{\lawllnl}{Lawrence Livermore National Laboratory, Livermore, CA 94550, USA}
\newcommand{\losalamos}{Los Alamos National Laboratory, Los Alamos, NM 87545, USA}
\newcommand{\lpc}{LPC, Universit{\'e} Blaise Pascal, CNRS-IN2P3, Clermont-Fd, 63177 Aubiere Cedex, France}
\newcommand{\lund}{Department of Physics, Lund University, Box 118, SE-221 00 Lund, Sweden}
\newcommand{\muenster}{Institut f\"ur Kernphysik, University of Muenster, D-48149 Muenster, Germany}
\newcommand{\myongji}{Myongji University, Yongin, Kyonggido 449-728, Korea}
\newcommand{\nagasaki}{Nagasaki Institute of Applied Science, Nagasaki-shi, Nagasaki 851-0193, Japan}
\newcommand{\newmex}{University of New Mexico, Albuquerque, NM 87131, USA }
\newcommand{\nmsu}{New Mexico State University, Las Cruces, NM 88003, USA}
\newcommand{\ornl}{Oak Ridge National Laboratory, Oak Ridge, TN 37831, USA}
\newcommand{\orsay}{IPN-Orsay, Universite Paris Sud, CNRS-IN2P3, BP1, F-91406, Orsay, France}
\newcommand{\peking}{Peking University, Beijing, People's Republic of China}
\newcommand{\pnpi}{PNPI, Petersburg Nuclear Physics Institute, Gatchina, Leningrad region, 188300, Russia}
\newcommand{\riken}{RIKEN, The Institute of Physical and Chemical Research, Wako, Saitama 351-0198, Japan}
\newcommand{\rikjrbrc}{RIKEN BNL Research Center, Brookhaven National Laboratory, Upton, NY 11973-5000, USA}
\newcommand{\rikkyo}{Physics Department, Rikkyo University, 3-34-1 Nishi-Ikebukuro, Toshima, Tokyo 171-8501, Japan}
\newcommand{\saispbstu}{Saint Petersburg State Polytechnic University, St. Petersburg, Russia}
\newcommand{\saopaulo}{Universidade de S{\~a}o Paulo, Instituto de F\'{\i}sica, Caixa Postal 66318, S{\~a}o Paulo CEP05315-970, Brazil}
\newcommand{\seoulnat}{System Electronics Laboratory, Seoul National University, Seoul, Korea}
\newcommand{\stonybrkc}{Chemistry Department, Stony Brook University, Stony Brook, SUNY, NY 11794-3400, USA}
\newcommand{\stonycrkp}{Department of Physics and Astronomy, Stony Brook University, SUNY, Stony Brook, NY 11794, USA}
\newcommand{\subatech}{SUBATECH (Ecole des Mines de Nantes, CNRS-IN2P3, Universit{\'e} de Nantes) BP 20722 - 44307, Nantes, France}
\newcommand{\tenn}{University of Tennessee, Knoxville, TN 37996, USA}
\newcommand{\titech}{Department of Physics, Tokyo Institute of Technology, Oh-okayama, Meguro, Tokyo 152-8551, Japan}
\newcommand{\tsukuba}{Institute of Physics, University of Tsukuba, Tsukuba, Ibaraki 305, Japan}
\newcommand{\vandy}{Vanderbilt University, Nashville, TN 37235, USA}
\newcommand{\waseda}{Waseda University, Advanced Research Institute for Science and Engineering, 17 Kikui-cho, Shinjuku-ku, Tokyo 162-0044, Japan}
\newcommand{\weizmann}{Weizmann Institute, Rehovot 76100, Israel}
\newcommand{\yonsei}{Yonsei University, IPAP, Seoul 120-749, Korea}
\affiliation{\abilene}
\affiliation{\banaras}
\affiliation{\bnl}
\affiliation{\caucr}
\affiliation{\charlesczech}
\affiliation{\ciae}
\affiliation{\cns}
\affiliation{\colorado}
\affiliation{\columbia}
\affiliation{\czechtech}
\affiliation{\dapnia}
\affiliation{\debrecen}
\affiliation{\elte}
\affiliation{\fit}
\affiliation{\fsu}
\affiliation{\gsu}
\affiliation{\hiroshima}
\affiliation{\ihepprot}
\affiliation{\illuiuc}
\affiliation{\instpasczech}
\affiliation{\isu}
\affiliation{\jinrdubna}
\affiliation{\kaeri}
\affiliation{\kek}
\affiliation{\kfki}
\affiliation{\korea}
\affiliation{\kurchatov}
\affiliation{\kyoto}
\affiliation{\labllr}
\affiliation{\lawllnl}
\affiliation{\losalamos}
\affiliation{\lpc}
\affiliation{\lund}
\affiliation{\muenster}
\affiliation{\myongji}
\affiliation{\nagasaki}
\affiliation{\newmex}
\affiliation{\nmsu}
\affiliation{\ornl}
\affiliation{\orsay}
\affiliation{\peking}
\affiliation{\pnpi}
\affiliation{\riken}
\affiliation{\rikjrbrc}
\affiliation{\rikkyo}
\affiliation{\saispbstu}
\affiliation{\saopaulo}
\affiliation{\seoulnat}
\affiliation{\stonybrkc}
\affiliation{\stonycrkp}
\affiliation{\subatech}
\affiliation{\tenn}
\affiliation{\titech}
\affiliation{\tsukuba}
\affiliation{\vandy}
\affiliation{\waseda}
\affiliation{\weizmann}
\affiliation{\yonsei}
\author{A.~Adare}	\affiliation{\colorado}
\author{S.~Afanasiev}	\affiliation{\jinrdubna}
\author{C.~Aidala}	\affiliation{\columbia}
\author{N.N.~Ajitanand}	\affiliation{\stonybrkc}
\author{Y.~Akiba}	\affiliation{\riken} \affiliation{\rikjrbrc}
\author{H.~Al-Bataineh}	\affiliation{\nmsu}
\author{J.~Alexander}	\affiliation{\stonybrkc}
\author{A.~Al-Jamel}	\affiliation{\nmsu}
\author{K.~Aoki}	\affiliation{\kyoto} \affiliation{\riken}
\author{L.~Aphecetche}	\affiliation{\subatech}
% ...
\collaboration{PHENIX Collaboration} \noaffiliation

\date{\today}

\begin{abstract}
Azimuthal angle ($\Delta\phi$) correlations are presented for a
broad range of transverse momentum ($0.4 < \pt < 10$~GeV/$c$) and
centrality (0-92\%) selections for charged hadrons from di-jets in
Au+Au collisions at $\sqrt{s_{\rm{NN}}}$ = 200 GeV. With increasing
\pt, the away-side $\Delta\phi$ distribution evolves from a broad
and relatively flat shape to a concave shape, then to a convex
shape. Comparisons to $p+p$ data suggest that the away-side
distribution can be divided into a partially suppressed ``head''
region centered at $\Delta\phi\sim\pi$, and an enhanced
``shoulder'' region centered at $\Delta\phi\sim \pi\pm1.1$. The \pt
spectrum for the associated hadrons in the head region softens
toward central collisions. The spectral slope for the shoulder
region is independent of centrality and trigger \pt. The properties
of the near-side distributions are also modified relative to those
in $p+p$ collisions, reflected by the broadening of the jet shape
in $\Delta\phi$ and $\Delta\eta$, and an enhancement of the
per-trigger yield. However, these modifications seem to be limited
to $p_{\rm T}\lesssim4$ GeV/$c$, above which both the hadron pair
shape and per-trigger yield become similar to $p+p$ collisions.
These observations suggest that both the away- and near-side
distributions contain a jet fragmentation component which dominates
for $p_{\rm T}\gtrsim5$ GeV and a medium-induced component which is
important for $p_{\rm T}\lesssim4$ GeV/$c$. We also quantify the
role of jets at intermediate and low \pt through the yield of
jet-induced pairs in comparison to binary scaled $p+p$ pair yield.
The yield of jet-induced pairs is suppressed at high pair proxy
energy (sum of the \pt magnitudes of the two hadrons) and is
enhanced at low pair proxy energy. The former is consistent with
jet quenching; the latter is consistent with the enhancement of
soft hadron pairs due to transport of lost energy to lower \pt.
\end{abstract}

\pacs{25.75.Dw}

\maketitle

\tableofcontents

\section{INTRODUCTION}

High transverse momentum (\pt) partons are informative probes of
the high energy density matter created in nuclear collisions at the
Relativistic Heavy-Ion Collider (RHIC). These partons lose a large
fraction of their energy in the matter prior to forming final state
hadrons. Such an energy loss is predicted to lead to a reduction of
both single hadron and correlated dihadron yields at high
\pt~\cite{Baier:1996sk,Gyulassy:2003mc,Kovner:2003zj}, a phenomenon
known as jet-quenching. Indeed, current results for high \pt have
revealed a strong suppression of inclusive hadron
yields~\cite{Adler:2003qi,Adler:2003au,Adams:2003kv}, as well as
the suppression of correlated away-side hadron
pairs~\cite{Adler:2002tq}.

Despite this strong suppression, particle production for $p_{\rm
T}\gtrsim 5$ GeV/$c$ appears to have a significant contribution
from in-vacuum jet fragmentation. This is suggested by a
\pt-independent suppression factor for single
hadrons~\cite{Adcox:2001jp, Adler:2003qi, Adler:2003au,
Adams:2003kv}, which implies a $p+p$-like power law spectral shape
in Au+Au collisions, and similar $\pi^0$ to $\eta$
meson~\cite{Adler:2006hu,Adler:2006bv} and proton to
pion~\cite{Adler:2003au,Abelev:2006jr} ratios between Au+Au and p+p
collisions. More direct evidence has been provided by high-\pt
dihadron azimuthal angle ($\Delta\phi$) correlations measurements.
In particular, our current measurements, as well as prior
ones~\cite{Adams:2006yt,Adare:2007vu}, reveal characteristic
jet-like peaks for the near-side ($\Delta\phi\sim0$) and the
away-side ($\Delta\phi\sim\pi$) at high \pt.

In most energy loss models, the stopping power of the medium is
normally characterized by the transport coefficient $\hat{q}$,
defined as the squared average momentum transfer from the medium to
the hard parton per unit path length. However, due to the steeply
falling parton spectra and strong jet quenching, the observed
high-\pt single hadrons and hadron pairs mainly come from (di)jets
that suffer minimal interaction with the medium. Thus, the overall
suppression factor is sensitive to the full energy loss probability
distribution instead of just the average energy loss itself. In
fact, simple calculations~\cite{Renk:2006pk} with different energy
loss probability distributions have been shown to match the data
quite well. However, the extracted $\langle\hat{q}\rangle$ values
are sensitive to the theoretical models and their associated
assumptions~\cite{Majumder:2007iu}. Additional experimental
constraints on the dynamics of the energy loss processes are
clearly needed.

In order to improve our understanding of the parton-medium
interactions, it is important to study the fate of partons that
suffer energy loss in the medium. These partons are quenched by the
medium and their energy is believed to be transported to lower-\pt
hadrons ($p_{\rm T}\lesssim4$ GeV/$c$). Prior
measurements~\cite{Adler:2002tq, Adams:2004pa, Adams:2005ph,
Adler:2005ee, Adare:2006nr, Adams:2006tj} in this \pt region, as
well as the present study, indicate strong modifications of the
near- and the away-side $\Delta\phi$ distributions. The near-side
jet-induced pairs peak at $\Delta\phi\sim0$, but the peak is
broadened and enhanced with respect to $p+p$ collisions. The
away-side jet-induced pairs are observed to peak at
$\Delta\phi\sim\pi\pm1.1$~\cite{Adler:2005ee, Adare:2006nr,
Adare:2007vu} with a local minimum at $\Delta\phi\sim\pi$. These
modification patterns reflect characteristics of the energy
transport of the quenched partons in both \pt and $\Delta\phi$.
Many mechanisms for this energy transport have been proposed for
the near-side~\cite{Armesto:2004pt, Voloshin:2004th,Chiu:2005ad,
Romatschke:2006bb,Majumder:2006wi,Shuryak:2007fu,Wong:2007pz,Pantuev:2007sh}
and away-side~\cite{Armesto:2004vz,Majumder:2006wi,
Romatschke:2006bb, Chiu:2006pu,
Armesto:2004pt,Vitev:2005yg,Polosa:2006hb,Dremin:1979yg,
Koch:2005sx,Stoecker:2004qu,Casalderrey-Solana:2004qm}.

Such energy transport is expected to enhance jet contributions to
the production of low-\pt hadrons. However, jet-induced hadron pair
correlations can be affected by soft processes such as
hydrodynamical flow~\cite{Kolb:2002ve} and quark
coalescence~\cite{Molnar:2003ff,Hwa:2004ng,
Fries:2003kq,Greco:2003xt}, which dominate the hadron production in
the intermediate \pt region. The coupling of partons with
hydrodynamical flow could modify the jet shape and yield.
Similarly, quark coalescence could modify the particle composition
in the near- and away-side
jets~\cite{Fries:2004hd,Afanasiev:2007wi}. Therefore, detailed
correlation studies for $p_{\rm T}\lesssim4$ GeV/$c$ can improve
our knowledge of the interplay between soft and hard processes for
hadron production.

In this paper we present a detailed survey of the trigger \pt,
partner \pt and centrality dependence of the near- and away-side
jet shapes and yields from Au+Au collisions. These measurements
provide a comprehensive overview of the different physical features
that come into play for different \pt ranges, and provide new
insights on the interplay between the processes leading to jet
energy loss and the response of the medium to the lost energy. In
addition, they allow a detailed study of the similarities and
differences between the correlation patterns for the near- and
away-side jets. When coupled with inclusive hadron production,
these measurements also allow quantification of the role of jets at
intermediate \pt, where the particle production is believed to be
dominated by the soft processes.

The results reported here comprise significant extensions to
results published earlier~\cite{Adare:2006nr,Adare:2007vu}. In
Section II, we introduce variables used to quantify the jet
properties and their in-medium modifications. In Section III, we
present data analysis details, jet signal extraction and background
subtraction, and several sources of systematic errors related to
the measurements. The main results are presented in Section IV and
model comparisons and discussions are given in Section V. Several
technical issues related to the correlation analysis are addressed
in Appendices A-C, and tabulated data are given in Appendix D.

\section{JET AZIMUTHAL CORRELATIONS}
\label{sec:2} The defining characteristic of a jet is the
collimated production of hadrons in the direction of a fragmenting
parton. Traditionally, such energetic jets have been identified
using standard jet reconstruction algorithms. However, direct jet
reconstruction in heavy ion collisions is difficult due to the
large amount of soft background. Measurements in a relatively
limited acceptance also pose additional challenges because of a
possible leakage of the jet fragments outside of the detectors
acceptance.

The two-particle (dihadron) relative azimuthal angle ($\Delta\phi$)
correlation technique provides an alternative approach for
accessing the properties of jets. Two classes of hadrons, trigger
hadrons (denoted as type {\em a}) and partner hadrons (denoted as
type {\em b}), typically from different \pt ranges, are correlated
with each other. Jet properties are extracted on a statistical
basis from the $\Delta\phi$ distribution built of many events. This
approach overcomes problems due to background and limited
acceptance, and allows the study of jets to be extended to low \pt
where soft processes dominate.

\begin{figure}[thb]
\includegraphics[width=1.0\linewidth]{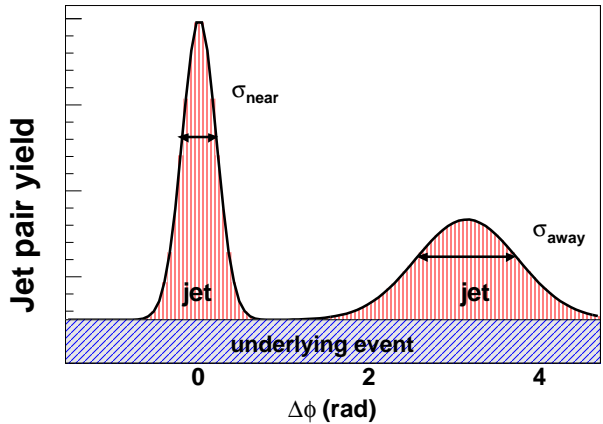}
\caption{\label{fig:cartoon} (Color online) Cartoon of hadron pair distribution in $\Delta\phi$ for $p+p$ collisions.
It has two peaks corresponding to near- and away-side jet, and a flat component representing the underlying event pairs.}
\end{figure}

To leading order in QCD, high-\pt jets are produced back-to-back in
azimuth. This back-to-back correlation is, however, smeared by the
fragmentation process and initial and final state radiation, to
give a characteristic $\Delta\phi$ distribution schematically shown
in Fig.~\ref{fig:cartoon}~\cite{Adler:2006sc}. Hadron pairs from
the same jet (near-side) dominates at $\Delta\phi=\phi^{\rm
a}-\phi^{\rm b}\sim0$ and those from back-to-back dijets
(away-side) tend to appear at $\Delta\phi\sim\pi$.

Two observables which are commonly exploited in dihadron
correlation studies are the hadron-pair yield (the rate of
jet-induced hadron pairs per-event, JPY) and the per-trigger yield
(jet-induced hadron-pair yield divided by trigger yield, $Y_{\rm
jet\_ind}$) in a given event sample. The former is related to the
two-particle cross-section for jet production; the latter is
related to the ratio of the two-particle to single-particle
cross-sections:
\begin{eqnarray}
\label{eq:jpy}
{\rm JPY}(p_{\rm T}^{\rm a},p_{\rm T}^{\rm b},\Delta\phi)&\equiv& \frac{1}{N_{\rm evts}}\frac{d^3N^{\rm{ab}}}{dp_{\rm T}^{\rm a}dp_{\rm T}^{\rm b}d\Delta\phi}\nonumber \\
&=&\frac{1}{\sigma_{_{\rm tot}}}\frac{d^3\sigma_{\rm{jet\_ind}}}{dp_{\rm T}^{\rm a}dp_{\rm T}^{\rm b}d\Delta\phi},\\
\label{eq:pty}
Y_{\rm jet\_ind}(p_{\rm T}^{\rm a},p_{\rm T}^{\rm b},\Delta\phi) &\equiv&
{{{\rm JPY}(p_{\rm T}^{\rm a},p_{\rm T}^{\rm b},\Delta\phi)}
\mathord{\left/
 {\vphantom { { {\rm JPY}(p_{\rm T}^{\rm a},p_{\rm T}^{\rm b},\Delta\phi)} {\frac{dN^{\rm a}}{N_{\rm evts}dp_{\rm T}^{\rm a}}}}} \right.
 \kern-\nulldelimiterspace} { \frac{dN^{\rm a}}{N_{\rm evts}dp_{\rm T}^{\rm a}}}}\nonumber \\
&=& {{\frac{d^3\sigma_{\rm{jet\_ind}}}{dp_{\rm T}^{\rm a}dp_{\rm T}^{\rm b}d\Delta\phi} }
\mathord{\left/
 {\vphantom { { \frac{d^3\sigma_{\rm{jet\_ind}}}{dp_{\rm T}^{\rm a}dp_{\rm T}^{\rm b}d\Delta\phi} } { \frac{d\sigma}{dp_{\rm T}^{\rm a}} }}} \right.
 \kern-\nulldelimiterspace} { \frac{d\sigma}{dp_{\rm T}^{\rm a}} } },
\end{eqnarray}
where $N^{\rm{ab}}$ is the number of jet-induced hadron pairs,
$N_{\rm evts}$ is the number of events and $\sigma_{\rm tot}$ is
the semi-inclusive cross-section for that event sample. Thus, JPY
is simply the product of the per-trigger yield and the number of
triggers. To first order, the two-particle cross-section for the
near-side jet is governed by the dihadron fragmentation function.
By contrast, the cross-section for the away-side jet is governed by
two independent fragmentation functions, i.e one parton produces a
hadron with $p_{\rm T}^{\rm a}$ and the other scattered parton
produces a hadron with $p_{\rm T}^{\rm b}$.

In A+A collisions, the single and dihadron cross-sections can be
modified by the medium. This modification can be quantified by
comparing the yield in A+A collisions to that for $p+p$ collisions.
Thus, modification to the single hadron cross-section is
characterized by the nuclear modification factor, $R_{\rm AA}$
\begin{eqnarray}
R_{\rm{AA}}(p_{\rm T}) =\frac{1/\sigma_{_{\rm
A+A}}d\sigma^{\rm A+A}/dp_{\rm T}}{\langle N_{\rm{coll}}\rangle/\sigma_{_{\rm
p+p}}d\sigma^{p+p}/dp_{\rm T}}
\end{eqnarray}
where  $\sigma_{_{\rm A+A}}$ and $\sigma_{_{\rm p+p}}$ are the
semi-inclusive cross-section in A+A and $p+p$ collisions,
respectively. $\langle N_{\rm{coll}}\rangle$ is the average number
of binary collisions for a given centrality selection in A+A
collisions. Modification of the dihadron cross-section can be
characterized by $J_{\rm{AA}}$, which is defined as,
\begin{eqnarray}
\label{eq:jaa} &&J_{\rm{AA}}(p_{\rm T}^{\rm a},p_{\rm T}^{\rm b},\Delta\phi) =
\frac{\rm{JPY}^{\rm{A+A}}}{\langle N_{\rm{coll}}\rangle\;
\rm{JPY}^{p+p}}\\\nonumber &&={\frac{1}{\sigma_{\rm
A+A}}\frac{d^3\sigma_{\rm
jet\_ind}^{\rm{A+A}}}{dp_{\rm T}^{\rm a}dp_{\rm T}^{\rm b}d\Delta\phi}} \mathord{\left/
 {\vphantom {\frac{d^3\sigma_{\rm{jet\_ind}}}{dp_{\rm T}^{\rm a}dp_{\rm T}^{\rm b}d\Delta\phi} }} \right.
 \kern-\nulldelimiterspace}
{\frac{\langle N_{\rm{coll}}\rangle}{\sigma_{ p+p}}\frac{d^3\sigma_{\rm
jet\_ind}^{p+p}}{dp_{\rm T}^{\rm a}dp_{\rm T}^{\rm b}d\Delta\phi}} \label{eq:6}
\end{eqnarray}
In the absence of nuclear effects, both the single and dihadron
cross-sections from jets are expected to scale with $\langle N_{\rm
coll}\rangle$. Therefore, $R_{\rm AA}$ and $J_{\rm AA}$ should be
equal to unity.

The medium modifications of jets are also characterized by the
per-trigger yield and its corresponding modification factor,
$I_{\rm AA}$.
\begin{equation} \label{eq:iaadef}
I_{\rm{AA}}(p_{\rm T}^{\rm a},p_{\rm T}^{\rm b}) = \frac{Y_{\rm
jet\_ind}^{\rm A+A}(p_{\rm T}^{\rm a},p_{\rm T}^{\rm b})}{Y_{\rm{jet\_ind}}^{p+p}(p_{\rm T}^{\rm a},p_{\rm T}^{\rm b})}.
\end{equation}
In general, the value of $I_{\rm AA}$ depends on modifications to
both the hadron-pair yield and the trigger yield. For high-\pt
correlation measurements, the per-trigger yield is a convenient
choice since each jet typically produces at most one high-\pt
trigger. Because of the steeply falling parton spectrum, the
probability of having a high-\pt parton that produces multiple
trigger hadrons is small. Thus the per-trigger yield effectively
represents the per-jet yield in $p+p$ collisions, and $I_{\rm AA}$
represents the modification of the partner yield per-jet. For
intermediate and low \pt, however, jet fragmentation is not the
only source of triggers, and this can lead to an artificial
reduction of the per-trigger yield (see discussion in
Section~IV.E). For such situations, $J_{\rm AA}$ is a more robust
variable for correlation analysis since it is only sensitive to the
modification of jet-induced hadron pairs.

JPY and $J_{\rm AA}$ are symmetric with respect to $p_{\rm T}^{\rm
a}$ and $p_{\rm T}^{\rm b}$. By contrast, the per-trigger yield and
$I_{\rm AA}$ are not, due to the appearance of the normalization
factor $N^{\rm a}$ in Eq.~\ref{eq:pty}. This normalization factor
is the only distinction between triggers and the partners in this
analysis. In addition, JPY can be expressed in terms of the
per-trigger yield and the inclusive yield as
\begin{eqnarray}
\label{eq:ptyiy}
{\rm JPY}(p_{\rm T}^{\rm a},p_{\rm T}^{\rm b}) &=&
Y_{\rm{jet\_ind}}(p_{\rm T}^{\rm a},p_{\rm T}^{\rm b})\frac{dN^{\rm a}}{N_{\rm evts}dp_{\rm T}^{\rm a}}\\\nonumber
&=&Y_{\rm{jet\_ind}}(p_{\rm T}^{\rm b},p_{\rm T}^{\rm a})\frac{dN^{\rm b}}{N_{\rm evts}dp_{\rm T}^{\rm b}}\\\nonumber
\end{eqnarray}
Similarly, $J_{\rm AA}$ can be expressed in terms of $R_{\rm AA}$
and $I_{\rm AA}$ as
\begin{eqnarray}
\label{eq:iaajaa}
J_{\rm{AA}}(p_{\rm T}^{\rm a},p_{\rm T}^{\rm b}) &=&
I_{\rm{AA}}(p_{\rm T}^{\rm a},p_{\rm T}^{\rm b})R_{\rm{AA}}(p_{\rm T}^{\rm a})\\\nonumber
&=&I_{\rm{AA}}(p_{\rm T}^{\rm b},p_{\rm T}^{\rm a})R_{\rm{AA}}(p_{\rm T}^{\rm b})
\end{eqnarray}
Thus, $I_{\rm AA}(p_{\rm T}^{\rm b},p_{\rm T}^{\rm a})$ can be
calculated from $I_{\rm AA}(p_{\rm T}^{\rm a},p_{\rm T}^{\rm b})$,
$R_{\rm{AA}}(p_{\rm T}^{\rm a})$ and $R_{\rm{AA}}(p_{\rm T}^{\rm
b})$.

In the current analysis, the in-medium modifications of the jet
shape and yield are characterized via comparisons of the
per-trigger yield and hadron-pair yield in Au+Au and $p+p$
collisions i.e, via $I_{\rm AA}$ and $J_{\rm AA}$. As discussed
earlier, these quantities are defined in their differential form in
$\Delta\phi$, $p_{\rm T}^{\rm a}$ and $p_{\rm T}^{\rm b}$.
Operationally, this means that the hadron-pair yields and the
per-trigger yields are measured in a finite $p_{\rm T}$ range
and/or integrated over a limited $\Delta\phi$ range. $I_{\rm AA}$
and $J_{\rm AA}$ are then obtained from these integrated yields.

\section{EXPERIMENTAL ANALYSIS}
\label{sec:3}
\subsection{Dataset and Centrality}
\label{sec:3.1} The results presented in this article are based on
three datasets collected with the PHENIX
detector~\cite{Adcox:2003zm} at $\sqrt{s_{\rm{NN}}}$=200 GeV,
during the 2004-2005 RHIC running periods. The first is comprised
of a minimum-bias (MB) Au+Au dataset triggered by the Beam-Beam
Counters (BBC) and the Zero-Degree Calorimeters (ZDC) and taken in
2004. The second is a MB $p+p$ dataset triggered by the BBC and
taken in 2005, and the third is a level-1 triggered (LVL1) $p+p$
dataset also obtained in 2005. The level-1 trigger requirement is
an energy threshold of 1.4 GeV in 4$\times$4 electromagnetic
calorimeter (EMC) towers in coincidence with the BBC
trigger~\cite{Adare:2006hc}. The MB and LVL1 $p+p$ datasets serve
as baseline measurements for the Au+Au dataset; they are used to
select triggers for $p_{\rm T}<5$ and $p_{\rm T}>$ 5 GeV/$c$,
respectively.

The collision vertex along the beam direction, $z$, was measured by
the BBCs. After an offline vertex cut of $|z| <$~30 cm and
selecting good runs, a total of 840 million or 136$\mu$b$^{-1}$
Au+Au events were obtained. This is a factor x30 higher than
obtained in a previous analysis~\cite{Adler:2005ee}. The total
statistics for MB $p+p$ and LVL1 $p+p$ datasets are equivalent to
73nb$^{-1}$ and 2.5pb$^{-1}$ sampled luminosities respectively.

The event centrality was determined via cuts in the space of BBC
charge versus ZDC energy~\cite{Adcox:2003zp}. The efficiency of the
MB triggered events is estimated to be $92.2^{+2.5}_{-3.0}\%$ of
the total Au+Au inelastic cross section (6.9 barn)
\cite{Adler:2003au}. To optimize the \pt reach of our results,
relatively {\it coarse} centrality selections of 0-20\%, 20-40\%,
40-60\%, 60-92.2\% were chosen. However for $p_{\rm T}<4$ GeV/$c$,
excellent statistical significance of the measurements allows the
results to be presented in {\it fine} centrality selections of
0-5\%, 5-10\%, 10-20\%, 20-30\%, 30-40\%, 40-50\%, 50-60\%,
60-70\%, 70-92.2\%.

A Glauber model Monte-Carlo simulation~\cite{Adcox:2000sp,
Adcox:2003nr} that includes the responses of the BBC and ZDC was
used to estimate the average number of binary collisions $\langle
N_{\rm{coll}}\rangle$, and participating nucleons $\langle
N_{\rm{part}}\rangle$ for each centrality class. These values are
listed in Table.~\ref{tab:sys0}.
\begin{table}
\caption{\label{tab:sys0} Average number of nucleon-nucleon
collisions $\langle N_{\rm{coll}}\rangle$ and  participant nucleons
$\langle N_{\rm{part}}\rangle$ for several centrality classes.
$\langle N_{\rm{coll}}\rangle$ and $\langle N_{\rm{part}}\rangle$
are obtained from a Glauber Monte-Carlo simulation of the response
of the BBC and ZDC in Au+Au collisions at $\sqrt{s_{\rm{NN}}}$ =
200 GeV. The errors for these centrality classes are correlated.}
\begin{ruledtabular} \begin{tabular}{ccc}
Centrality     &   $\langle N_{\rm{coll}}\rangle$  &$\langle N_{\rm{part}}\rangle$       \\ \hline
 0 - 5\%       &       $1065  \pm 105.5$       &       $351.4 \pm 2.9$\\
 5 - 10\%      &       $854.4 \pm 82.1$        &       $ 299  \pm 3.8$\\
10 - 20\%      &       $602.6 \pm 59.3$        &       $234.6 \pm 4.7$\\
20 - 30\%      &       $373.8 \pm 39.6$        &       $166.6 \pm 5.4$\\
30 - 40\%      &       $219.8 \pm 22.6$        &       $114.2 \pm 4.4$\\
40 - 50\%      &       $120.3 \pm 13.7$        &       $ 74.4 \pm 3.8$\\
50 - 60\%      &       $ 61.0 \pm 9.9$         &       $ 45.5 \pm 3.3$\\
60 - 70\%      &       $ 28.5 \pm 7.6$         &       $ 25.7 \pm 3.8$\\
70 - 92\%      &       $ 8.3  \pm 2.4$         &       $ 9.5  \pm 1.9$\\
min. bias      &       $257.8 \pm 25.4$        &       $109.1 \pm 4.1$\\\hline
0 - 20\%       &       $779   \pm 75.2$        &       $279.9 \pm 4.0$\\
20 - 40\%      &       $297   \pm 30.8$        &       $140.4 \pm 4.9$\\
40 - 60\%      &       $90.6  \pm 11.8$        &       $60    \pm 3.5$\\
60 - 92\%      &       $ 14.5 \pm 4$           &       $ 14.5 \pm 2.5$\\
\end{tabular}  \end{ruledtabular}
\end{table}
\subsection{Tracking and Background Estimation}
\label{sec:3.2}

Charged hadrons were reconstructed in the two central arms of
PHENIX, each covering -0.35 to 0.35 in pseudo-rapidity and
$90^{\circ}$ in azimuth. Tracks were measured outside the PHENIX
central magnetic field by the drift chambers, located at a radius
of 2.0~m from the vertex, and two layers of multi-wire proportional
chamber (PC1 and PC3), located 2.5 and 5.0~m, respectively, from
the vertex~\cite{Adcox:2003zp}. The momentum resolution was
determined to be $0.7\% \bigoplus
1.0\%p$~(GeV/$c$)~\cite{Adler:2003au}.

To reduce background particles that do not originate from the event
vertex, such as weak decays and conversion electrons, tracks were
required to have a matching hit within a $\pm 2.3 \sigma$ window in
PC3. For $p_{\rm T}>4$ GeV/$c$, an additional matching hit at the
EMC was required to suppress background tracks that randomly
associate with a hit in PC3~\cite{Adler:2003au}. For triggers with
$p_{\rm T}>5$ GeV/$c$, a \pt-dependent energy cut in the EMC and a
tight $\pm1.5\sigma$ matching cut at the PC3 were applied to reduce
the physical background from post-field photon conversions and weak
decays to a level $<10$\% of real tracks~\cite{Adler:2005ad}. This
energy cut helps to suppress any level-1 bias for the LVL1 $p+p$
dataset. We checked the consistency between the MB and LVL1 $p+p$
dataset for triggers with $p_{\rm T}>5$ GeV/$c$ by performing the
same analysis separately on the two $p+p$ datasets. Any remaining
biases due to level-1 trigger selection were found to be within the
quoted errors. With these cuts, the background level for triggers
was estimated to be $\alt 5$\% for $p_{\rm T} \alt 3$~GeV/$c$, and
increases to $\sim 10$\% for $p_{\rm T}>4$
GeV/$c$~\cite{Adler:2003au}. A \pt-dependent correction to the
per-trigger yield was used to account for this background.

For partner hadrons, the same matching cuts used for trigger
hadrons were applied. However, the \pt-dependent energy cut for
$5-10$ GeV/$c$ partners was found to be unnecessary. The jet
associated partner charged hadron spectrum is much flatter than
that for the trigger hadrons. Thus, the background contamination of
partners is much reduced relative to that for the trigger hadrons.
In addition, the background tracks contributing to combinatoric
pairs do not affect the jet signal and can be subtracted out.

A full GEANT simulation of PYTHIA jet events in the PHENIX
detector, as detailed in~\cite{Adler:2005ad}, was used to evaluate
the effects of this high-\pt background for partner hadrons.
Partner yields were extracted following the same procedures used
for the actual data analysis. These were then compared to the input
partner hadron spectra. For high-\pt triggered events, the
background contributions to the partner hadrons were found to be
less than 10\% for integrated partner yields in the 5-10 GeV/$c$
cut.

The single particle efficiency for triggers and partners,
$\epsilon^{\rm a}$ and $\epsilon^{\rm b}$ in
Eq.~\ref{eq:jetpairyield} and Eq.~\ref{eq:jetyield}, were
determined such that the single unidentified hadron \pt spectra
reproduce the previously published data for
Au+Au~\cite{Adler:2003au} and $p+p$~\cite{Adler:2005in}. It
includes detector acceptance, reconstruction efficiency, occupancy,
and background~\cite{Adler:2003au}. The detector acceptance and
reconstruction efficiency were estimated with a Monte-Carlo
simulation in which simulated single tracks were reconstructed in
the PHENIX detector, using the same analysis chain employed for the
real data. The efficiency loss due to detector occupancy in Au+Au
collisions was estimated by reconstructing simulated single tracks
embedded into real events. More details can be found in Au +
Au~\cite{Adler:2003au} and $p+p$~\cite{Adler:2005in} analyses.

\subsection{Jet Signal Extraction}
The dihadron correlation technique is commonly employed in PHENIX
for jet measurements because it surmounts the challenges posed by
the detector's limited azimuthal acceptance for single hadrons.
Even so, physical correlations due to anisotropic production of
hadrons relative to the reaction plane in Au+Au collisions, i.e the
elliptic flow, need to be distinguished from the jet correlations.
In what follows, we layout the framework for our correlation
analysis and an associated decomposition procedure used to separate
the elliptic flow and jet correlation contributions.

We define the azimuthal correlation function as
\begin{eqnarray}
\label{eq:jet1} C(\Delta\phi)
\equiv\frac{N^{\rm{same}}\left(\Delta\phi\right)}{N^{\rm{mixed}}\left(\Delta\phi\right)}
\end{eqnarray}
where $N^{\rm{same}}(\Delta\phi)$ and $N^{\rm{mixed}}(\Delta\phi)$
are pair distributions from the same- and mixed-events,
respectively. Each mixed-event is constructed by combining triggers
from a real event with partners from a different, randomly selected
event with similar centrality and collision vertex as the real
event.

The shape of the mixed-event pair distribution reflects the pair
$\Delta\phi$ acceptance of PHENIX detector, but it does not contain
physical correlations. The integral of mixed-event pairs reflects
the rate of the combinatoric pairs,
\begin{eqnarray}
\label{eq:jet2} \int d\Delta\phi N^{\rm mixed}(\Delta\phi) =
N_{\rm{evts}}\langle n^{\rm a}\rangle\langle n^{\rm b}\rangle
\end{eqnarray}
where $N_{\rm{evts}}$ is the number of events and $\langle n^{\rm
a}\rangle$, $\langle n^{\rm b}\rangle$ represent the average number
of triggers and partners per-event in the PHENIX acceptance. Both
$N^{\rm same}(\Delta\phi)$ and $N^{\rm{mixed}}(\Delta\phi)$ are
affected by the pair efficiency, which cancels out in the ratio
(see Appendix A). Therefore, the correlation function
Eq.~\ref{eq:jet1} contains only physical correlations.

The elliptic flow correlation leads to a harmonic modulation of the
combinatoric pair distribution by a factor that is proportional to
$(1+2v_2^{\rm a}v_2^{\rm b}\cos2\Delta\phi)$, where $v_2^{\rm a}$
and $v_2^{\rm b}$ are the average elliptic flow values for triggers
and partners respectively. To extract the jet-induced pairs, we
follow a two-source ansatz where each particle is assumed to come
from a jet-induced source and an underlying event containing
elliptic flow. The pair distribution can be expressed as
\begin{eqnarray}
\label{eq:jet2c}
\nonumber N^{\rm{same}}\left(\Delta\phi\right) &=& \xi(1+2v_2^{\rm a}v_2^{\rm b}\cos2\Delta\phi) N^{\rm{mixed}}\left(\Delta\phi\right) \\
&& + {\rm Jet}(\Delta\phi)
\end{eqnarray}
Where the Jet($\Delta\phi$) represents all pairs from (di-)jets.
The integral of $N^{\rm{same}}$ can be written as,
\begin{eqnarray}
\label{eq:jet2b} \int d\Delta\phi N^{\rm same}(\Delta\phi) =
N_{\rm{evts}}\langle n^{\rm a} n^{\rm b}\rangle + \int d\Delta\phi {\rm Jet}(\Delta\phi)\nonumber\\
\end{eqnarray}
Comparing to Eq.~\ref{eq:jet2}, we obtain
\begin{eqnarray}
\label{eq:xi}\xi = \frac{\langle n^{\rm a}n^{\rm
b}\rangle}{\langle n^{\rm a}\rangle\langle n^{\rm b}\rangle} .
\end{eqnarray}
Thus, $\xi$ is simply the ratio of the trigger-partner combinatoric
rate in the same-event to that in mixed-events, which can be bigger
than one due to centrality smearing (see discussion in
Section~\ref{sec:3.4}). An alternative approach used to fix $\xi$
is to assume that the jet function has zero yield at its minimum
$\Delta\phi_{\rm min}$ (ZYAM) \cite{Ajitanand:2005jj,Adler:2005ee}.

Finally, the ratio of jet-induced pairs to combinatoric pairs from
mixed events, JPR (jet-induced hadron-pair ratio) is given by,
\begin{eqnarray} \label{eq:jet3}
\rm{JPR}(\Delta\phi)&\equiv&\frac{{\rm Jet}(\Delta\phi)}{N^{\rm{mixed}}\left(\Delta\phi\right)}\\\nonumber
&=&\frac{N^{\rm{same}}\left(\Delta\phi\right)}{N^{\rm{mixed}}\left(\Delta\phi\right)}-\xi(1+2v_2^{\rm a}v_2^{\rm b}\cos2\Delta\phi)
\end{eqnarray}

\begin{figure}[thb]
\includegraphics[width=1.0\linewidth]{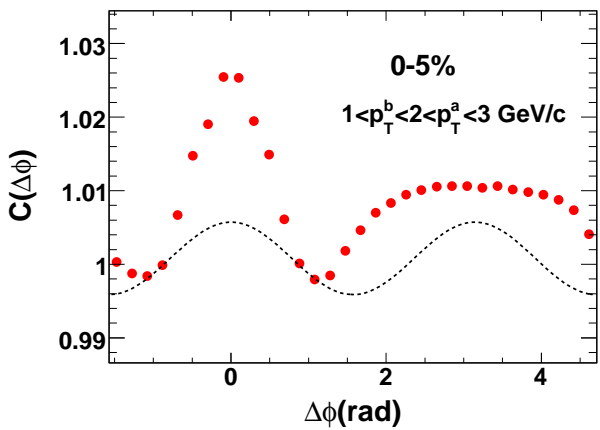}
\caption{\label{fig:ep} (Color online) The correlation function for $2<p_{\rm T}^{\rm a}<3$,
$1<p_{\rm T}^{\rm b}<2$ GeV/$c$ in 0-5\% Au+Au collisions. The dashed line represents the estimated elliptic flow modulated combinatoric background using zero yield at minimum (ZYAM) method
(see Section.\ref{sec:3.4}).}
\end{figure}

A representative correlation function is given in Fig.~\ref{fig:ep}
for 0-5\% Au+Au collisions and for triggers and partners in 2-3 and
1-2 GeV/$c$, respectively. It shows a peak around $\Delta\phi\sim0$
and a broad structure around $\Delta\phi\sim\pi$. The dashed line
indicates the estimated elliptic flow modulated background via ZYAM
method. The area between the data points and the dashed line
reflects the jet-induced pair ratio. It is only a few percent
relative to the background level.

We define $\varepsilon^{\rm a}, \varepsilon^{\rm b}$ as the single
particle efficiency within the PHENIX pseudo-rapidity acceptance
($|\eta|<0.35$). The true numbers of triggers and partners are
given by
\begin{eqnarray} \label{eq:jet4}
\langle n^{\rm a}_0 \rangle =\langle n^{\rm
a}\rangle/\varepsilon^{\rm a}\; ; \; \langle n^{\rm b}_0\rangle =
\langle n^{\rm b}\rangle/\varepsilon^{\rm b}
\end{eqnarray}
For uncorrelated sources, the triggers and partners are uniform in
azimuth. Thus the true combinatoric pair distribution for mixed
events is flat with $\Delta\phi$ with a density of $\langle n^{\rm
a}_0\rangle \langle n^{\rm b}_0 \rangle/(2\pi)$. The yield of
jet-induced pairs per event, JPY, is given as the product of the
combinatoric pair rate and the hadron-pair ratio,
\begin{eqnarray}
\label{eq:jetpairyield} \nonumber &&\rm{JPY}(\Delta\phi) =
\frac{\langle n^{\rm a}_0\rangle\langle n^{\rm b}_0\rangle}{2\pi}
\rm{JPR}(\Delta\phi)= \frac{\langle n^{\rm a}\rangle\langle n^{\rm
b}\rangle}{2\pi\varepsilon^{\rm a}\varepsilon^{\rm
b}}\rm{JPR}(\Delta\phi)\\\nonumber&&= \frac{\int d\Delta\phi
N^{\rm{mixed}}(\Delta\phi)}{2\pi N_{\rm{evts}}\varepsilon^{\rm a}
\varepsilon^{\rm
b}}\\&&\times\left[\frac{N^{\rm{same}}\left(\Delta\phi\right)}{N^{\rm{mixed}}\left(\Delta\phi\right)}-\xi(1+2v_2^{\rm
a}v_2^{\rm b}\cos2\Delta\phi)\right]
\end{eqnarray}

Thus far, we have not made any distinction between trigger and
partner hadrons. As discussed earlier in Section~\ref{sec:2}, the
correlation function, hadron-pair ratio and hadron-pair yield are
symmetric between the trigger and partner \pt, i.e.
\begin{eqnarray}
C(p_{\rm T}^{\rm a}, p_{\rm T}^{\rm b}) &=& C(p_{\rm T}^{\rm b},p_{\rm T}^{\rm a})\quad,\nonumber\\
{\rm JPR}(p_{\rm T}^{\rm a},p_{\rm T}^{\rm b}) &=& {\rm JPR}(p_{\rm T}^{\rm b}, p_{\rm T}^{\rm a})\quad,\quad
{\rm and}\nonumber\\ {\rm JPY}(p_{\rm T}^{\rm a}, p_{\rm T}^{\rm b}) &=& {\rm JPY}(p_{\rm T}^{\rm b},
p_{\rm T}^{\rm a}).
\end{eqnarray}

The associated partner yield per trigger,
$Y_{\rm{jet\_ind}}\left(\Delta\phi\right)$ is obtained by dividing
the hadron-pair yield per event with the number of triggers per
event,
\begin{eqnarray}\label{eq:jetyield}
\nonumber &&Y_{\rm{jet\_ind}}(\Delta\phi) =
\frac{\rm{JPY}(\Delta\phi)}{n^{\rm a}_0}= \frac{\int
d\Delta\phi N^{\rm{mixed}}(\Delta\phi)}{2\pi N^{\rm a}
\varepsilon^{\rm b}}\\ && \;\;\;\;\;\times\left[\frac{N^{\rm{same}}\left(\Delta\phi\right)}{N^{\rm{mixed}}\left(\Delta\phi\right)}-\xi(1+2v_2^{\rm a}v_2^{\rm b}\cos2\Delta\phi)\right]
\end{eqnarray}
$Y_{\rm jet\_ind}$ is often referred to as the per-trigger yield or
conditional yield. It is clearly not invariant to the exchange of
trigger and partner \pt.

The analysis proceeds in the following steps. We first measure the
correlation function Eq.~\ref{eq:jet1}. We then obtain the
efficiency for partner hadrons ($\varepsilon^{\rm b}$) and the
elliptic flow coefficients for the two hadron categories ($v_2^{\rm
a}$,$v_2^{\rm b}$). We then determine the background level ($\xi$)
via ZYAM background subtraction method (see Section.\ref{sec:3.4}),
followed by a calculation of the per-trigger yield according to
Eq.~\ref{eq:jetyield}. Subsequently, we obtain the hadron-pair
yield by multiplying the per-trigger yield with the inclusive
charged hadron yield~\cite{Adler:2003au} integrated in the
corresponding trigger \pt range.

\begin{figure}[thb]
\includegraphics[width=1.0\linewidth]{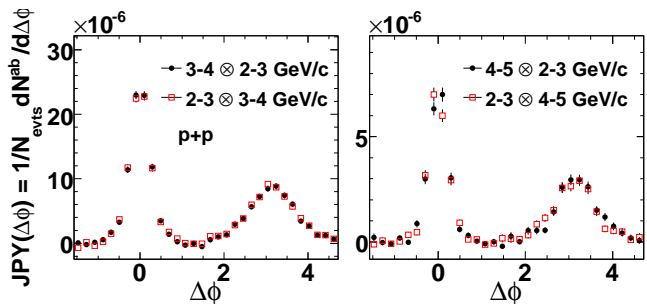}
\caption{\label{fig:checkpaper18} (Color online) The $p+p$ jet-induced hadron-pair yield $\Delta\phi$ distributions calculated from
the per-trigger yield using low-\pt hadrons as triggers (solid symbols) and high-\pt hadrons as triggers (open symbols).}
\end{figure}

\begin{figure}[thb]
\includegraphics[width=1.0\linewidth]{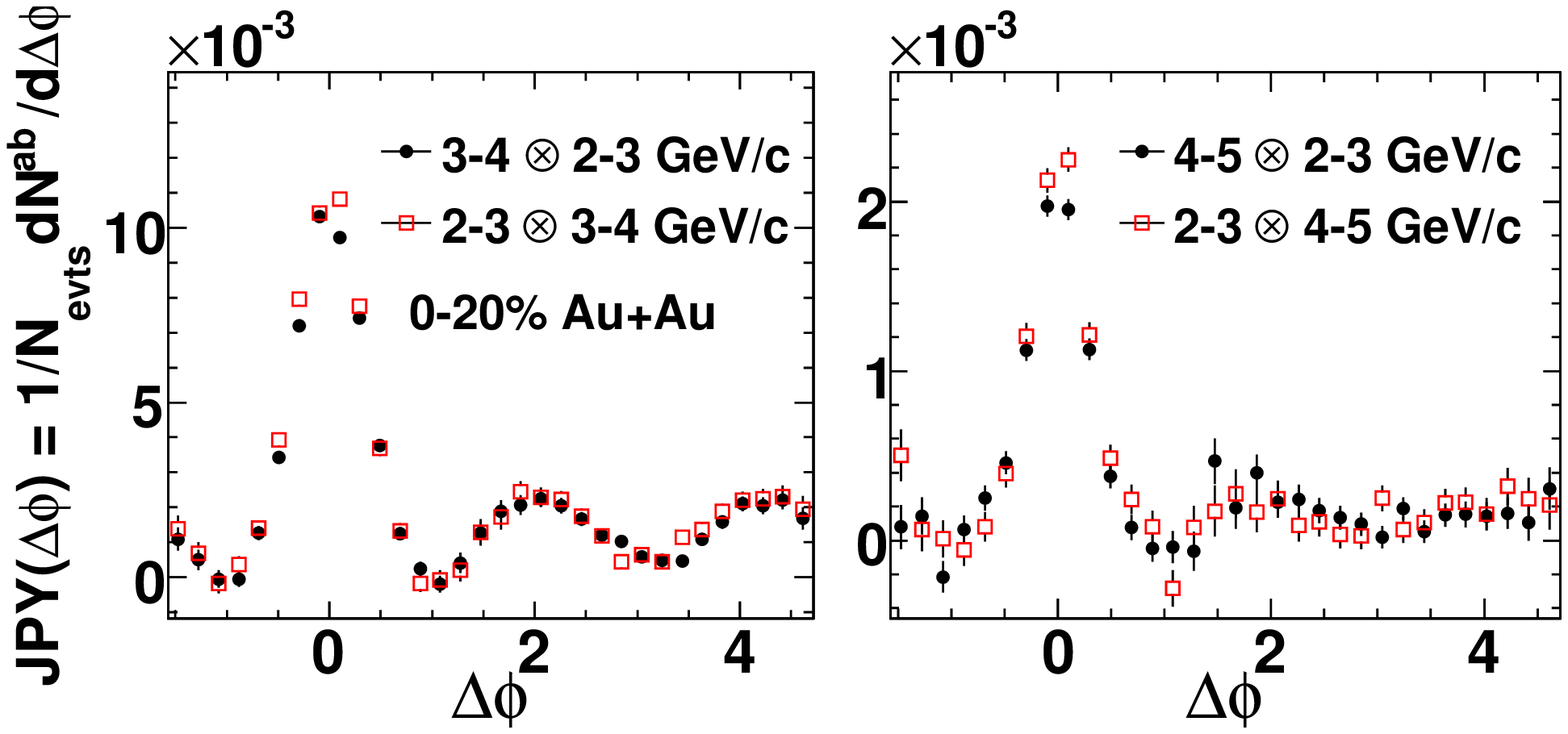}
\caption{\label{fig:checkpaper10} (Color online) The 0-20\% Au+Au jet-induced hadron-pair yield $\Delta\phi$ distributions calculated from
the per-trigger yield using low-\pt hadrons as triggers (solid symbols) and high-\pt hadrons as triggers (open symbols).}
\end{figure}

According to Eq.~\ref{eq:ptyiy}, the hadron-pair yields calculated
from the per-trigger yields are independent of which hadron, from
the pair, is used as trigger. We have used this fact to cross check
the efficacy of our analysis.   Figure~\ref{fig:checkpaper18} compares
the hadron-pair yields obtained when the trigger and partner \pt is
exchanged in $p+p$ collisions (in $p_{\rm T}^{a}\otimes p_{\rm
T}^{b}$). The open symbols indicate the results for low-\pt trigger
hadrons in association with high-\pt partners. The filled symbols
show the converse. A similar comparison for 0-20\% Au+Au collisions
is shown in Fig.~\ref{fig:checkpaper10}. Overall good agreement is
indicated by these distributions. It is important to emphasize here
that there is no a priori reason for these distributions to be
identical, since the cuts on trigger and partner hadrons are a
little different (cf. Section.~\ref{sec:3.2}) and therefore could
lead to somewhat different systematic errors for each measurement.


\subsection{Elliptic Flow Measurement}
\label{sec:3.3}

The differential elliptic flow measurements for charged hadrons
were carried out with the reaction plane
method~\cite{Poskanzer:1998yz}. The event plane (EP), which is the
experimental estimate of the reaction plane (RP), is determined via
the two BBCs positioned symmetrically along the beam line. They
cover full azimuth and $3 < |\eta| < 3.9$ in pseudo-rapidity. The
BBCs allow an unbiased measurement of the event plane, and ensures
that there are no residual distortions on the correlation function
that could result from the limited azimuthal coverage of PHENIX
central arms. A detailed analytical proof of this latter point is
provided in Appendix B.

We determine the value of elliptic flow, $v_2$, as
\begin{eqnarray}
v_{2} = \frac{v_{2,\rm{raw}}}{c_{v_2}}
=\frac{\langle\langle\cos2(\phi-\Phi_{\rm{EP}})\rangle\rangle}{\langle\cos2(\Phi_{\rm{EP}}-\Phi_{\rm{RP}})\rangle}
\end{eqnarray}
where $\Phi_{\rm{EP}}$ is the event plane angle and
$\Phi_{\rm{RP}}$ is the true reaction plane angle, $v_{2,\rm{raw}}
= \langle\langle\cos2(\phi-\phi_{\rm{EP}})\rangle\rangle$ is the
raw $v_{2}$ and $c_{v_2} =
\langle\cos2(\Phi_{\rm{EP}}-\Phi_{\rm{RP}})\rangle$ is the
estimated reaction plane resolution. The former is obtained by
averaging over all tracks and all events, the latter is obtained by
averaging over all events. The resolution is estimated from the
event plane angle of the north and south BBC as $c_{v_2} =
\sqrt{2\langle\cos2(\Phi_{\rm{EP,North}}-
\Phi_{\rm{EP,South}})}$~\cite{Adler:2003kt,Adler:2005ab}. It is 0.3
for minimum bias events, and reaches a maximum of 0.42 in the
20-30\% centrality bin. Further details are given
in~\cite{Adler:2005ab}.

\begin{figure}[thb]
\includegraphics[width=1.0\linewidth]{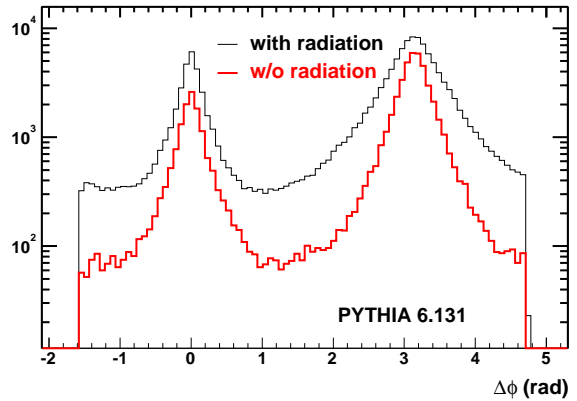}
\caption{\label{fig:rad} (Color online) PYTHIA simulation showing jet-induced hadron pair $\Delta\phi$
distribution for $3<p_{T}^{\rm a}, p_{\rm T}^{\rm b}<5$ GeV/$c$ with (top histogram) and
without (bottom histogram) initial and final state radiation. The radiation
accounts for the increase of the background level.}
\end{figure}

Reliable extraction of the jet signal requires accurate
determination of $v_2^{\rm a}$ and $v_2^{\rm b}$.  To this end,
non-flow effects that lead to azimuthal correlations unrelated to
the true RP direction, need to be studied. These effects include
various long- or short-range correlations among clusters of
particles, such as momentum conservation effects, resonance decays,
HBT correlations and jets~\cite{Borghini:2000cm,Borghini:2001vi}.
While jets potentially bias the $v_2$ measurement at high \pt,
other non-flow effects may be important at low and intermediate
\pt. The values of $v_2$ are also sensitive to event by event
fluctuations of the collision
geometry~\cite{Miller:2003kd,Zhu:2005qa,Manly:2005zy,Bhalerao:2006tp}
(so called $v_2$ fluctuations), which affect all \pt regions.

A bias to $v_2$ resulting from jets has been reported for high-\pt
hadrons~\cite{Adams:2004wz}. However, the relative significance of
other non-flow effects and $v_2$ fluctuations is still under
debate. Recent studies from PHOBOS~\cite{Alver:2007qw} and
STAR~\cite{Sorensen:2006nw} suggest that the fluctuations dominate
over the non-flow effects for the \pt integral $v_2$.

Following the two-source assumption in Eq.~\ref{eq:jet2b}, any
correlations other than jets are attributed to the background term,
i.e. $1+2v_2^{\rm a}v_2^{\rm b}\cos2\Delta\phi$. These naturally
include most non-flow correlations and $v_2$ fluctuations. In order
to estimate the potential biases from jets and dijets, we carried
out a detailed study in Appendix~\ref{appendix:C}, in which we
embedded dijet PYTHIA events into flow modulated HIJING events. Our
study shows that the large rapidity separation between the PHENIX
BBCs and central arms greatly reduces the influence of jets on our
$v_2$ measurements. Consequently, we use the BBC reaction plane
$v_2$ measurements to evaluate and subtract the elliptic flow
modulated background.

\subsection{Combinatoric Background Subtraction}
\label{sec:3.4}

The background level $\xi$ can be determined precisely if we know
the exact functional form for the near- and away-side jets, or if
we can independently measure the underlying event rate. However,
due to in-medium modifications, the near- and away-side jets are
not necessarily Gaussian, especially for $\Delta\phi$ values away
from 0 and $\pi$. Even in $p+p$ collisions, the underlying event
can include contributions from multiple-parton interaction, beam
remnants, initial and final state radiation~\cite{Affolder:2001xt},
which are related to the hard-scattering but not necessarily
correlated in $\Delta\phi$. Such effects have been studied at the
Tevatron~\cite{Acosta:2004wq,Affolder:2001xt} and
RHIC~\cite{Adler:2006sc} energies. For illustration purposes,
Figure~\ref{fig:rad} shows the dihadron correlation from
PYTHIA~\cite{Sjostrand:2001yu} with and without initial and final
state radiation effects. The difference between the two is clearly
significant.

Rigorous decomposition of the jet from its underlying event
currently requires assumptions about the jet shape or the physics
of the underlying event. As discussed earlier, a simple approach to
fix $\xi$ is to follow the subtraction procedure outlined in
Refs.~\cite{Ajitanand:2005jj,Adler:2005ee}. That is, one assumes
that the jet function has zero yield at its minimum
$\Delta\phi_{\rm min}$ (ZYAM), after subtraction of the underlying
event. The uncertainty on $\xi$ from this procedure is related to
the statistical accuracy of the data around $\Delta \phi_{\rm
min}$. In the present analysis, this uncertainty is negligible at
low \pt, but becomes important for $p_{\rm T}^{\rm{A,B}}>4$ GeV/$c$
in central collisions.

The ZYAM procedure, by definition, provides only a lower limit on
the jet yield. To estimate the possible over-subtraction of jet
yield at $\Delta\phi_{\rm{min}}$, we also made independent
estimates of $\xi$ via an absolute combinatoric background
subtraction method (ABS)~\cite{Adler:2004zd} and by a fitting
method. In the ABS method, $\xi$, as defined by Eq.~\ref{eq:xi}, is
assumed to reflect only a residual multiplicity smearing effect
caused by intrinsic positive correlations between the $n^{\rm a}_0$
and $n^{\rm b}_0$ in real events, i.e. a larger $n^{\rm a}_0$
implies a larger $n^{\rm b}_0$ and vice versa. Because of this
positive correlation, the average of the product can become larger
than the product of the average, i.e. $\langle n^{\rm a}_0n^{\rm
b}_0\rangle>\langle n^{\rm a}_0\rangle\langle n^{\rm b}_0\rangle$
or $\xi>1$.

To estimate $\xi$, we parameterize the centrality dependence of the
trigger and partner rate from the measured single particle spectra
in relevant momentum range, as a function of either $N_{\rm{part}}$
or $N_{\rm{coll}}$
\begin{eqnarray}
\label{eq:abs1} \langle n^{\rm{a,b}}_0\rangle = f(N_{\rm{part}}) =
g\left(N_{\rm{coll}}\right).
\end{eqnarray}
We then assume the event-by-event fluctuation of trigger and
partner hadrons to follow a Poisson distribution around their mean
values,
\begin{eqnarray}
\label{eq:abs2}n^{\rm{a,b}}_0 = \rm{Poisson} (\langle
n^{\rm{a,b}}_0\rangle).
\end{eqnarray}
However, we have verified that that our estimates are not very
sensitive to the functional forms of the fluctuations.

For each centrality bin, we determine the distribution of
$N_{\rm{part}}$ and $N_{\rm{coll}}$ from standard PHENIX Glauber
calculation~\cite{Adcox:2000sp,Adcox:2003nr}. For each simulated
event, we sample randomly from the $N_{\rm{part}}$ distribution,
calculate the corresponding mean value $\langle
n^{\rm{a,b}}_0\rangle$ and then the actual value $n^{\rm{a,b}}_0$
after taking into account the fluctuation. The same exercise is
repeated for the $N_{\rm{coll}}$ distributions. The final $\xi$ is
given by the average of the two and their difference is taken as
the systematic error. The correction modifies the background level
by 0.2\% in the most central and 25\% in 60-92\% centrality bin.
The ABS method and ZYAM methods give consistent $\xi$ values in
central collisions, but the ABS method gives somewhat lower values
in peripheral collisions.

\begin{figure}[thb]
\includegraphics[width=1.0\linewidth]{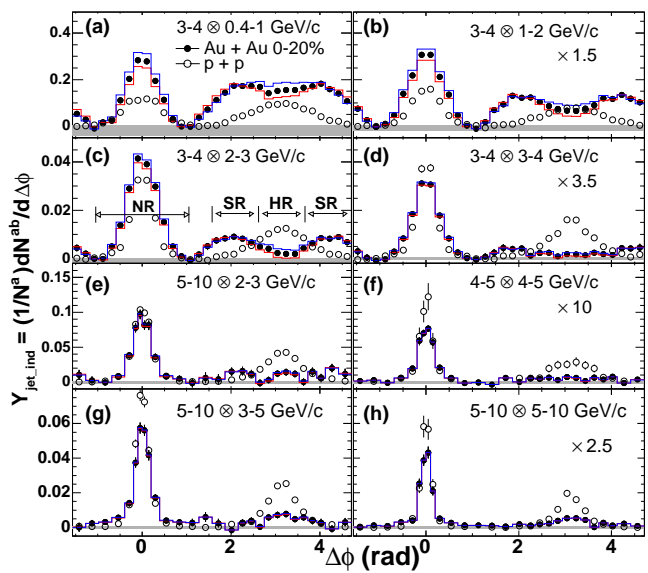}
\caption{\label{fig:shape} (Color online) Per-trigger yield versus $\Delta\phi$ for
various trigger and partner \pt ($p_{\rm T}^{\rm a}\otimes p_{T}^{\rm b}$),
arranged by increasing pair proxy energy (sum of $p_{\rm T}^{\rm a}$ and
$p_{\rm T}^{\rm b}$), in $p+p$ and 0-20\% Au+Au collisions. The data in
several panels are scaled as indicated. Solid histograms (shaded bands)
indicate elliptic flow (ZYAM) uncertainties. Arrows in Fig.~\ref{fig:shape}c
depict the ``Head'' region ({\bf HR}), the ``Shoulder'' region
({\bf SR}) and the ``Near-side'' region ({\bf NR}).}
\end{figure}

In the fit method, $Y_{\rm{jet\_ind}}(\Delta\phi)$ is fitted with a
function comprised of one near- and two symmetric away-side
Gaussians, following a procedure similar to that reported in
Ref.~\cite{Adare:2006nr}. One important difference is that a region
around $\pi$ ($|\Delta\phi-\pi|<1$) is excluded to avoid the
punch-through jet contributions (see Fig.~\ref{fig:shape}). Thus,
the fit uses the near-side and the falling edge of the away-side to
estimate the overlap of the near- and away-side Gaussians at
$\Delta\phi_{\rm{min}}$. This approach gives systematically lower
$\xi$ values than those obtained from the ZYAM and ABS methods.

Table.~\ref{tab:xi} summarizes the $\xi$ values from the three
methods. The results for the ZYAM and ABS methods are close, but
the values from the fitting method are systematically lower. This
could be due to the correlations between the fitting parameters or
a limitation in the Gaussian assumptions for the jet shape. To
avoid a possible overestimation of the jet yield in the
$\Delta\phi$ region where the near- and away-side Gaussians
overlap, we constrain the $\xi$ to be $\geq 1$. This is a
reasonable assumption in the absence of anti-correlation in trigger
and partner hadron multiplicity. We assign the differences with the
ZYAM method as a one-sided systematic error on $\xi$. This error is
important in central collisions and for $p_{\rm T}^{\rm{a,b}}<3$
GeV/$c$.

\begin{table}[h]
\caption{\label{tab:xi} Comparison of the $\xi$ values obtained for
three different normalization methods for several centrality
selections. They are calculated for $2.5<p_{\rm T}^{\rm a}<4.0$
GeV/$c$ and $1.0<p_{\rm T}^{\rm b}<2.0$ GeV/$c$ bin.}
\begin{tabular}{cccc}\hline
 Cent.  & ZYAM               & ABS & Constrained fit\\\hline
 0-5\%       & $1.0018\pm0.0004$  &$1.0023\pm0.0002$& $0.998\pm0.002$\\
20-30\%      & $1.015\pm0.0015$  &$1.012\pm0.003$&$1.004\pm0.006$\\
50-60\%      & $1.076\pm0.009$  &$1.07\pm0.02$
&$1.054\pm0.009$\\\hline
\end{tabular}
\end{table}

\subsection{Systematic Uncertainties}
\label{sec:3.5}

We classify the systematic errors associated with the jet yield
into three main categories: (1) Uncertainties in the single
particle efficiency correction, $\epsilon^{\rm a}$ and
$\epsilon^{\rm b}$, for Au+Au and $p+p$; (2) Statistical and
systematic uncertainties associated with the determination of the
elliptic flow values, $v_2^{\rm a}$ and $v_2^{\rm b}$, in Au+Au
collisions; (3) Uncertainties associated with the determination of
the combinatoric background level $\xi$ in both Au+Au and $p+p$
collisions.

The uncertainties associated with the efficiency corrections
include contributions from the detector acceptance (5\%), matching
cuts (4\%), momentum scale and momentum resolution (5\%). The
background contamination is estimated to be 5\% for $p_{\rm T}<5$
GeV/$c$, increasing to 10\% for the 5-10 GeV/$c$ bin. This leads to
an overall systematic error of $\sim10$\% for $p_{\rm T}<5$ GeV/$c$
and 13\% for the 5-10 GeV/$c$ bin. For central Au+Au collisions,
there is an additional maximally 5\%, centrality-dependent
uncertainty due to occupancy effect.

The propagation of uncertainties arising from the single particle
efficiency are different for different jet variables. For the
per-trigger yield, it depends on the errors associated with the
efficiency estimated for the partners ($\varepsilon^{\rm b}$). For
the hadron-pair yields, it depends on the errors related to
efficiencies for both trigger ($\varepsilon^{\rm a}$) and partner
($\varepsilon^{\rm b}$) hadrons. Since the efficiency correction
uncertainties are similar for trigger and partner hadrons in both
Au+Au and $p+p$ collisions, we can use a single variable
$\varepsilon$ to represent them. If the uncertainties are
independent between Au+Au and $p+p$, the total uncertainty would be
$\sim\sqrt{2} \delta \varepsilon/\varepsilon$ for JPY and $I_{\rm
AA}$ and $2 \delta \varepsilon/\varepsilon$ for $J_{\rm AA}$.
However, some systematic errors partially cancel between Au+Au and
$p+p$, especially those for the matching cut, momentum scale and
momentum resolution. The total uncertainties are estimated to be
12\% for $I_{\rm AA}$ and JPY, and 17\% for $J_{\rm AA}$.

The statistical uncertainties for $v_2$ are important in the most
central and most peripheral centrality bins for $p_{\rm T}>4$
GeV/$c$. The systematic uncertainties are however driven by the
uncertainty associated with the determination of the reaction plane
resolution; they are estimated to be $\sim 6$\% for central and
mid-central collisions, and $\sim 10$\% for peripheral
collisions~\cite{Adler:2005ee}. This error is nearly independent of
\pt, i.e  $\delta v_2^{\rm a} /v_2^{\rm a} \approx \delta v_2^{\rm
b} /v_2^{\rm b}$, and the resulting error for the hadron-pair ratio
is,
\begin{eqnarray}
\label{eq:4} \nonumber\delta \rm{JPR}(\Delta\phi) &=& 2\xi (\delta v_2^{\rm a}
v_2^{\rm b} + \delta v_2^{\rm b} v_2^{\rm a})\cos
2\Delta\phi\\ &\approx&  (\delta v_2/v_2)4 v_2^{\rm a}
v_2^{\rm b}\cos 2\Delta\phi
\end{eqnarray}
where the last approximation takes into account the fact that $\xi$
is close to 1. Additional systematic errors related to $v_4$ and
the factorization assumption that $\langle v_2^{\rm a}v_2^{\rm
b}\rangle =\langle v_2^{\rm a}\rangle\langle v_2^{\rm b}\rangle$,
were also estimated and found to be small~\cite{Adare:2006nr}.

The uncertainty due to $\xi$ (ZYAM uncertainty) can be expressed
as,
\begin{eqnarray}
\label{eq:v21} \delta {\rm JPR}(\Delta\phi) =
\delta\xi\left(1+2v_2^{\rm a} v_2^{\rm b}\cos 2\Delta\phi\right) \approx
\delta\xi
\end{eqnarray}
where we ignored the $\cos2\Delta\phi$ term since $2v_2^{\rm a}
v_2^{\rm b}\ll1$. The uncertainty of $\xi$ includes both the
statistical error of the data points around $\Delta\phi_{\rm{min}}$
and the systematic error from the fitting procedure.

Table.~\ref{tab:t1} summarizes the systematic errors for the jet
yield in 0-20\% Au+Au collisions due to $v_2$ and ZYAM subtraction.
Errors for several combinations of trigger and partner \pt (in
$p_{\rm T}^{\rm a}\otimes p_{\rm T}^{\rm b}$) are given. The
uncertainties are $\Delta\phi$ dependent, so we present them
separately for the three regions used in this analysis: a ``head''
region ($|\Delta\phi-\pi|<\pi/6$, HR), a ``shoulder'' region
($\pi/6<|\Delta\phi-\pi|<\pi/2$, SR) and the ``near-side`` region
($|\Delta\phi|<\pi/3$, NR). These regions are indicated in
Figure~\ref{fig:shape}c.
\begin{table*}
\caption{\label{tab:t1} Systematic errors for the per-trigger yield
in 0-20\% Au+Au collisions for several combinations of trigger and
partner \pt (in trigger $p_{\rm T}\otimes$ partner \pt). The errors
are in percentage and are shown separately for near-side
($|\Delta\phi|<\pi/3$), away-side ($|\Delta\phi-\pi|<\pi/2$),
away-side head region ($|\Delta\phi-\pi|<\pi/6$), and away-side
shoulder region ($\pi/6<|\Delta\phi-\pi|<\pi/2$).}
\begin{ruledtabular}
\begin{tabular}{cccccc}\hline\hline
 Errors in \%  & $2-3\otimes0.4-1$ GeV/$c$ & $2-3\otimes2-3$ GeV/$c$ & $3-4\otimes$$3-4$ GeV/$c$  &  $4-5\otimes$$4-5$ GeV/$c$ &  $5-10\otimes$$5-10$ GeV/$c$ \\\hline
&\multicolumn{5}{c}{near-side}\\\hline
$v_2$ err.                    &$\pm$18&$\pm$9.5&$\pm$3.8&$\pm$1&$<1$\\
ZYAM err. stat.               &$\pm$0.9&$\pm$1.1&$\pm$4.1&$\pm$9&$\pm$8\\
ZYAM over-sub. &+30&+9.5&$<1$&$<1$&$<1$\\\hline
&\multicolumn{5}{c}{away-side}\\\hline
$v_2$ err.                    &$\pm$10&$\pm$10&$\pm$9.3&$\pm$3&$\pm$1\\
ZYAM err. stat.               &$\pm$0.8&$\pm$2&$\pm$17&$\pm$39&$\pm$28\\
ZYAM over-sub.           &+28&+17&$<1$&$<1$&$<1$\\\hline
&\multicolumn{5}{c}{away-side head region}\\\hline
$v_2$ err.                    &$\pm$26&+42-39&+36-34&$\pm$5&$\pm$1\\
ZYAM err. stat.               &$\pm$1&$\pm$3&$\pm$27&$\pm$32&$\pm$16\\
ZYAM over-sub.           &+28&+29&$<1$&$<1$&$<1$\\\hline
&\multicolumn{5}{c}{away-side shoulder region}\\\hline
$v_2$ err.                    &$\pm$2.6&$\pm$2.3&$\pm$2&$\pm$1&$\pm$1\\
ZYAM err. stat.               &$\pm$0.8&$\pm$1.6&$\pm$15&$\pm$43&$\pm$45\\
ZYAM over-sub.           &+27&+15&$<1$&$<1$&$<1$\\\hline
\end{tabular}\end{ruledtabular}
\end{table*}

The three types of systematic errors impacts the jet shape and jet
yield differently. The single particle efficiency correction is a
multiplicative factor, so its uncertainty influences the
normalization of the jet yield, but does not influence its shape.
The uncertainties associated with the elliptic flow varies with
$\Delta\phi$. It is largest for regions around 0 and $\pi$, but
reaches a minimum in the shoulder region. The influence of the
$\xi$ uncertainty on the jet yield also depends on $\Delta\phi$. It
is the dominant uncertainty for the away-side yield at high \pt.

\section{RESULTS}
\label{sec:4}

\subsection{Jet-induced Dihadron Azimuthal ($\Delta\phi$) Distributions}

\label{sec:4.1}

Figure~\ref{fig:shape} shows a representative subset
of the per-trigger yield distributions,
$Y_{\rm{jet\_ind}}(\Delta\phi)$ for various combinations of trigger
and partner \pt ($p_{\rm T}^{a}\otimes p_{\rm T}^{b}$) for $p+p$
and 0-20\% Au+Au collisions, arranged by increasing pair proxy
energy, i.e. by $p_{\rm T}^{\rm sum}=p_{\rm T}^{\rm a}+p_{T}^{\rm
b}$.  The comprehensive array of results, covering the momentum
range of 0.4 to 10 GeV/$c$ from which this subset is derived, are
summarized in Appendix D
(Figs.\ref{fig:shape1}-\ref{fig:shape4}). From Eq.~\ref{eq:ptyiy},
one can see that the distributions for $p_{\rm T}^{\rm a}\otimes
p_{\rm T}^{\rm b}$ and $p_{\rm T}^{\rm b}\otimes p_{\rm T}^{\rm a}$
are related to each other by a normalization factor $n^{\rm
a}_0/n^{\rm b}_0$, i.e, the ratio of the number of hadrons in the
two \pt ranges. We have checked that these distributions, when
re-scaled by $n^{\rm a}_0$ or $n^{\rm b}_0$, are consistent with
each other. These $\Delta\phi$ distributions not only carry
detailed jet shape and yield information, they serve as a basis for
our systematic study of the \pt dependence of the contributions
from various physical processes.

The results in Fig.~\ref{fig:shape} constitutes one of many
possible ways of illustrating the evolution from low \pt to high
\pt in the two dimensional space of $p_{\rm T}^{\rm a}$ and $p_{\rm
T}^{\rm b}$. It is designed to highlight the main features of an
evolution from the soft-process dominated low-\pt region to the
hard-process dominated high-\pt region. As shown in the figure, the
$p+p$ data show essentially Gaussian away-side peaks centered at
$\Delta\phi\sim\pi$ for all $p_{\rm T}^{\rm a}$ and $p_{\rm T}^{\rm
b}$. By contrast, the Au+Au data show substantial modifications
relative to those for $p+p$ collisions, and these modifications
vary non-trivially with $p_{\rm T}^{\rm a}$ and $p_{\rm T}^{\rm
b}$. For a fixed value of $p_{\rm T}^{\rm a}$,
Figs.~\ref{fig:shape}a - \ref{fig:shape}d reveal a striking
evolution from a broad, roughly flat away-side peak to a local
minimum at $\Delta\phi\sim\pi$ with side-peaks at $\Delta\phi
\sim\pi\pm1.1$. Interestingly, the location of the side-peaks in
$\Delta\phi$ is found to be roughly constant with increasing
$p_{\rm T}^{\rm b}$ (see Fig.~\ref{fig:compfit4}). Such \pt
independence is compatible with an away-side jet modification
expected from a medium-induced Mach
shock~\cite{Casalderrey-Solana:2004qm} but provides a challenge for
models which incorporate large angle gluon
radiation~\cite{Vitev:2005yg,Polosa:2006hb}, Cherenkov gluon
radiation\cite{Koch:2005sx} or deflected jets
\cite{Armesto:2004pt,Chiu:2006pu}.

For relatively large values of $p_{\rm T}^{\rm a}\otimes p_{\rm
T}^{\rm b}$, Figs.~\ref{fig:shape}e - \ref{fig:shape}h (also
Fig.\ref{fig:shape1}) show that the away-side jet shape for Au+Au
gradually becomes peaked as for $p+p$, albeit suppressed. This
``re-appearance'' of the away-side peak seems to be due to a
reduction of the yield centered at $\Delta\phi \sim\pi\pm1.1$
relative to that at $\Delta\phi\sim \pi$, rather than a merging of
the peaks centered at $\Delta\phi \sim\pi\pm1.1$. This is
consistent with the dominance of dijet fragmentation at large
$p_{\rm T}^{\rm a}\otimes p_{\rm T}^{\rm b}$, possibly due to jets
that ``punch-through'' the medium~\cite{Renk:2006pk} or those
emitted tangentially to the medium's
surface~\cite{Loizides:2006cs}.

The evolution pattern of the away-side jet shape with \pt suggests
separate contributions from a medium-induced component at
$\Delta\phi \sim \pi\pm1.1$ and a fragmentation component centered
at $\Delta\phi \sim \pi$. A model independent study of these
contributions can be made by dividing the away-side jet function
into equal-sized ``head'' ($|\Delta\phi-\pi|<\pi/6$, HR) and
``shoulder'' ($\pi/6<|\Delta\phi-\pi|<\pi/2$, SR) regions, as
indicated in Fig.~\ref{fig:shape}c.

Figure~\ref{fig:shape} also shows significant modifications of the
near-side $\Delta\phi$ distributions. For the $p_{\rm T}^{\rm
a}\otimes p_{\rm T}^{\rm b}$ bins where the away-side has a concave
shape, the near-side jet also shows a clear enhancement in the
yield and a modification of the width relative to $p+p$. To
facilitate a more detailed investigation, we define a ``near-side''
region ($|\Delta\phi|<\pi/3$, NR) as indicated in
Fig.~\ref{fig:shape}c. In the following, we focus on the
jet-induced pairs in these three $\Delta\phi$ regions, and discuss
in detail the \pt and centrality dependence of their shapes and
yields.

\subsection{Medium Modification of Away-side Jets}
\label{sec:4.2}
\subsubsection{Away-side jet shape}
\label{sec:4.2.1}

We characterize the relative importance of the jet yields in the HR
and the SR by the ratio, $R_{\rm{HS}}$,
\begin{eqnarray}
R_{\rm{HS}}  = {{\frac{{\int_{\Delta\phi\in\rm{HR}} {d\Delta \phi
Y_{\rm{jet\_ind}} (\Delta \phi )} }} {\int_{\Delta\phi\in\rm{HR}}d\Delta\phi}}
\mathord{\left/
 {\vphantom {{\frac{{\int_{\rm{HR}} {d\Delta \phi Y_{\rm{jet\_ind}} (\Delta \phi )} }}
{\rm{HR}}} {\frac{{\int_{\rm{SR}} {d\Delta \phi Y_{\rm{jet\_ind}}
(\Delta \phi )} }} {\rm{SR}}}}} \right.
 \kern-\nulldelimiterspace} {\frac{{\int_{\Delta\phi\in\rm{SR}} {d\Delta \phi Y_{\rm{jet\_ind}} (\Delta \phi )} }}{\int_{\Delta\phi\in\rm{SR}}d\Delta\phi}}}
\end{eqnarray}
i.e. it is a ratio of area-normalized jet yields in the HR and the
SR. This ratio reflects the away-side jet shape and is symmetric
with respect to $p_{\rm T}^{\rm a}$ and $p_{\rm T}^{\rm b}$, i.e.
$R_{\rm{HS}}(p_{\rm T}^{\rm a},p_{\rm T}^{\rm
b})=R_{\rm{HS}}(p_{\rm T}^{\rm b},p_{\rm T}^{\rm a})$. For concave
and convex shapes, one expects $R_{\rm{HS}} < 1$ and $R_{\rm{HS}}>
1$ respectively; for a flat distribution, $R_{\rm{HS}} = 1$.

\begin{figure}[thb]
\includegraphics[width=1.0\linewidth]{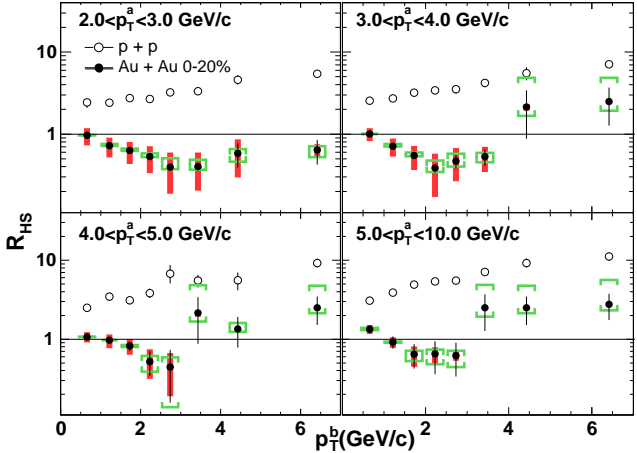}
\caption{\label{fig:ratio} (Color online) $R_{\rm{HS}}$ versus $p_{\rm T}^{\rm b}$ for $p+p$
(open) and Au+Au (filled) collisions for four trigger selections.
Shaded bars (brackets) represent
\pt-correlated uncertainties due to elliptic flow (ZYAM
procedure).}
\end{figure}

Figure~\ref{fig:ratio} shows the $p_{\rm T}^{\rm b}$ dependence of
$R_{\rm{HS}}$ for both $p+p$ and central Au+Au collisions for four
$p_{\rm T}^{\rm a}$ bins. The uncertainty for efficiency
corrections drops out in the ratio; the $v_2$ errors (shaded bars)
and ZYAM errors (brackets) are correlated in the two regions, thus
they partially cancel.

The $R_{\rm HS}$ values for $p+p$ are always above one and increase
with $p_{\rm T}^{\rm{b}}$. This reflects the narrowing of a peaked
away-side jet shape with increasing $p_{\rm T}^{\rm{b}}$. In
contrast, the ratios for Au+Au show a non-monotonic dependence on
$p_{\rm T}^{\rm{a,b}}$. They evolve from $R_{\rm{HS}} \sim 1$ for
$p_{\rm T}^{\rm{a}}$ or $p_{\rm T}^{\rm{b}}\lesssim1$ GeV/$c$,
through $R_{\rm{HS}}< 1$ for $1\alt p_{\rm T}^{\rm{a,b}}\alt 4$
GeV/$c$, followed by $R_{\rm{HS}}> 1$ for $p_{\rm T}^{\rm{a,b}}
\agt 5 $ GeV/$c$. These trends reflect the competition between
medium-induced modification and jet fragmentation in determining
the away-side jet shape, and suggest that the latter dominates for
$p_{\rm T}^{\rm{a,b}} \agt 5 $ GeV/$c$.

$R_{\rm{HS}}$ values for Au+Au are smaller than those for ${p+p}$
even at the highest \pt. This difference could be due to some
medium modification of the punch-through jets. However, the HR
yield dominates the SR yield ($R_{\rm{HS}} \gg 1$) in this \pt
region, and the values for $R_{\rm{HS}}$ becomes very sensitivity
to the SR yield. For instance, a small enhancement of the SR yield
can significantly reduce the value of $R_{\rm{HS}}$, without
significantly affecting the overall away-side feature.

\begin{figure}[thb]
\includegraphics[width=1.0\linewidth]{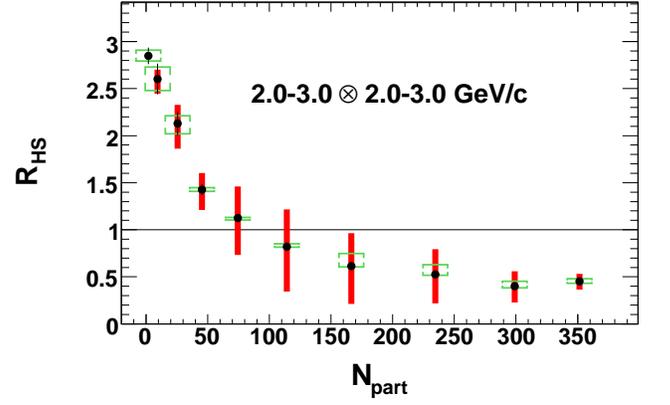}
\caption{\label{fig:ratio1} (Color online) $R_{\rm{HS}}$ versus $N_{\rm{part}}$ for
2-3 $\otimes$ 2-3 GeV/$c$. Shaded bars (brackets) represent
\pt-correlated uncertainties due to elliptic flow (ZYAM
procedure). The most left point is from $p+p$.}
\end{figure}
\begin{table}[th]
\caption{\label{tab:rhs2} Centrality dependence of $R_{\rm HS}$ for
$2<p_{\rm T}^{\rm a},p_{\rm T}^{\rm b}<3$ GeV/$c$
(Fig.~\ref{fig:ratio1}).}
\begin{ruledtabular} \begin{tabular}{ccccc}
$\langle N_{\rm part}\rangle$ & $R_{\rm HS}$ & Stat. & $v_2$ Err. & Norm Err.\\
351.4 & 0.451 & 0.030 & +0.080-0.087 & +0.028-0.021\\
299.0 & 0.402 & 0.024 & +0.158-0.175 & +0.051-0.019\\
234.6 & 0.526 & 0.018 & +0.268-0.308 & +0.102-0.011\\
166.6 & 0.614 & 0.021 & +0.351-0.402 & +0.133-0.009\\
114.2 & 0.821 & 0.030 & +0.395-0.477 & +0.031-0.006\\
74.4 & 1.126 & 0.045 & +0.333-0.394 & +0.005-0.022\\
45.5 & 1.427 & 0.060 & +0.175-0.215 & +0.020-0.018\\
25.7 & 2.130 & 0.125 & +0.196-0.267 & +0.081-0.110\\
9.5 & 2.603 & 0.164 & +0.099-0.158 & +0.125-0.124\\
2.0 & 2.848 & 0.090 & -- & +0.060-0.056\\
\end{tabular}  \end{ruledtabular}
\end{table}
$R_{\rm{HS}}$ values reach their minimum around 2-3 GeV/$c$.
Additional information can be obtained from their centrality
dependence, as shown in Fig.~\ref{fig:ratio1} for $2-3\otimes2-3$
GeV/$c$ bin. $R_{\rm{HS}}$ starts at around 3 for $p+p$ collisions
but quickly drops and crosses 1 at $N_{\rm{part}}\sim80$. It then
slowly decreases with $N_{\rm{part}}$ to a level of about 0.5 in
central collisions. This trend implies a quick change of the HR
and/or SR yield in relatively peripheral collisions. The saturation
of the $R_{\rm{HS}}$ for $N_{\rm{part}}>200$ may suggest that the
HR yield is dominated by the feed-in of the SR yield (see further
discussion in Section.\ref{sec:4.2.2}).

Although the \pt and centrality dependence of $R_{\rm HS}$ suggests
that the away-side yield contains separate contributions from a
fragmentation component (in the HR) and a medium-induced component
(in the SR), $R_{\rm HS}$ does not constrain the shape of the two
components directly. An alternative approach for quantifying the
away-side shape, is to assume a specific functional form for these
two components and carry out a model-dependent fit. Such a fit was
performed with the following two functional forms,
\begin{eqnarray} \label{eq:pty1}
Y^{\rm{FIT1}}_{\rm{jet\_ind}}(\Delta\phi) &=& G_1(\Delta\phi) +
G_2(\Delta\phi-\pi+D) +
\\\nonumber&& G_2(\Delta\phi-\pi-D)
+\kappa\\
\label{eq:pty3} Y^{\rm{FIT2}}_{\rm{jet\_ind}}(\Delta\phi) &=&
G_1(\Delta\phi) + G_2(\Delta\phi-\pi+D) +
\\\nonumber&&G_2(\Delta\phi-\pi-D) + G_3(\Delta\phi-\pi)+\kappa
\end{eqnarray}

The first (FIT1) assumes two Gaussian shoulder components located
symmetrically about $\pi$, each separated by the distance $D$ from
$\pi$. The second (FIT2) assumes the same shoulder components but
also includes an additional Gaussian component centered at $\pi$;
the latter represents the jet fragmentation contribution, and is
parameterized to have the same width as that for the $p+p$
away-side jet. FIT1 has six free parameters: background level
($\kappa$), near-side peak integral and width, shoulder peak
location ($D$), integral and width. In addition to the parameters
of FIT1, FIT2 has a parameter which controls the integral of the
fragmentation component.

\begin{figure}[thb]
\includegraphics[width=1.0\linewidth]{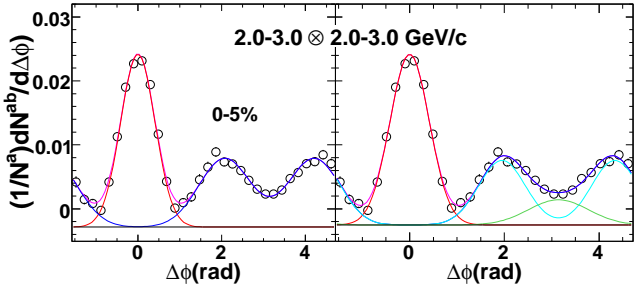}
\caption{\label{fig:compfit0} (Color online) Per-trigger yield $\Delta\phi$
distribution and corresponding fits for $2-3 \otimes 2-3$ GeV/$c$
in 0-5\% Au+Au collisions. FIT1 (FIT2) is shown in the left panel (right
panel). The total fit function and individual components are shown
relative to the $\kappa$ level indicated by the
horizontal line.}
\end{figure}

The separate contributions of FIT1 and FIT2 are illustrated for a
typical $p_{\rm T}^{\rm a}\otimes p_{\rm T}^{\rm b}$ in
Fig.~\ref{fig:compfit0}.

The two fits treat the region around $\Delta\phi=\pi$ differently.
FIT2 tends to assign the yield around $\pi$ to the center Gaussian,
while FIT1 tends to split that yield into the two shoulder
Gaussians. Note, however, that a single Gaussian centered at $\pi$
can be treated as two shoulder Gaussians with $D = 0$. Thus FIT1
does a good job at low \pt and high \pt, where the away-side is
dominated by shoulder and head component, respectively. It does not
work as well for intermediate \pt, where both components are
important. The center Gaussian and shoulder Gaussians used in FIT2
are strongly anti-correlated. That is, a small shoulder yield
implies a large head yield and vice versa. In addition, the center
Gaussian tends to ``push'' the shoulder Gaussians away from $\pi$,
and this results in larger D values than obtained with FIT1.

\begin{figure}[thb]
\includegraphics[width=1.0\linewidth]{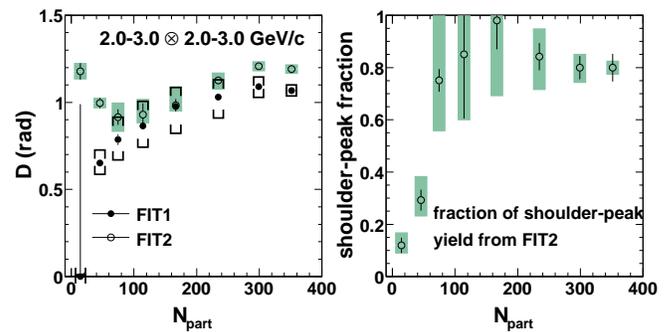}
\caption{\label{fig:compfit1a} (Color online) a) D versus $N_{\rm{part}}$ from
FIT1 (solid circles) and FIT2 (open circles) for $2-3\otimes 2-3$
GeV/$c$ bin. The error bars are the statistical errors; The shaded
bars and brackets are the systematic errors due to $v_2$. b) The
fraction of the shoulder Gaussian yield relative to the total
away-side yield as function of $N_{\rm{part}}$ determined from
FIT2.}
\end{figure}

Figure~\ref{fig:compfit1a}a shows the D values obtained from the
two fitting methods as a function of centrality for the \pt
selection $2-3\otimes 2-3$ GeV/$c$. The systematic errors from
$v_2$ are shown as brackets (shaded bars) for FIT1 (FIT2). The
values of D for FIT1 are consistent with zero in peripheral
collisions, but grow rapidly to $\sim 1$ for
$N_{\rm{part}}\sim100$, approaching $\sim1.05$ in the most central
collisions. The D values obtained from FIT2 are slightly larger
($\sim 1.2$~radians) in the most central collisions. They are also
relatively stable to variations of $v_2$ because most of the yield
variation is ``absorbed'' by the center Gaussian (cf.
Fig.~\ref{fig:compfit1a}b). Thus, the associated systematic errors
are also smaller than those for FIT1.

For $N_{\rm part}<100$, the centrality dependence of D is also
quite different for FIT1 and FIT2. As seen in
Fig.~\ref{fig:compfit1a}, the D values for FIT2 are above 1.
However the away-side yield in the SR, associated with these D
values, are rather small and the away-side distribution is
essentially a single peak centered around $\pi$. For such cases,
the values of D are prone to fluctuations and non-Gaussian tails.
The deviation between the D values obtained with FIT1 and FIT2 for
$N_{\rm{part}}<100$ simply reflects the weak constraint of the data
on D in peripheral collisions.

\begin{figure}[thb]
\includegraphics[width=1.0\linewidth]{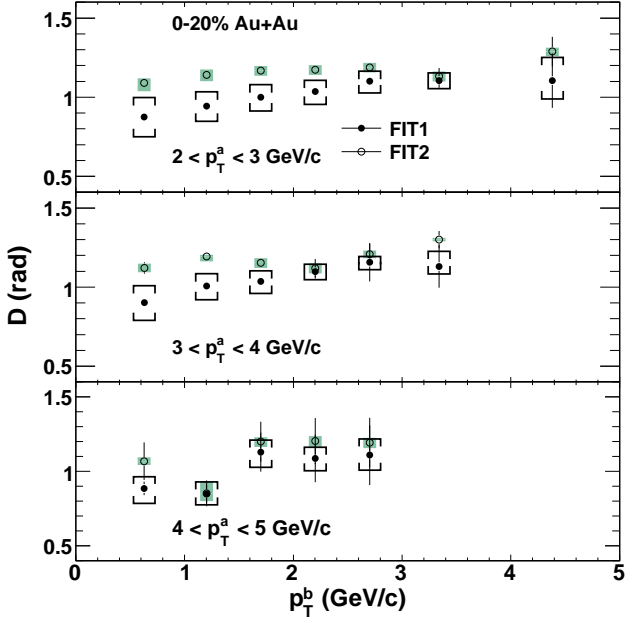}
\caption{\label{fig:compfit4} (Color online) Values of D determined from FIT1
(solid circles) and FIT2 (open circles) as function of partner
\pt for three trigger \pt ranges in 0-20\% Au+Au collisions.
The error bars are the statistical errors. The shaded bars and brackets are the systematic
errors due to elliptic flow.}
\end{figure}

Figure~\ref{fig:compfit4} shows the \pt dependence of $D$ in 0-20\%
central Au+Au collisions. The values from FIT2 are basically flat
with $p_{\rm T}^{\rm b}$. Those from FIT1 show a small increase
with $p_{\rm T}^{\rm b}$, but with a larger systematic error. At
low \pt, the values from FIT1 are systematically lower than those
from FIT2. However, they approach each other at large $p_{\rm
T}^{\rm a,b}$. From FIT1 and FIT2, it appears that the values of D
covers the range 1-1.2 radians for $p_{\rm T}^{\rm a}, p_{\rm
T}^{\rm b}\lesssim4$ GeV/$c$. This trend ruled out a Cherenkov
gluon radiation model~\cite{Koch:2005sx} (with only transition from
scalar bound states), which predicts decreasing D with increasing
momentum.
\subsubsection{Away-side jet per-trigger yield}
\label{sec:4.2.2}

\begin{figure}[thb]
\includegraphics[width=1.0\linewidth]{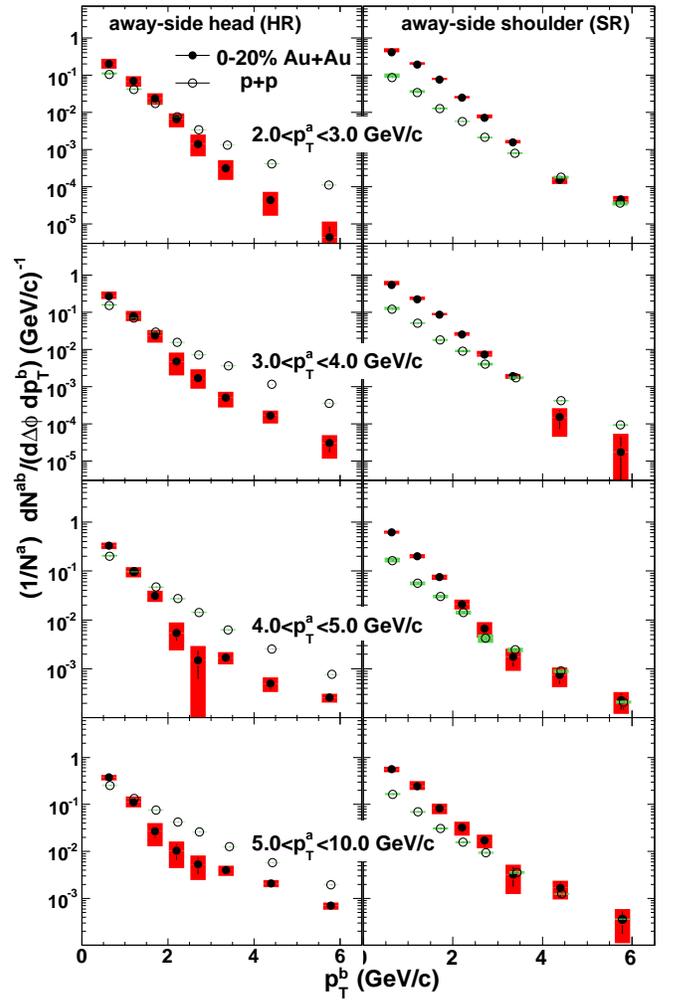}
\caption{\label{fig:spec1} (Color online) Per-trigger yield as function of
partner \pt for the HR (left panels) and the SR region (right
panels). Filled (open) circles represent the 0-20\%
Au+Au ($p+p$) collisions. Results for four trigger \pt
in 2-3, 3-4, 4-5, 5-10 GeV/$c$ are shown.
The shaded bars represent the total systematic uncertainties.}
\end{figure}

Relative to $p+p$, the Au+Au yield is suppressed in the HR but is
enhanced in the SR (cf. Fig.~\ref{fig:shape}). A more detailed
mapping of this modification pattern is obtained by comparing the
jet yields in the HR and SR as a function of partner \pt. Such a
comparison is given in Fig.~\ref{fig:spec1} for central Au+Au and
for $p+p$ collisions. The figure shows that, relative to $p+p$, the
Au+Au data are enhanced in the SR for low \pt, and suppressed in
the HR for high \pt. The shape of the Au+Au spectra in the HR is
also quite different from that for $p+p$. For $p_{\rm
T}^{\rm{a,b}}\lesssim4$ GeV/$c$, the spectra for Au+Au are steeper
than those for $p+p$. For higher \pt, both spectra have the same
shape (parallel to each other) but the yield for Au+Au is clearly
suppressed.

\begin{figure}[thb]
\includegraphics[width=1.0\linewidth]{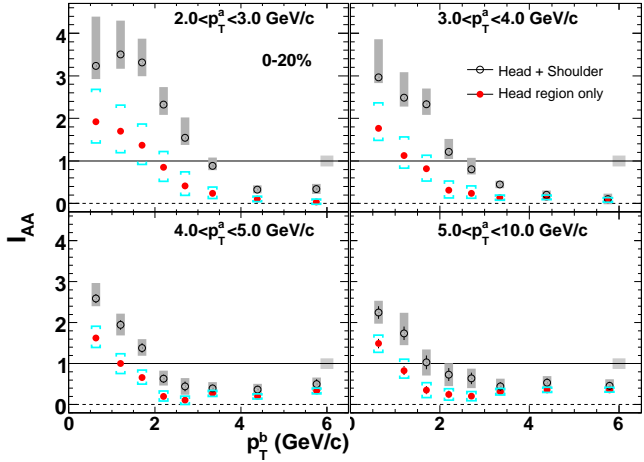}
\caption{\label{fig:iaaaway1} (Color online) $I_{\rm{AA}}$ versus $p_{\rm T}^{\rm b}$ for four
trigger \pt bins in HR+SR ($|\Delta\phi-\pi|<\pi/2$) and HR ($|\Delta\phi-\pi|<\pi/6$) in 0-20\% Au+Au collisions.
The shaded bars and brackets represent the total systematic errors in the two regions.
They are strongly correlated. Grey bands around $I_{\rm{AA}}=1$ represent 12\% combined
uncertainty on the single particle efficiency in Au+Au and $p+p$.}
\end{figure}

To quantify this suppression/enhancement, we use the per-trigger
yield ratio $I_{\rm{AA}}$, the ratio of per-trigger yield for Au+Au
collisions to that for $p+p$ collisions (cf. Eq.~\ref{eq:iaadef}).
Such ratios for the HR and the HR+SR are shown as a function of
$p_{\rm T}^{\rm b}$ for four different $p_{\rm T}^{a}$ selections
in Fig.~\ref{fig:iaaaway1}. For triggers of $2<p_{\rm T}^{\rm a}<3$
GeV/$c$, $I_{\rm{AA}}$ for HR+SR exceeds one at low $p_{\rm T}^{\rm
b}$, but falls with $\pt^b$ and crosses one around 3.5 GeV/$c$. A
similar trend is observed for the higher \pt triggers, but the
enhancement for low $p_{\rm T}^{\rm b}$ is smaller and the
suppression for high $p_{\rm T}^{\rm b}$ is stronger. The
$I_{\rm{AA}}$ values in the HR are also lower relative to HR+SR,
for all $p_{\rm T}^{\rm a,b}$. For the low-\pt triggers, the HR
suppression sets in for $1\alt p_{\rm T}^{\rm b} \alt 3$ GeV/$c$,
followed by a fall-off for $p_{\rm T}^{\rm b} \agt 4$ GeV/$c$. For
the higher-\pt triggers, a constant level $\sim 0.2-0.3$ is
observed above $\sim 2$ GeV/$c$, similar to the suppression level
of inclusive hadrons~\cite{Adler:2003au}.

\begin{figure}[thb]
\includegraphics[width=1.0\linewidth]{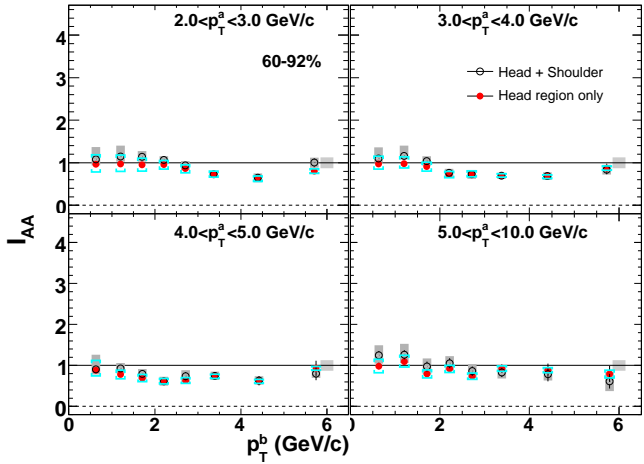}
\caption{\label{fig:iaaaway4} (Color online) $I_{\rm{AA}}$ versus $p_{\rm T}^{\rm b}$ for four
trigger \pt bins in HR+SR ($|\Delta\phi-\pi|<\pi/2$) and HR ($|\Delta\phi-\pi|<\pi/6$)
in 60-92\% Au+Au collisions. The shaded bars and brackets represent the total systematic errors in the two regions.
They are strongly correlated. Grey bands around $I_{\rm{AA}}=1$ represent 12\% combined
uncertainty on the single particle efficiency in Au+Au and $p+p$.}
\end{figure}

For comparison, Fig.~\ref{fig:iaaaway4} shows the $I_{\rm{AA}}$ for
peripheral collisions. They indicate that, in contrast to the
values for central collisions, there is only a small suppression in
both the HR and the HR+SR for low-\pt triggers at large $p_{\rm
T}^{\rm b}$. Moreover, the overall modifications are much smaller
than observed for the more central Au+Au collisions.

\begin{figure}[thb]
\includegraphics[width=1.0\linewidth]{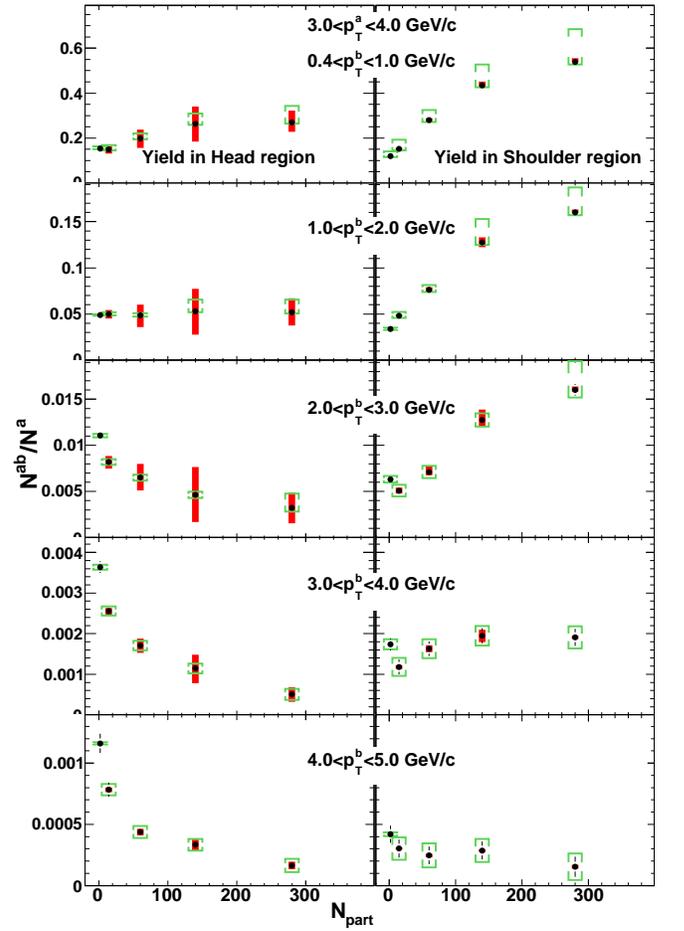}
\caption{\label{fig:spec1c} (Color online) The per-trigger yield versus
$N_{\rm{part}}$ in the HR (left panels) and the SR (right panels)
for partner \pt in 0.4-1, 1-2, 2-3, 3-4, 4-5 GeV/$c$ and fixed
trigger in 3-4 GeV/$c$. The left most points are from $p+p$ collisions.
The shaded bars (brackets) represent
uncertainties due to elliptic flow (ZYAM procedure).}
\end{figure}

A more detailed view of the enhancement/suppression patterns in the
SR/HR can be provided by investigating their centrality dependence.
Figure~\ref{fig:spec1c} shows the per-trigger yield in the SR and HR
as function of $N_{\rm{part}}$ for trigger \pt of 3-4 GeV/$c$ and
five partner \pt bins ranging from 0.4 to 5 GeV/$c$. With
increasing partner \pt, both the SR and HR yields show a
characteristic evolution with $N_{\rm{part}}$. That is, they first
show an increase, followed by an essentially flat dependence,
followed by a decrease (in the HR only). However, the values of
$p_{\rm T}^{\rm b}$ at which the centrality dependence becomes flat
is quite different for the SR and the HR ($p_{\rm T}^{\rm b}\sim 4$
GeV/$c$ for the SR and $p_{\rm T}^{\rm b}\sim 1-2$ GeV/$c$ for the
HR, respectively).

Figures \ref{fig:iaaaway1} and \ref{fig:spec1c} provide clear
evidence that in central Au+Au collisions, there is significant
yield suppression in the HR and an enhancement in the SR. The
suppression for the HR is consistent with a jet quenching scenario
in which the HR yield at high \pt is dominated by radiated gluons
and jets which survive passage through the medium. The enhancement
for the SR could reflect the dissipative processes that
redistribute the energy lost in the medium.

Several previous jet correlation measurements were carried out for
an intermediate range of $p_{\rm T}^{\rm a}\otimes p_{\rm T}^{\rm
b}$~\cite{Adler:2002tq,Adams:2005ph,Adler:2004zd,
Adler:2005ee,Adare:2006nn,Adare:2006nr} and/or in a limited
away-side integration window roughly equal to the HR~\cite{
Adler:2002tq,Adler:2004zd, Adare:2006nn,Adams:2006yt}. However,
Fig.~\ref{fig:spec1c} shows that away-side yield modifications are
sensitive to both \pt and the $\Delta\phi$ integration range. By
choosing a certain \pt and $\Delta\phi$ range, the combined effect
of SR enhancement and HR suppression can result in away-side yields
that are almost independent of centrality, while their shapes still
vary with centrality. Thus a detailed survey of the jet yield in a
broad \pt and more differential study in $\Delta\phi$ for the
away-side is important to obtain the full picture.
\subsection{Medium Modification of Near-side Jets}
\label{sec:4.3} In this section, we map out the \pt and centrality
dependence of the shape and yields of the near-side jets.
\subsubsection{Near-side jet shape}
\label{sec:4.3.1} We characterize the near-side shape in
$\Delta\phi$ by the Gaussian fit functions FIT1 and FIT2 described
in Section~\ref{sec:4.2.1}. The systematic uncertainties include
the differences between FIT1 and FIT2 and the uncertainties
associated with the elliptic flow subtraction. To account for a
possible influence from feed-in of the shoulder component, we also
performed fits to the near-side distribution with a single Gaussian
function in $\pm1$$\sigma$, $\pm2$$\sigma$ and $\pm3$ $\sigma$
windows, where $\sigma$ is the width of the near-side peak obtained
with FIT2. Deviations from $\sigma$ were added in quadrature to the
total systematic errors. For $p+p$, a simple fit with a near- and
an away-side Gaussian plus a constant background was used.

\begin{figure}[thb]
\includegraphics[width=1.0\linewidth]{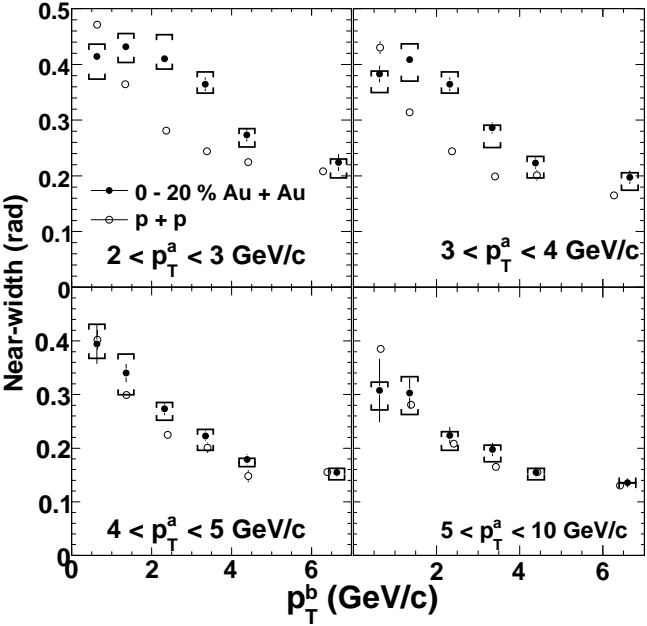}
\caption{\label{fig:compfit3b} Near-side Gaussian widths versus partner \pt for
four trigger \pt ranges compared between 0-20 \% Au+Au (solid circles)
and $p+p$ (open circles).}
\end{figure}

Figure~\ref{fig:compfit3b} compares the near-side Gaussian widths
obtained for $p+p$ and 0-20\% central Au+Au collisions. The results
are shown as a function of partner \pt for four trigger \pt bins as
indicated. The $p+p$ widths show the expected decrease with partner
\pt for all trigger bins, as expected from a narrowing of the jet
cone as $p_{\rm T}^{\rm b}$ increases. The Au+Au widths also
decrease with partner \pt except at low $p_{\rm T}^{\rm a,b}$. For
low-\pt triggers of $2<p_{\rm T}^{\rm a}<3$ and $3<p_{\rm T}^{\rm
a}<4$ GeV/$c$, the values for the widths are slightly below those
for $p+p$ for $p_{\rm T}^{\rm b}<1$ GeV/$c$; they are, however,
significantly broader for $1\lesssim p_{\rm
T}^{b}\lesssim4$GeV/$c$. For higher trigger \pt, the extracted
widths are similar for Au+Au and $p+p$.

\begin{figure}[thb]
\includegraphics[width=1.0\linewidth]{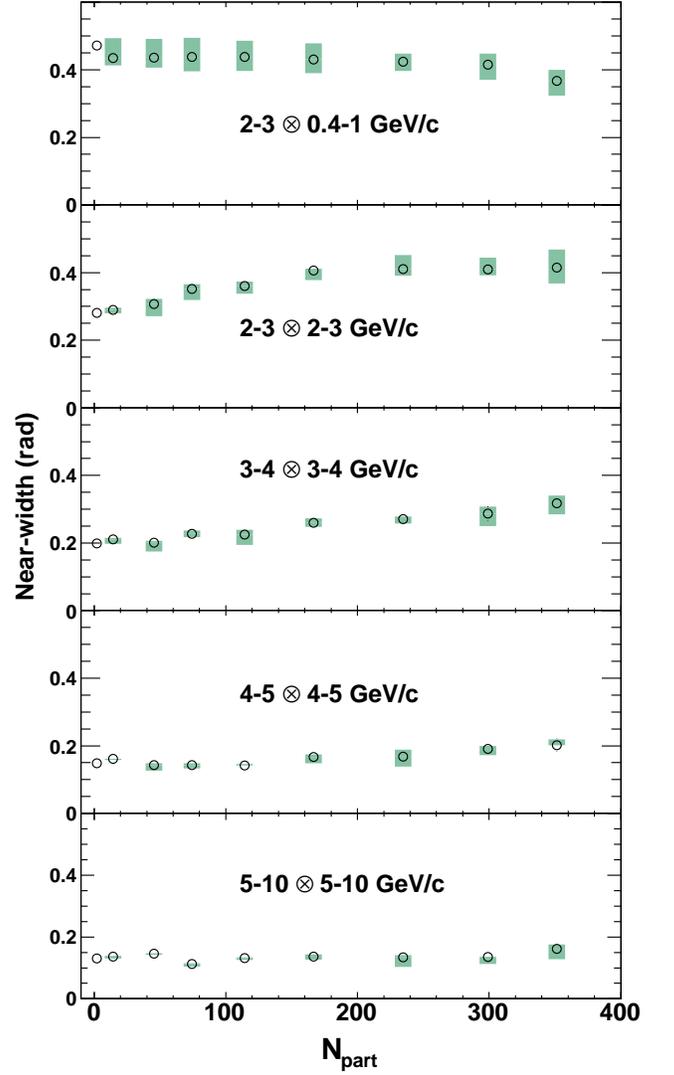}
\caption{\label{fig:compfit3} (Color online) Near-side Gaussian widths versus $N_{\rm
part}$ for five successively increasing $p_{\rm T}^{\rm a}\otimes p_{\rm T}^{\rm b}$.
The most left point in each panel represents the value from $p+p$.}
\end{figure}

Figure~\ref{fig:compfit3} shows the centrality dependence of the
near-side widths for successively higher $p_{\rm T}^{\rm a}\otimes
p_{\rm T}^{\rm b}$. For the lowest \pt bin of $2-3\otimes 0.4-1$
GeV/$c$, the width decreases slightly with $N_{\rm part}$ and
approaches a value of 0.4 radian in central collisions. A similar
trend has been reported for measurements at low \pt
\cite{Adams:2004pa}. For the $2-3\otimes 2-3$ GeV/$c$ selection,
the near-side width grows with $N_{\rm{part}}$, and approaches a
value about 40\% larger than the $p+p$ value for
$N_{\rm{part}}>200$. For higher $p_{\rm T}^{\rm a}\otimes p_{\rm
T}^{\rm b}$ bins, the near-side widths are narrower and their
centrality dependence is flatter. For the $5-10\otimes5-10$ GeV/$c$
bin, the near-side widths ($\sim 0.14$ radian) are essentially
independent of $N_{\rm{part}}$, as might be expected if the
near-side correlations are dominated by jet fragmentation.
\subsubsection{Near-side jet per-trigger yield}
\label{sec:4.3.2} 
\begin{figure}[thb]
\includegraphics[width=1.0\linewidth]{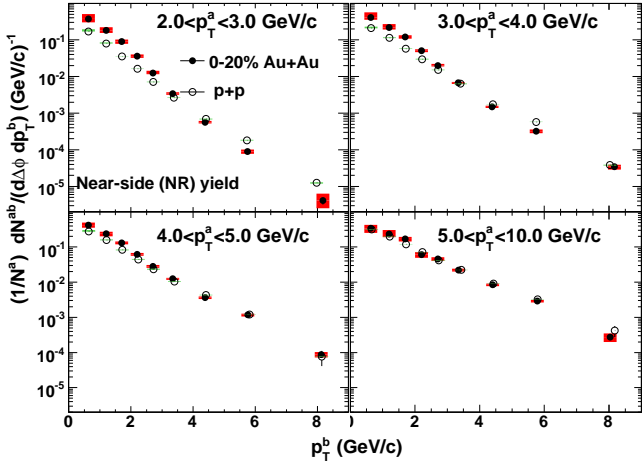}
\caption{\label{fig:spec2} (Color online) Near-side yield in $|\Delta\phi|<\pi/3$
versus partner \pt for four trigger \pt selections. The filled and open
circles are for 0-20\% Au+Au and $p+p$, respectively.}
\end{figure}

\begin{figure}[thb]
\includegraphics[width=1.0\linewidth]{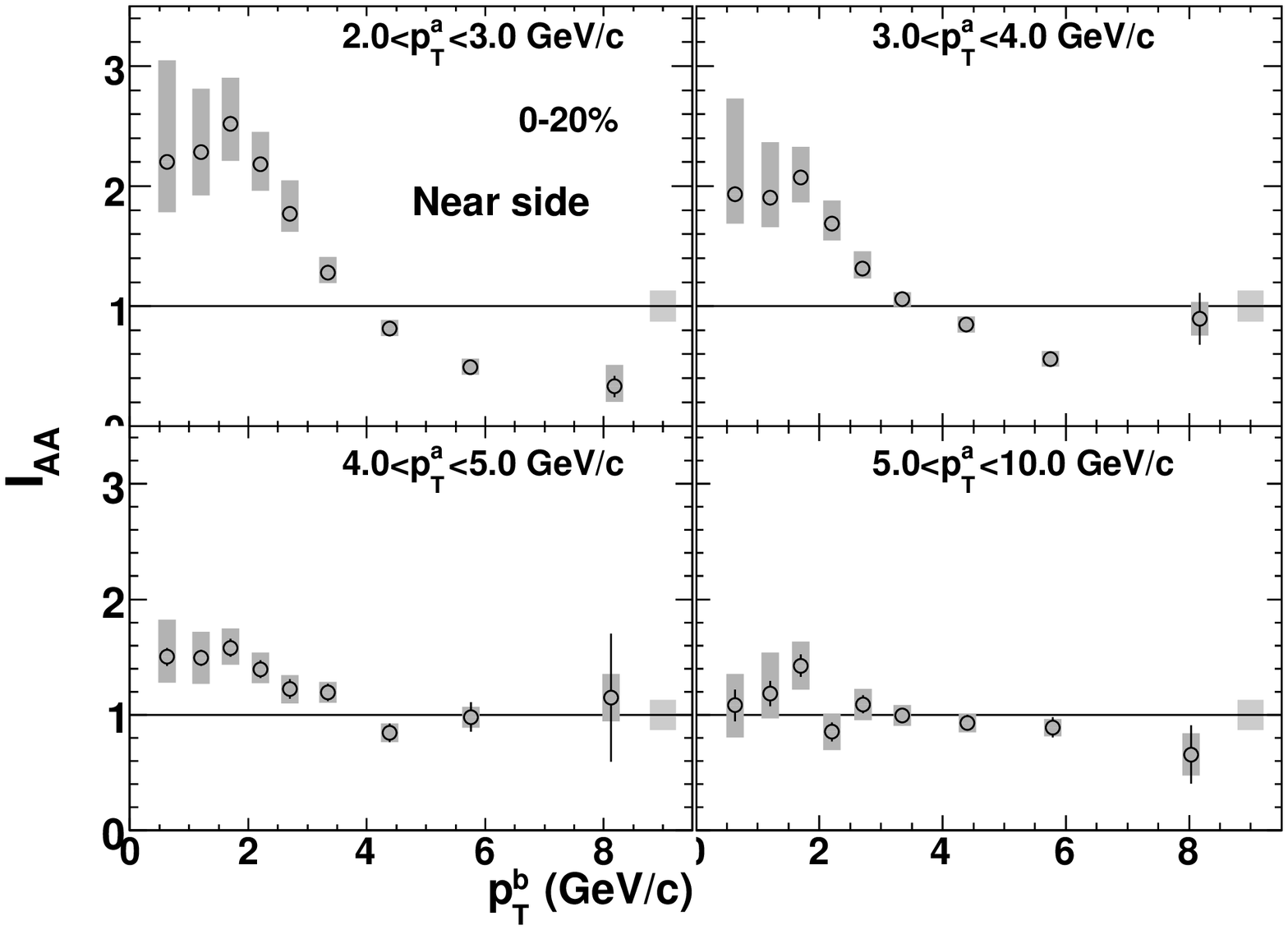}
\caption{\label{fig:iaanear1} Near-side $I_{\rm{AA}}$ in 0-20\%
Au+Au versus partner \pt for four trigger \pt bins. The shaded
bars around the data points are the total systematic errors. Grey
bands around $I_{\rm{AA}}=1$ represent 12\% combined uncertainty
on the single particle efficiency in Au+Au and $p+p$.}
\end{figure}

\begin{figure}[thb]
\includegraphics[width=1.0\linewidth]{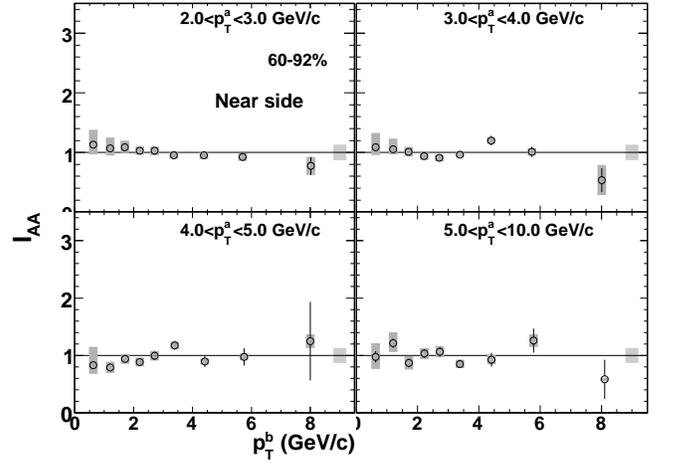}
\caption{\label{fig:iaanear4} Near-side $I_{\rm{AA}}$ in 60-92\%
Au+Au versus partner \pt for four trigger \pt bins. The shaded
bars around the data points are the total systematic errors. Grey
bands around $I_{\rm{AA}}=1$ represent 12\% combined uncertainty
on the single particle efficiency in Au+Au and $p+p$.}
\end{figure}

The near-side yield in $|\Delta\phi|<\pi/3$ (NR)
as a function of partner \pt, is shown in Fig.~\ref{fig:spec2} for
$p+p$ and Au+Au collisions, for four trigger \pt bins. The
corresponding results for the modification factor, $I_{\rm{AA}}$,
are shown in Fig.~\ref{fig:iaanear1}. For triggers of 2-3 GeV/$c$,
$I_{\rm{AA}}$ is enhanced by more than a factor of two for $p_{\rm
T}^{B}<2$ GeV/$c$, followed by a fall-off below unity for $p_{\rm
T}^{b}\gtrsim4$ GeV/$c$. The overall deviation from $I_{\rm{AA}}=1$
decreases with increasing trigger \pt. For the highest \pt trigger,
the near-side yield is close to that for $p+p$ over the full range
of $p_{\rm T}^{\rm b}$. As a comparison, the $I_{\rm{AA}}$ values
for the 60-92\% centrality bin, shown in Fig.~\ref{fig:iaanear4},
are close to 1 for all $p_{\rm T}^{\rm{a,b}}$, suggesting a rather
weak medium modification of the near-side yield in peripheral
collisions.

The patterns of the near-side jet shape and yields in
Figs.\ref{fig:compfit3b}-\ref{fig:iaanear4} suggest an influence
from both medium modification and jet fragmentation at intermediate
\pt. The influence of the medium has also been linked to a long
range correlation component in $\Delta\eta$~\cite{Adams:2004pa,
Adams:2005ph,Adams:2006tj}. This so called $\eta$ ``ridge'' has
been shown to be flat up to $|\Delta\eta|\sim2$. The PHENIX
$\Delta\eta$ acceptance is limited to $|\Delta\eta|<0.7$. However,
if contributions from the ridge are significant, they should show
up in $\Delta\eta$ distributions.

\begin{figure}[thb]
\includegraphics[width=1.0\linewidth]{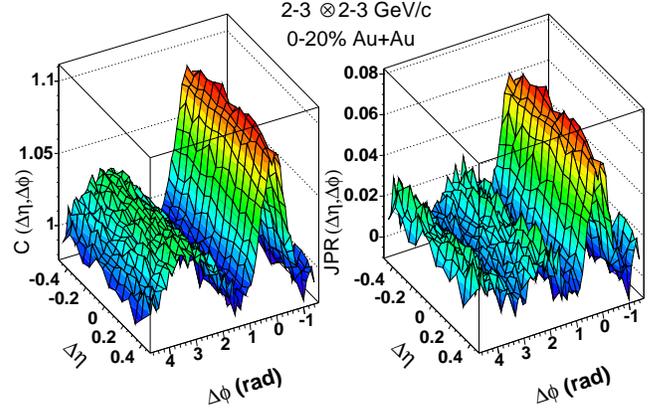}
\caption{\label{fig:phieta1} (Color online) Correlation function (left panel, Eq.~\ref{eq:jet1}) and background
subtracted hadron-pair ratio (right panel, Eq.~\ref{eq:jet3}) in $\Delta\eta$ and $\Delta\phi$
space for $2-3 \otimes 2-3$ GeV/$c$ in 0-20 \% Au+Au collisions.}
\end{figure}

\begin{figure}[thb]
\includegraphics[width=1.0\linewidth]{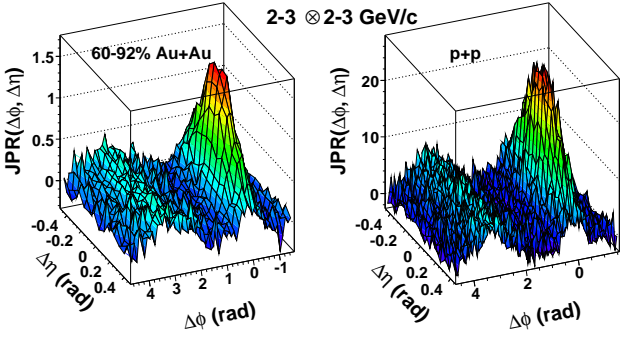}
\caption{\label{fig:phieta2} (Color online) Background
subtracted hadron-pair ratio (via Eq.~\ref{eq:jet3}) for $2-3 \otimes 2-3$ GeV/$c$ in 60-92 \% Au+Au (left panel) and p+p (right panel) collisions.}
\end{figure}

Figure~\ref{fig:phieta1}a shows a two-dimensional
$\Delta\eta$-$\Delta\phi$ correlation function for $2-3\otimes 2-3$
GeV/$c$ in 0-20\% central Au+Au collisions. The $\Delta\eta$ range
is displayed for $|\Delta\eta|<0.5$ to suppress the relatively
large statistical fluctuations at the edge of $\Delta\eta$
acceptance. The correlation function in Fig.~\ref{fig:phieta1}a
peaks along $\Delta\phi\sim 0$ and $\pi$, largely because of the
elliptic flow modulation of the combinatoric pairs. To subtract the
flow term, we assume that $\xi$ and $v_2$ are identical to those
used in our 1-D $\Delta\phi$ correlation analysis, and are constant
for $|\eta|<0.35$. The distribution after $v_2$ subtraction is
shown in Fig.~\ref{fig:phieta1}b. One can clearly see one near-side
and two shoulder peaks (in $\Delta\phi$), which extend over the
full range of $\Delta\eta$. However both p+p and peripheral Au+Au
collisions for the same $p_{\rm T}$ selections
(Fig.\ref{fig:phieta2}) show one near-side peak centered around
$\Delta\eta\sim0$ and one away-side peak elongated over
$\Delta\eta$. These features are expected for fragmentation of
back-to-back dijets in vacuum.

\begin{figure}[thb]
\includegraphics[width=1.0\linewidth]{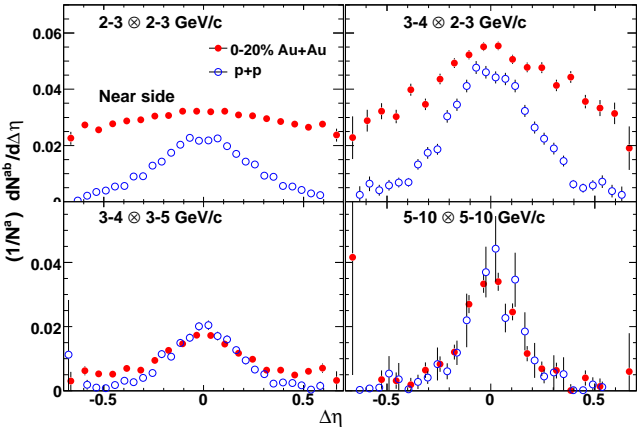}
\caption{\label{fig:etacent} (Color online) Per-trigger yield versus
$\Delta\eta$ for $p+p$ (open symbols) and 0-20\% central Au+Au (filled
symbols) collisions. Results are shown for four
$p_{\rm T}^{\rm a}\otimes p_{\rm T}^{\rm b}$ selections as indicated.}
\end{figure}

To facilitate further detailed investigation, we focus on a
near-side region defined by $|\Delta\phi|<0.7$ and
$|\Delta\eta|<0.7$ and study the projected distributions in
$\Delta\phi$ and $\Delta\eta$. Figure~\ref{fig:etacent} compares the
the $\Delta\eta$ distributions for $p+p$ and 0-20\% central Au+Au
collisions. The $p+p$ data indicate a relatively narrow jet-like
peak for all four \pt selections. For the $2-3\otimes2-3$ GeV/$c$
bin, the Au+Au data are enhanced and broadened relative to $p+p$.
However, these differences gradually decrease toward higher \pt and
essentially disappear for the $5-10\otimes5-10$ GeV/$c$ bin. This
possibly suggests that the ridge component at high \pt either
disappears or becomes overwhelmed by the jet fragmentation
component~\cite{footnote1}.

\begin{figure}[thb]
\includegraphics[width=1.0\linewidth]{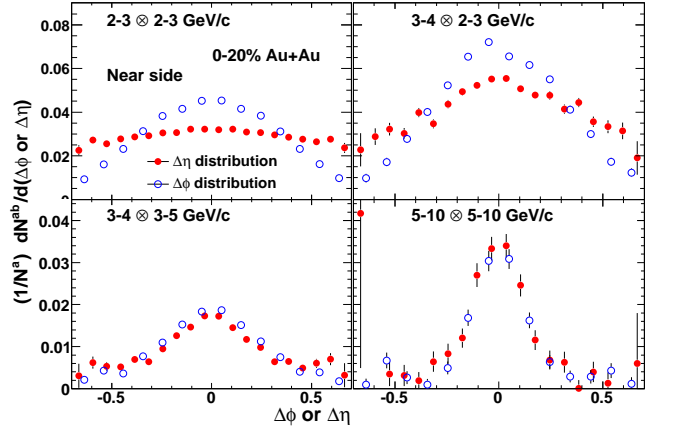}
\caption{\label{fig:phieta} (Color online) The projection of 2-D per-trigger
yield in $|\Delta\eta|<0.7\otimes|\Delta\phi|<0.7$ and 0-20\% central Au+Au collisions onto
$\Delta\phi$ (open symbols) and $\Delta\eta$ (filled symbols) for
four $p_{\rm T}^{\rm a}\otimes p_{\rm T}^{\rm b}$ selections.}
\end{figure}

Figure~\ref{fig:phieta} show the comparison of the projected
distributions in $\Delta\eta$ and $\Delta\phi$ for 0-20\% central
Au+Au collisions. By construction, the integrals of the two
distributions are the same. For the $2-3\otimes2-3$ and
$2-3\otimes3-4$ GeV/$c$ bins, the distributions in $\Delta\eta$ are
broader than in $\Delta\phi$. For the $3-4\otimes3-5$ and
$5-10\otimes5-10$ GeV/$c$ bins, the distributions become similar
between $\Delta\phi$ and $\Delta\eta$. These observations suggest
that the medium modifications are limited to $p_{\rm T}^{\rm
a,b}\lesssim4$ GeV/$c$, a similar \pt range in which the away-side
is also strongly modified.

The evolution of the enhancement and broadening with the $p_{\rm
T}^{\rm{a,b}}$ reflects the competition between contributions from
the medium response and jet fragmentation. The former is important
at $p_{\rm T}^{\rm a,b}\lesssim4$ GeV/$c$ and manifests itself as
an enhanced and broadened distribution in $\Delta\eta$ and
$\Delta\phi$. The latter dominates at higher \pt, reflected by
$I_{\rm{AA}}\approx 1$ and a near-side width similar to $p+p$. The
strong modifications at intermediate \pt may reflect the remnants
of those jets that interact with the medium, appearing as low \pt
hadron pairs with modified width and multiplicity. Possible physics
mechanisms for this parton-medium interaction include jet
interaction with a longitudinal flowing
medium~\cite{Armesto:2004pt,Chiu:2005ad,Wong:2007pz},
position-momentum correlations induced by radial
flow~\cite{Voloshin:2004th}, correlation between radial flow
boosted beam jet and medium suppressed transverse
jet~\cite{Shuryak:2007fu}, plasma
instability~\cite{Romatschke:2006bb,Majumder:2006wi}, or
back-splash caused by the quenched jets~\cite{Pantuev:2007sh}.
However, the modification mechanisms for the near- and away-side
may be related: both near- and away-side distributions show
enhancement and broadening in the lower range $\pt<4$ GeV/$c$,
above which the jet characteristics qualitatively approach the jet
fragmentation.

\subsection{Away- and Near-side Spectral Slopes} \label{sec:4.4}
To further explore the differences between the NR, HR and SR, we
compare the shapes of the partner \pt spectra in these three
$\Delta\phi$ regions. To do this, we characterize the local inverse
slope of the spectra via a truncated mean \pt in a given \pt range,
\begin{equation}
\langle p_{_T}^{\rm{\prime}}\rangle\equiv\langle
p_{\rm T}^{\rm b}\rangle_{p_{\rm T}^{\rm{min}}<p_{\rm T}^{\rm b}<p_{\rm T}^{\rm{max}}} -
p_{\rm T}^{\rm{min}}.
\end{equation}
where $\langle p_{_T}^{\rm{\prime}}\rangle$ is calculated from the
jet yields shown in Fig.~\ref{fig:spec1} and Fig.~\ref{fig:spec2}.
For an exponential spectrum with an inverse slope of $T$ and $T\ll
p_{\rm T}^{\rm{max}}-p_{\rm T}^{\rm{min}}$, $\langle
p_{_T}^{\rm{\prime}} \rangle\approx T$.

\begin{figure}[thb]
\includegraphics[width=1.0\linewidth]{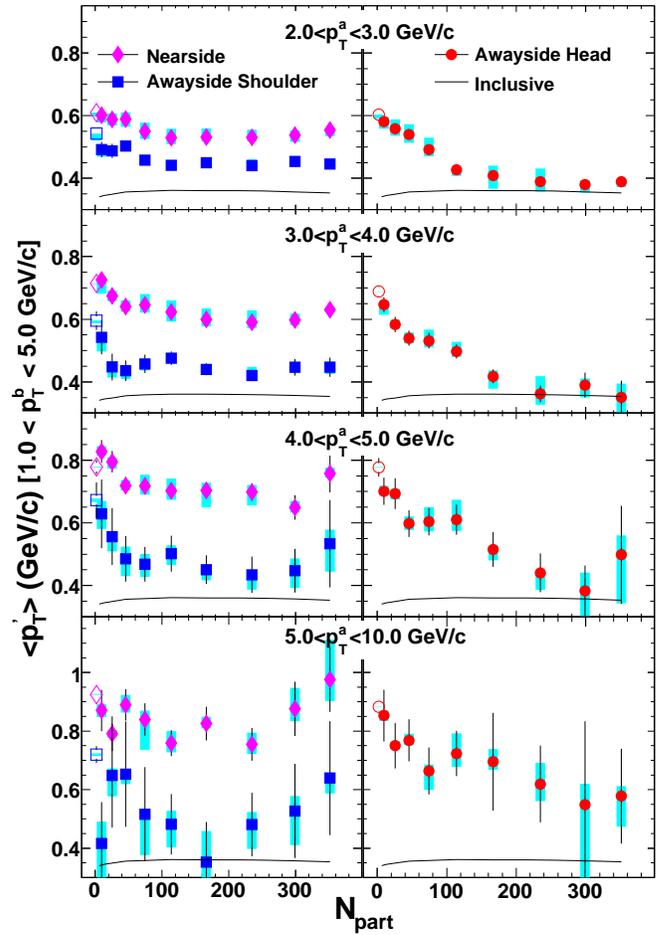}
\caption{\label{fig:slope}
(Color online) Truncated mean \pt, $\langle p_{\rm
T}^{\prime}\rangle$, in $1<p_{\rm T}^{\rm b}<5$ GeV/$c$ versus
$N_{\rm{part}}$ for the near-side (diamonds), away-side shoulder
(squares) and head (circles) regions for Au+Au (filled) and $p+p$
(open) for four trigger \pt bins. Solid lines represent values
for inclusive charged hadrons ($\sim0.36$ GeV/$c$)~\cite{Adler:2003au}.
Error bars represent the statistical errors. Shaded bars
represent the sum of $N_{\rm{part}}$-correlated
elliptic flow and ZYAM error.}
\end{figure}

First, we focus on an intermediate \pt region, $1<p_{\rm T}^{b}<5$
GeV/$c$, where the medium-induced contributions are important for
the near- and away-side yields.  Figure~\ref{fig:slope} shows
values of $\langle p_{_T}^{\rm{\prime}} \rangle$ for the HR, SR and
NR as a function of $N_{\rm{part}}$. For all trigger \pt bins, the
values for the NR drop slightly with centrality to a lower level
relative to $p+p$. This can be understood from the shape difference
in Fig.~\ref{fig:iaanear1}, where the Au+Au spectra drop faster
with increasing $p_{\rm T}^{\rm b}$. For $3<p_{\rm T}^{\rm a}<4$
GeV/$c$, a factor of two decrease in 1 - 5 GeV/$c$ amounts to a
reduction of $\sim 0.1$ GeV/$c$ in $\langle p_{_T}^{\rm{\prime}}
\rangle$.

Despite the small decrease with $N_{\rm part}$, the overall average
level of $\langle p_{_T}^{\rm{\prime}} \rangle$ for the NR for
$N_{\rm part}>100$ increases with trigger \pt. They are
$0.533+0.024-0.016$, $0.605+0.033-0.023$, $0.698+0.03-0.04$ GeV/$c$
and $0.797+0.052-0.042$ GeV/$c$ for triggers in 2-3, 3-4, 4-5 and
5-10 GeV/$c$, respectively. This trend is consistent with the
dominance of jet fragmentation on the near-side, i.e. a harder
spectrum for partner hadrons is expected for higher \pt trigger
hadrons.

Values for the SR also show an almost flat centrality dependence
for $N_{\rm{part}}\gtrsim100$. In this case, the values for
$\langle p_{_T}^{\rm{\prime}} \rangle$ are lower ($\approx 0.45$
GeV/$c$, see Table.~\ref{tab:meanpt}) and they do not depend on the
trigger \pt. They are, however, larger than the values obtained for
inclusive charged hadrons (0.36 GeV/$c$ as indicated by the solid
lines)~\cite{Adler:2003au}. The relatively sharp change in $\langle
p_{_T}^{\rm{\prime}} \rangle$ for $N_{\rm{part}}\lesssim100$ may
reflect the dominance of jet fragmentation contribution in
peripheral collisions.

The values of $\langle p_{_T}^{\rm{\prime}} \rangle$ for the HR
show a gradual decrease with $N_{\rm{part}}$. They start at a level
close to the values for the near-side, and gradually decrease with
increasing $N_{\rm part}$, consistent with a softening of the
partner spectrum in central collisions. For 2-3 and 3-4 GeV/$c$
trigger bins, the values of $\langle p_{_T}^{\rm{\prime}}\rangle$
for $N_{\rm{part}}\gtrsim 150$ approach those for inclusive
spectra. For higher trigger \pt bins, the drop with $N_{\rm part}$
is less dramatic, possibly due to punch-through jet fragmentation
contribution at high \pt.

\begin{figure}[thb]
\includegraphics[width=1.0\linewidth]{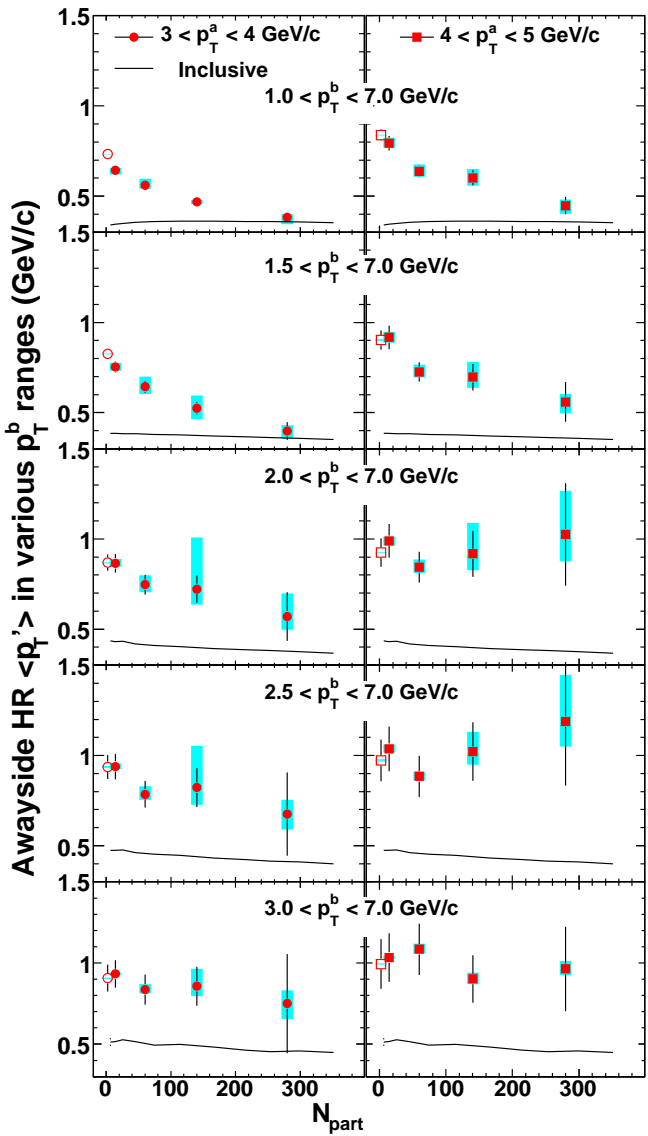}
\caption{\label{fig:slope1} (Color online) Truncated mean \pt, $\langle
p_{\rm T}^{\prime}\rangle$, calculated for five partner \pt ranges
(from top to bottom) in the HR. The left panels (right panels) shows the
results for triggers in $3-4$ ($4-5$) GeV/$c$. Solid lines
represent values for inclusive charged hadrons. Error bars
represent the statistical errors. Shaded bars represent the sum of
$N_{\rm{part}}$-correlated elliptic flow and ZYAM error.}
\end{figure}

To further investigate the onset of the jet fragmentation in the
HR, we study the dependence of $\langle
p_{_T}^{\rm{\prime}}\rangle$ on partner momentum.
Figure~\ref{fig:slope1} shows the centrality dependence of $\langle
p_{_T}^{\rm{\prime}}\rangle$ calculated in various ranges of
$p_{\rm T}^{\rm b}$ for triggers in 3-4 GeV/$c$ (left panels) and
4-5 GeV/$c$ (right panels). These results are compared with values
for inclusive hadron spectra calculated in the same $p_{\rm T}^{\rm
b}$ ranges. For $1<p_{\rm T}^{\rm b}<7$ GeV/$c$ bin, $\langle
p_{_T}^{\rm{\prime}}\rangle$ decreases with $N_{\rm{part}}$. As the
$p_{\rm T}^{\rm b}$ range shifts upward, the centrality dependence
becomes flatter. $\langle p_{_T}^{\rm{\prime}} \rangle$ for
$3<p_{\rm T}^{\rm b}<7$ GeV/$c$ is essentially constant with
$N_{\rm part}$. The flattening of the spectral slope with
$N_{\rm{part}}$ starts at a lower $p_{\rm T}^{\rm b}$ for $4<p_{\rm
T}^{\rm a}<5$ GeV/$c$ than that for $3<p_{\rm T}^{\rm a}<4$
GeV/$c$. This implies (1) a similar spectra shape for Au+Au and
$p+p$ at high $p_{\rm T}^{\rm a,b}$, and (2) the jet fragmentation
contribution dominates the HR yield at large $p_{\rm T}^{\rm a,b}$.

\begin{table}[th]
\caption{\label{tab:meanpt} Truncated mean \pt, $\langle
p_{_T}^{\rm{\prime}}\rangle$, calculated for $1<p_{\rm T}^{b}<5$
GeV/$c$ and averaged for $N_{\rm{part}}>100$ in the NR and SR for
various bins of trigger \pt.}
\begin{tabular}{ccc}\hline
$p_{\rm T}^{\rm a}$ range (GeV/$c$) & NR  $\langle p_{_T}^{\rm{\prime}}\rangle$ (GeV/$c$) & SR $\langle p_{_T}^{\rm{\prime}}\rangle$ (GeV/$c$)\\\hline
2-3                 &$0.533+0.024-0.016$  & $0.445+0.013-0.007$\\
3-4                 &$0.605+0.033-0.023$  & $0.443+0.018-0.018$\\
4-5                 &$0.698+0.030-0.040$  & $0.461+0.031-0.051$ \\
5-10                &$0.797+0.052-0.042$  & $0.478+0.079-0.139$\\\hline
\end{tabular}
\end{table}

The different patterns observed for the yields in the HR and SR
suggest a different origin for these yields. The suppression of the
HR yield and the softening of its spectrum are consistent with jet
quenching. The observed HR yield could be comprised of
contributions from ``punch-through'' jets, radiated gluons and
feed-in from the SR. By contrast, the enhancement of the SR yield
for $p_{\rm T}^{\rm a,b}\lesssim4$ GeV/$c$ suggests a remnant of
the lost energy from quenched jets. The very weak dependence on \pt
and centrality (for $N_{\rm{part}}\gtrsim100$) for its peak
location and mean \pt may reflect an intrinsic response of the
medium to the lost energy. These observations are challenging for
simple deflected jet scenarios~\cite{Armesto:2004pt,Chiu:2006pu},
since both the deflection angle and jet spectral slope would depend
on $p_{\rm T}^{\rm a}$ or $p_{\rm T}^{\rm b}$. On the other hand,
they are consistent with expectations for Mach-shock in a
near-ideal hydrodynamical
medium~\cite{Renk:2005si,Casalderrey-Solana:2004qm}, and thus they
can be used to constrain the medium transport properties such as
speed of sound and viscosity to entropy ratio within these models.
\subsection{Medium Modification of Hadron Pair Yield}
\label{sec:4.5}

In $p+p$ collisions at 200 GeV, it is generally believed that
hadrons for $p_{\rm T}>2$ GeV/$c$ are dominated by jet
fragmentation~\cite{Owens:1977sj}. By contrast, particle production
in heavy-ion collisions is complicated by final-state medium
effects. Due to strong jet quenching, jet fragmentation
contribution only dominates for $p_{\rm T}\gtrsim5-7$
GeV/$c$~\cite{Adler:2003au,Abelev:2006jr}. The bulk of hadrons,
i.e. those at $p_{\rm T}\lesssim4$ GeV/$c$, are dominated by soft
processes such as the hydrodynamic flow of locally thermalized
partonic medium~\cite{Teaney:2000cw,Huovinen:2002rn, Kolb:2002ve}
which subsequently hadronize via coalescence of constituent
quarks~\cite{Molnar:2003ff,Hwa:2004ng,Fries:2003kq, Greco:2003xt}.
The \pt of $4\lesssim p_{\rm T}\lesssim7$ GeV/$c$ is a transition
region where both soft and hard processes contribute.

Dihadron correlations provide new tools for separating the hard and
soft contributions at low and intermediate \pt, albeit for hadron
pair production instead of single hadron production. If jets are
quenched by the medium and their energy is transported to lower
\pt, a significant fraction of the low and intermediate \pt hadron
pairs may retain some correlation with the original
jet~\cite{Fries:2004hd}. Consequently, they can contribute to a
dihadron correlation analysis. However such pairs can also be
influenced by soft processes which dominate the inclusive hadron
production in the same \pt region. For instance, they may couple
with hydrodynamical flow at the partonic
stage~\cite{Armesto:2004pt} or reflect the effects of the
coalescence between shower parton and thermal partons during
hadronization~\cite{Hwa:2004ng}.

Thus far, we have quantified the jet modifications via per-trigger
yields. While sensitive to modifications of jet-induced pairs,
these yields are also sensitive to modifications in the number of
triggers. For high-\pt triggers, however, the per-trigger yield is
roughly equal to per-jet yield, because most jets fragment into at
most one trigger hadron due to the steeply falling jet
spectra~\cite{Adler:2006sc}. For intermediate and low-\pt triggers,
a large fraction of triggers may come from soft processes, and
hence ``dilute'' the per-trigger yield.

\begin{figure}[thb]
\includegraphics[width=1.0\linewidth]{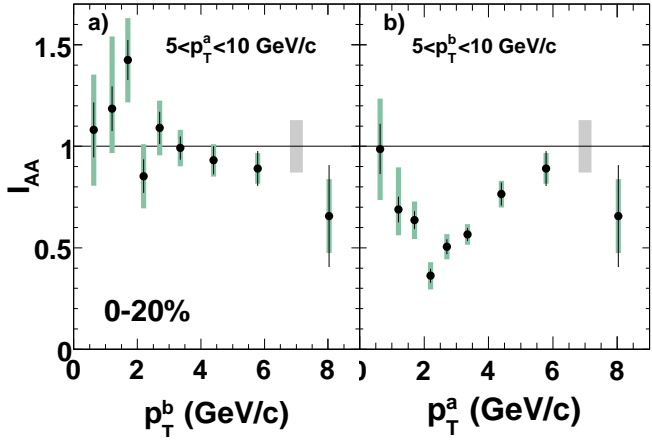}
\caption{\label{fig:near1} (Color online) a) $I_{\rm{AA}}$ versus partner \pt when
5-10 GeV/$c$ hadrons are designated triggers. b) $I_{\rm{AA}}$ versus trigger \pt
when 5-10 GeV/$c$ hadrons are used as partners.}
\end{figure}

To illustrate this dilution effect, we focus on the near-side
jet-induced pairs in which the first hadron is fixed in the 5-10
GeV/$c$ \pt range, and the second hadron is varied in \pt from 0.4
to 7 GeV/$c$. We note here that the requirement of a high-\pt
hadron ensures that most of the near-side pairs come from
fragmentation of partons.  Figure~\ref{fig:near1}a shows the
per-trigger yield modification factor $I_{\rm AA}$ when hadrons in
the 5-10 GeV/$c$ \pt range are designated triggers;
Fig.~\ref{fig:near1}b shows the corresponding $I_{\rm AA}$ when the
lower-\pt hadrons are designated as triggers, which is calculated
according to Eq.~\ref{eq:iaajaa}.

Figure~\ref{fig:near1}a shows that $I_{\rm{AA}}$ is near unity for
all $p_{\rm T}^{\rm b}$ for the 5-10 GeV/$c$ hadron triggers. This
is consistent with each high-\pt trigger coming from one jet. On
the other hand, when low-\pt hadrons are used as triggers,
Fig.~\ref{fig:near1}b shows that $I_{\rm{AA}}$ has a non-trivial
dependence on $p_{\rm T}^{\rm a}$. That is, there is a strong
suppression of $I_{\rm{AA}}$ in the 2-4 GeV/$c$ \pt range, which
reflects an excess of trigger hadrons at low \pt with weaker jet
correlation strength. This dilution effect is also reflected in the
near-side $I_{\rm AA}$ values in Fig.~\ref{fig:iaanear1} and for
the away-side $I_{\rm AA}$ values in Fig.~\ref{fig:iaaaway1} (also
reflected in Fig.\ref{fig:shape1} for low $p_{\rm T}^{\rm a}$ and
high $p_{\rm T}^{\rm b}$). The former shows a suppression at large
$p_{\rm T}^{\rm b}$ for soft triggers ($2<p_{\rm T}^{\rm a}<3$ and
$3<p_{\rm T}^{\rm a}<4$ GeV/$c$). The latter shows a stronger
suppression for low-\pt triggers ($2<p_{\rm T}^{\rm a}<3$ and
$3<p_{\rm T}^{\rm a}<4$ GeV/$c$) than for high-\pt triggers
($4<p_{\rm T}^{\rm a}<5$ and $5<p_{\rm T}^{\rm a}<10$ GeV/$c$).
This dilution effect might be the result of the following two
scenarios: 1) a large fraction of low-\pt hadrons are from soft
processes such as coalescence of flow-boosted thermal quarks
related to the anomalous proton/pion ratio~\cite{Adler:2003au,
Abelev:2006jr}, or 2) jets are quenched and these hadrons are the
remnants of the quenched jets, and thus lack associated hadrons at
high \pt.

\begin{figure}[thb]
\includegraphics[width=1.0\linewidth]{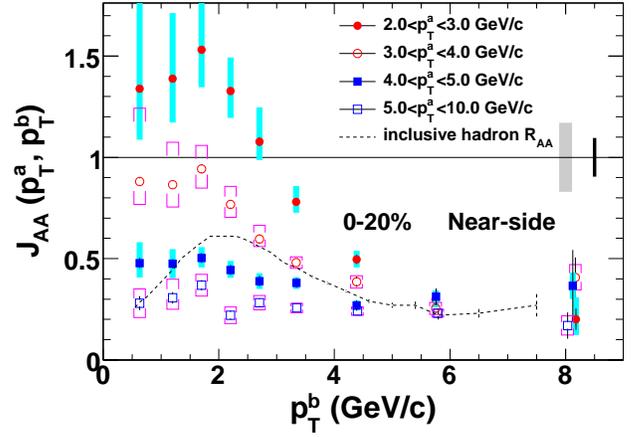}
\caption{\label{fig:near2} (Color online) The near-side pair
suppression factor $J_{\rm{AA}}$ in 0-20\% Au+Au collisions as
function of $p_{\rm T}^{\rm b}$ for various ranges of $p_{\rm
T}^{\rm a}$. The dashed line indicate the charged hadron
$R_{\rm{AA}}$. The grey band around one is the systematic error due
to the 20\% combined single particle efficiency of the triggers and
partners in Au+Au and $p+p$. The dark line around one indicate the
uncertainty on the $N_{\rm coll}$.}
\end{figure}

To gain more insights on intermediate-\pt correlations, we focus on
the pair suppression factor $J_{\rm{AA}}$ defined in
Eq.~\ref{eq:jaa}. We recall here that $J_{\rm{AA}}$ quantifies the
modification of jet-induced pairs in Au+Au relative to that in
$p+p$ normalized by $N_{\rm{coll}}$. It is symmetric with respect
to the interchange of $p_{\rm T}^{\rm a}$ and $p_{\rm T}^{\rm b}$
and equals one in the absence of medium effects.
Figure~\ref{fig:near2} shows the near-side $J_{\rm{AA}}$ as function
of $p_{\rm T}^{\rm b}$ for 0-20\% central Au+Au collisions and for
four different $p_{\rm T}^{\rm a}$ bins. Values of $J_{\rm{AA}}$
are above one for $p_{\rm T}^{\rm a,b}<2$ GeV/$c$. However, they
decrease with increasing $p_{\rm T}^{\rm a,b}$ and drop below one.
For high $p_{\rm T}^{\rm a,b}$, $J_{\rm{AA}}$ reaches a constant
value $\sim 0.2-0.3$, which is similar to the high-\pt single
particle suppression factor $R_{\rm AA}$.

To interpret these observations, we note that each high \pt pair at
the near-side comes from the same jet. Thus $J_{\rm{AA}}$ reflects
the modification of single jets, which at high \pt should be the
same as leading hadron $R_{\rm{AA}}$. Since the values of
$R_{\rm{AA}}$ is constant at high \pt, we expect high-\pt
$J_{\rm{AA}}$ to be constant and equal to $R_{\rm{AA}}$.

\begin{figure}[thb]
\includegraphics[width=1.0\linewidth]{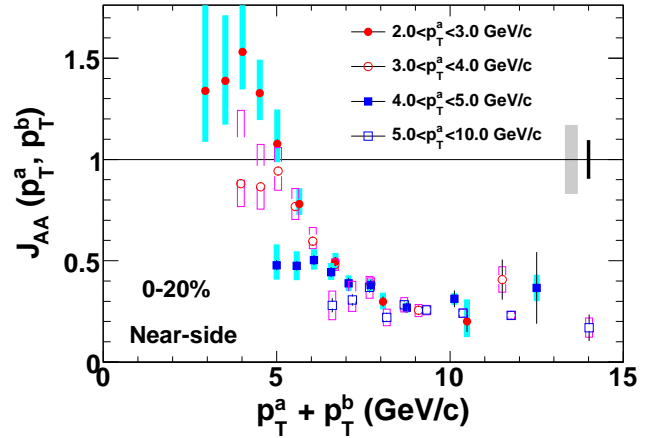}
\caption{\label{fig:nearsum1} (Color online) Near-side pair suppression
factor $J_{\rm{AA}}$ as function of
pair transverse momentum, $p_{\rm T}^{\rm{sum}} = p_{\rm T}^{\rm a}+p_{\rm T}^{\rm b}$ in 0-20\% Au+Au collisions.
The grey band around one is the systematic error due to the 20\%
combined single particle efficiency of the triggers and partners in
Au+Au and $p+p$. The dark line around one indicate the uncertainty
on the $N_{\rm coll}$.}
\end{figure}

Furthermore, if each high-\pt near-side pair comes from the same
jet, then the sum of their transverse momentum, $p_{\rm
T}^{\rm{sum}} = p_{\rm T}^{\rm a}+p_{\rm T}^{\rm b}$, should serve
as a better proxy for the original jet energy. With this in mind,
we re-plot in Fig.~\ref{fig:nearsum1} the near-side $J_{\rm{AA}}$
values as a function of this ``pair proxy energy'' $p_{\rm
T}^{\rm{sum}}$. Interestingly, the pair modification factors
roughly follow a single curve in $p_{\rm T}^{\rm{sum}}$. It is
above 1 below 5 GeV/$c$, followed by a decrease with $p_{\rm
T}^{\rm sum}$, and reaches a constant for $p_{\rm T}^{\rm{sum}}>8$
GeV/$c$. The approximate scaling behavior breaks when the \pt of
one hadron is $\lesssim2$ GeV/$c$, where $J_{\rm{AA}}$ is
systematically below the overall trend.

\begin{figure}[thb]
\includegraphics[width=1.0\linewidth]{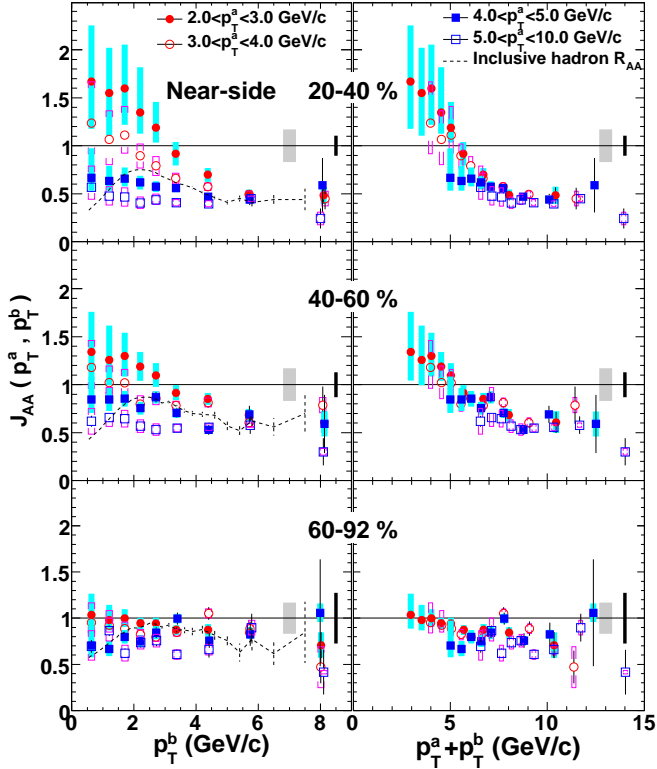}
\caption{\label{fig:jaacent} (Color online) The near-side
$J_{\rm{AA}}$ as function of $p_{\rm T}^{\rm b}$ (left panels) and
$p_{\rm T}^{\rm{sum}}$ (right panels) for four $p_{\rm T}^{\rm a}$
bins and three centrality bins. From top to bottom are 20-40\%,
40-60\% and 60-92 \%. The dashed lines in the left panels indicate
the charged hadron $R_{\rm{AA}}$. The grey bands around one is the
systematic error due to the 17\% combined single particle
efficiency of the triggers and partners in Au+Au and $p+p$. The
dark line around one indicate the uncertainty on the $N_{\rm
coll}$.}
\end{figure}

The fact that $J_{\rm AA}>1$ for $p_{\rm T}^{\rm{sum}}<5$ GeV/$c$
implies that the total jet-induced pair yield is enhanced relative
to the $N_{\rm coll}$ scaled $p+p$ collisions. This enhancement may
reflect the energy of the quenched jets being transported to low
\pt. $J_{\rm{AA}}$ is almost a factor of 6-7 larger than its high
$p_{\rm T}^{\rm{sum}}$ limit. By contrast, the enhancement shown
for $I_{\rm{AA}}$ in Fig.~\ref{fig:iaanear1} is only a factor of
2.5 at low $p_{\rm T}^{\rm a,b}$. This difference can be attributed
to the dilution effect on the triggers. For completeness,
Fig.~\ref{fig:jaacent} shows the values of $J_{\rm{AA}}$ versus
$p_{\rm T}^{\rm b}$ (left panels) and $p_{\rm T}^{\rm{sum}}$ (right
panels) for 20-40\%, 40-60\% and 60-92\% centrality bins.
$J_{\rm{AA}}$ versus $p_{\rm T}^{\rm{sum}}$ shows an approximate
scaling behavior for all centralities i.e, $J_{\rm{AA}}$ for large
$p_{\rm T}^{\rm{sum}}$ approaches a constant level roughly equal to
that for the high-\pt $R_{\rm{AA}}$ values.

\begin{figure}[thb]
\includegraphics[width=1.0\linewidth]{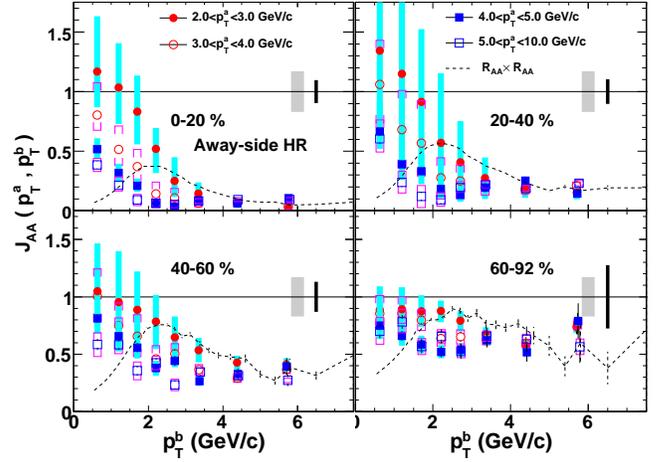}
\caption{\label{fig:awaysum1} (Color online) The pair suppression
factor $J_{\rm{AA}}$ for the away-side HR in Au+Au collisions as
function of $p_{\rm T}^{\rm b}$ for various ranges of $p_{\rm
T}^{\rm a}$. The dashed line indicate square of the charged hadron
$R_{\rm{AA}}$. The grey bands around one is the systematic error
due to the combined single particle efficiency of the two particles
in Au+Au and $p+p$. The dark line around one indicate the
uncertainty on the $N_{\rm coll}$.}
\end{figure}

In Fig.~\ref{fig:awaysum1} we show the $J_{\rm{AA}}$ for the
away-side HR as a function of $p_{\rm T}^{\rm b}$ in various
centrality bins. In central collisions, a possible enhancement at
low $p_{\rm T}^{\rm a,b}$ and a strong suppression at large $p_{\rm
T}^{\rm a,b}$ can be seen. This is consistent with the feedback of
lost energy to lower \pt on the away-side. The modifications
decrease for peripheral collisions, as expected for a weaker medium
effect. However, the suppression level seems to approach a constant
value for high $p_{\rm T}^{\rm b}$ for all centralities. This is
expected since $I_{\rm AA}\approx R_{\rm AA}$ at high \pt for the
away-side HR. This implies that $J_{\rm AA}=I_{\rm AA}R_{\rm
AA}\approx R_{\rm AA}^2$ when one of the hadrons is at high \pt, as
indicated by the dashed line.

To quantify this high-\pt scaling behavior, we calculate the ratio
$\frac{J_{\rm AA}}{R_{\rm AA}^2}$ for away-side head region for
$4<p_{\rm T}^{\rm b}<7$ GeV/$c$ and various $p_{\rm T}^{\rm a}$
selections. The values for the four centrality ranges used in
Fig.\ref{fig:awaysum1} are summarized in Table.\ref{tab:scal}.
Although the uncertainties are quite substantial, the HR $J_{\rm
AA}$ approximately equals $R_{\rm AA}^2$, suggesting a similar
suppression factor for inclusive hadrons and away-side jet at high
\pt. This, as pointed out in Ref.~\cite{Jia:2007qi,Zhang:2007ja},
could be a canceling effect between a stronger energy loss which
increase the suppression, and a harder away-side hadron spectra
associated with high-\pt triggers which decrease the suppression.

\begin{table}
\caption{\label{tab:scal} The average ratio $J_{\rm AA}/R_{\rm
AA}^2$ for the away-side HR for $4<p_{\rm T}^{\rm b}<7$ GeV/$c$
(see Fig.~\ref{fig:awaysum1}).}
\begin{tabular}{cc}
$\rm Centrality$  & $J_{\rm AA}/R_{\rm AA}^2$ for HR ($\pm$Stat+Sys-Sys)\\\hline
0-20\%  & $0.81\pm0.07+0.44-0.41$\\
20-40\% & $0.89\pm0.05+0.37-0.35$\\
40-60\% & $0.80\pm0.03+0.26-0.23$\\
60-92\% & $0.90\pm0.03+0.25-0.23$\\\hline
\end{tabular}
\end{table}

\section{DISCUSSION}
\label{sec:5}
\subsection{Insights from Identified Particle and Energy Dependent Correlations}
To elucidate the underlying physics of the medium-induced
component, we focus on intermediate \pt where the SR dominates, and
study the particle composition of the yield in the SR. PHENIX has
published results on correlations of a trigger hadron, at
intermediate transverse momentum ($2.5 < p_{\rm T}^{\rm a} < 4$
GeV/$c$), with identified partner mesons or baryons at lower \pt
\cite{Afanasiev:2007wi}. The away-side shape was found to be
similar for partner baryons and mesons; namely, the pairs peak at
$\Delta\phi\sim\pi\pm1.1$ with a local minimum at
$\Delta\phi\sim\pi$. The particle composition in the away-side jet,
as reflected by the baryon to meson ratio, was also found to grow
with increasing partner \pt. The trend is similar to that observed
for inclusive hadron production. These observations for
intermediate \pt correlations are consistent with strong
parton-medium interactions which induce correlations between soft
partons, followed by coalescence at hadronization.

\begin{figure}[thb]
\includegraphics[width=1.0\linewidth]{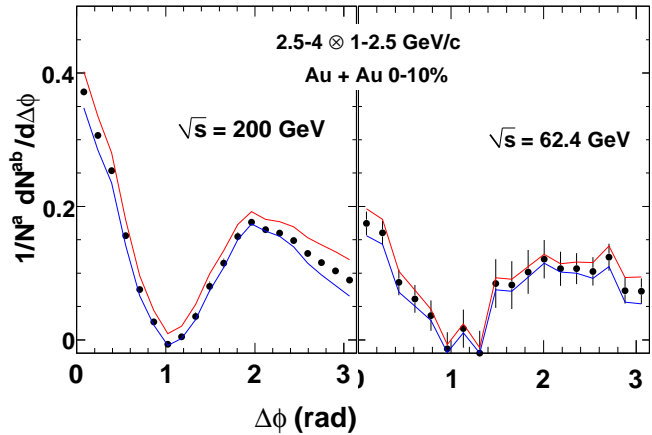}
\caption{\label{fig:jetsqrt} (Color online) Per-trigger yield in 0-10\% central
Au+Au collisions from PHENIX at (a) $\sqrt{s_{\rm{NN}}} =$ 200 GeV
and (b) $\sqrt{s_{\rm{NN}}} =$ 62.4 GeV.
The histograms show the combined uncertainty of the elliptic flow and ZYAM errors.}
\end{figure}

Further insight into the physics underlying the SR yield can be
obtained by studying its energy dependence. In particular, it is
interesting to see whether the two-component picture applies at
much lower collision energy.   Figure~\ref{fig:jetsqrt} compares the
per-trigger yield at $\sqrt{s_{\rm NN}}$ = 200 and 62.4 GeV for
$1<p_{T}^{\rm b}<2.5<p_{\rm T}^{\rm a}<4$ GeV/$c$. They are the
yields associated with the jet functions previously published
in~\cite{Adare:2006nr}. The away-side shapes are strongly
non-Gaussian in both cases. The 62.4 GeV data seems to be somewhat
flatter, however, its relatively large statistical uncertainties do
not allow a definite statement.

The CERES collaboration has recently released their high-statistics
preliminary results of per-trigger yields in 0-5\% and 5-10\% Pb+Au
collisions for $1<p_{T}^{\rm b}<2.5<p_{\rm T}^{\rm a}<4$
GeV/$c$~\cite{Ploskon:2007es}. This measurement was carried out at
$\sqrt{s_{\rm{NN}}} =$ 17.2 GeV at the SPS, for $0.1<\eta<0.7$ in
the center of mass (CM) frame. The equivalent pseudo-rapidity
window of 0.6 is close to the value of 0.7 for PHENIX. Thus, the
jet yields from both experiments can be compared after applying the
correction of 0.7/0.6 = 1.17. In contrast to the PHENIX results,
the CERES data show an essentially flat away-side jet shape. The
maximum of SR is about half that of the PHENIX value, whereas the
yield at the HR is close to the PHENIX value. The former might
suggest a weaker medium effect at lower energy; the latter could be
a combination of a lower jet multiplicity and a weaker jet
quenching at SPS energy. However, it is conceivable that other
nuclear effects, especially Cronin effect~\cite{Cronin:1974zm}, may
also broaden the away-side jet shape. Further detailed study of the
collision energy dependence of the HR and the SR components might
elucidate the onset of jet quenching and medium response.

\subsection{Comparison with models}

If jets are generated close to the surface, they exit and
subsequently fragment outside the medium. Otherwise, they lose
energy by radiating gluons. These shower gluons may be emitted at
large angles relative to the original partons~\cite{Vitev:2005yg,
Polosa:2006hb} and fragment into hadrons, or they can be deflected
to large angles by interactions with medium. Examples of the latter
include medium deflection in the
azimuthal~\cite{Chiu:2006pu,Armesto:2004pt} and the beam
directions~\cite{Majumder:2006wi} or excitation of collective Mach
shock~\cite{Stoecker:2004qu, Casalderrey-Solana:2004qm};

\begin{figure}[thb]
\includegraphics[width=1.0\linewidth]{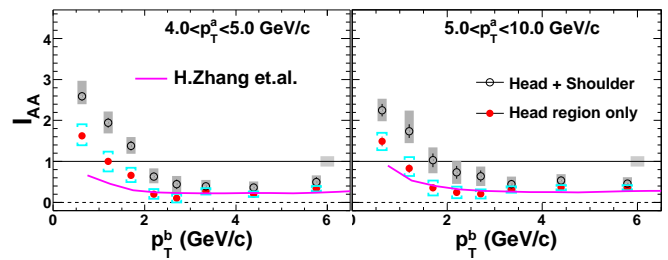}
\caption{\label{fig:comp1} (Color online) Comparison of $I_{\rm AA}$ with energy loss calculation from Ref.~\cite{Zhang:2007ja}.}
\end{figure}

Several calculations for radiative energy loss have been carried
out~\cite{Vitev:2005yg,Renk:2006pk,Zhang:2007ja} to describe
dihadron production at high \pt. They all describe the data fairly
well. As an example, Fig.~\ref{fig:comp1} shows a comparison of
data for the 4-5 and 5-10 GeV/$c$ triggers with recent calculations
from Ref.~\cite{Zhang:2007ja}. For $p_{\rm T}^{\rm a}>2$ GeV/$c$,
the calculated $I_{\rm AA}$ is approximately constant and agrees
well with the data. According to this calculation, both tangential
and punch-through jet emission are important, accounting for 3/4
and 1/4 of the away-side high-\pt pairs, respectively.

To extend the calculation to the low and intermediate \pt,
contributions from radiated gluons have to be taken into account.
Early energy loss model calculations suggested that these radiated
gluons are almost collinear with the jet
axis~\cite{Salgado:2003rv}. However, recent
calculations~\cite{Vitev:2005yg} favor large angle gluon emission
due to destructive interference that suppresses collinear emission.
By including gluon feedback, the calculation can reproduce
away-side per-trigger yields at low partner \pt, but the gluon
emission angle is too small to reproduce the away-side jet shape.

In an improved calculation that includes the Sudakov form
factors~\cite{Polosa:2006hb}, the authors can qualitatively
describe the away-side jet shape and its centrality dependence at
intermediate \pt, when the leading parton is assumed to split into
two semi-hard gluons which then fragment into hadrons. This model
assumes a transport coefficient of $\hat{q}\sim 5-10$ GeV$^{2}$/fm.
A smaller $\hat{q}$, for instance, would substantially reduce the
predicted split angle.

The away-side broadening may also arise from Cherenkov gluon
radiation~\cite{Koch:2005sx}. It could occur when the gluon is
scattered by colored bound states in such a way that the
permittivity for in-medium gluons becomes space-like. A first
calculation that includes only scalar bound states suggests that
the peak angle D should gradually decrease to zero with increasing
momentum; this trend seems to be ruled out by the present data.
More sophisticated calculations including other bound states are
needed in future studies.

It was suggested that shower partons could couple with the
longitudinal and transverse flow, and are broadened or deflected in
$\Delta\eta$~\cite{Armesto:2004pt,Voloshin:2004th,Chiu:2005ad,Wong:2007pz}
and/or $\Delta\phi$~\cite{Chiu:2006pu,Armesto:2004pt} directions.
The longitudinal deflection was argued to be responsible for the
$\Delta\eta$ ridge structure at the near-side. The transverse
deflection can lead to broadening in $\Delta\phi$. It was argued
in~\cite{Chiu:2006pu} that a random multiple scattering of the
leading parton, combined with energy loss, can result in the double
peaked structure of the away-side. However, in general, the
deflection angle decreases with the hadron momentum. This is not
compatible with the observation of \pt-independent D values and
universal spectral slopes of the SR.

Finally, it has been proposed that the lost energy can be absorbed
by the medium and converted into collective Mach
shock~\cite{Stoecker:2004qu, Casalderrey-Solana:2004qm}. In this
picture, the fluid elements are boosted along the Mach
angle~\cite{Renk:2005si} and then hadronize via coalescence. The
Mach angle depends only on the sound speed of the medium thus is
independent of \pt and consistent with the data. The boost effect
also produces a harder slope for the partners, qualitatively
consistent with experimental observations. Our results on the PID
dependence of the correlation pattern~\cite{Afanasiev:2007wi} are
consistent with the above picture. The propagation of the shock
wave requires hydrodynamical behavior of the medium with small
viscosity. If the Mach-shock is the underlying physics, the
observation can be used to constrain the value of $\eta/s$.

Many of the models discussed in this section are quite qualitative
in nature. These models typically focus on either jet shape or jet
yield, near-side or away-side, high \pt or low \pt. The fact that
both near- and away-side distributions are enhanced and broadened
at low \pt and that the modifications limited to $p_{\rm
T}\lesssim4$ GeV/$c$, above which the jet characteristics
qualitatively approach jet fragmentation, may suggest that the
modification mechanisms for the near- and away-side are related. A
model framework including both jet quenching and medium response,
which can describe the full $p_{\rm T}$ evolution of the jet shape
and yield at both near- and away-side is required to understand the
parton-medium interactions. Our data provide valuable guidance for
such future model developments.

\section{SUMMARY AND CONCLUSIONS}

We have analysed dihadron azimuthal correlations for $0.4 < p_{\rm
T} < 10$~GeV/$c$ unidentified charged hadrons in Au+Au collisions
at $\sqrt{s_{\rm{NN}}} = 200$ GeV. The results are presented as
functions of trigger \pt, partner \pt and centrality, and are
compared with correlations of identified hadrons as well as results
from lower energies. The evolution of the jet shape and yield with
\pt seems to suggest four distinct contributions to jet-induced
pairs: 1) a jet fragmentation component around $\Delta\phi\sim0$,
2) a punch-through jet fragmentation component around
$\Delta\phi\sim\pi$, 3) a medium-induced component around
$\Delta\phi\sim0$, and 4) a medium-induced components around
$\Delta\phi\sim\pi\pm1.1$.

The jet fragmentation components arise from jets that suffer small
energy loss due to the surface or punch-through jet emissions. They
dominate the near- and away-side pairs at large \pt. The near-side
pair suppression factor $J_{\rm AA}$ follows an approximate scaling
with the pair proxy energy $p_{\rm T}^{\rm sum}$. It reaches a
constant for $p_{\rm T}^{\rm sum}>8$ GeV/c, at a level similar to
the suppression for single jets at high \pt. In this \pt region,
the yield of both the single jets (near-side pairs) and
back-to-back jets (away-side pairs) are consistent with energy loss
calculations.

By contrast, the enhancement of medium-induced components may
reflect a remnant of the lost energy from quenched jets. This
enhancement is limited to $p_{\rm T}^{\rm A,B}\lesssim4$ GeV/$c$.
The near-side medium-induced component is responsible for
broadening in $\Delta\phi$ and significant elongation in
$\Delta\eta$, and is related to the ridge structure in
~\cite{Adams:2004pa,Adams:2005ph, Adams:2006tj}. The away-side
medium-induced component exhibits \pt- and centrality-independent
shape and mean \pt, and a bulk-medium like particle
composition~\cite{Afanasiev:2007wi}, possibly reflecting an
intrinsic property of the medium response to energetic jets.

We have also investigated the contribution of medium-induced
components to single particle production at intermediate \pt, where
soft processes such as hydrodynamical flow and coalescence are
important. The yields of jet-induced hadron pairs (via $J_{\rm
AA}$) are not suppressed at low pair proxy energy $p_{\rm T}^{\rm
sum}=p_{\rm T}^{\rm a}+p_{\rm T}^{\rm b}$. However, pair yields
divided by the yield of soft triggers show an apparent dilution
effect at large partner \pt. This suggests that these soft hadrons
either come from soft processes such as from coalescence of thermal
partons, or they are the remnant of quenched jets, and thus lack
high-\pt jet partners.


\section{ACKNOWLEDGEMENTS}

We thank the staff of the Collider-Accelerator
and Physics Departments at Brookhaven National Laboratory and the
staff of the other PHENIX participating institutions for their
vital contributions.
We thank Hanzhong Zhang and XinNian Wang for providing theoretical
calculation input.
We acknowledge support from the Department of
Energy, Office of Science, Office of Nuclear Physics, the National
Science Foundation, Abilene Christian University Research Council,
Research Foundation of SUNY, and Dean of the College of Arts and
Sciences, Vanderbilt University (U.S.A), Ministry of Education,
Culture, Sports, Science, and Technology and the Japan Society for
the Promotion of Science (Japan), Conselho Nacional de
Desenvolvimento Cient\'{\i}fico e Tecnol{\'o}gico and Funda\c
c{\~a}o de Amparo {\`a} Pesquisa do Estado de S{\~a}o Paulo
(Brazil), Natural Science Foundation of China (People's Republic of
China), Ministry of Education, Youth and Sports (Czech Republic),
Centre National de la Recherche Scientifique, Commissariat {\`a}
l'{\'E}nergie Atomique, and Institut National de Physique
Nucl{\'e}aire et de Physique des Particules (France), Ministry of
Industry, Science and Tekhnologies, Bundesministerium f\"ur Bildung
und Forschung, Deutscher Akademischer Austausch Dienst, and
Alexander von Humboldt Stiftung (Germany), Hungarian National
Science Fund, OTKA (Hungary), Department of Atomic Energy (India),
Israel Science Foundation (Israel), Korea Research Foundation and
Korea Science and Engineering Foundation (Korea), Ministry of
Education and Science, Rassia Academy of Sciences, Federal Agency
of Atomic Energy (Russia), VR and the Wallenberg Foundation
(Sweden), the U.S. Civilian Research and Development Foundation for
the Independent States of the Former Soviet Union, the US-Hungarian
NSF-OTKA-MTA, and the US-Israel Binational Science Foundation.

\appendix

\section{DIHADRON CORRELATION METHOD}
\label{appendix:A}

This section demonstrates that the shape of the
$\Delta\phi$ correlation function, constructed as the ratio of
same- to mixed-event, reproduces the shape of the true-pair
distribution in $\Delta\phi$. We demonstrate this is generally true
for a non-uniform experimental acceptance such as in PHENIX. Our
argument is not original with our analysis, but has been used
numerous times before in correlation analyses in heavy-ion
experiments.

We start by giving notation for the true distributions of type $a$
and type $b$ particles. They are
\begin{equation}
\frac{d^2 N_{0}^{ab}}{d\phi^{a} \; d\phi^{b}}, \;\;\;\;\;\;\;
\frac{d N_{0}^{a}}{d\phi^{a}}, \;\;\;\;\;\;\; \frac{d
N_{0}^{b}}{d\phi^{b}} \label{eq:define_true}
\end{equation}
for the true azimuthal distributions for $ab$ pairs and $a$, $b$
singles produced for PHENIX pseudo-rapidity acceptance and for
events in one centrality bin. We use superscript ``0'' to indicate
the true distributions. The {\it true} distributions are for
particles in PHENIX $\eta$ range, but with full azimuthal coverage.

The PHENIX Beam-Beam Counter (BBC) and Zero-Degree Calorimeter
(ZDC), which trigger on events and determine their centrality, are
uniform in azimuth. Therefore the {\it true} singles distributions
are uniform, and the {\it true} pair distribution depends only on
the difference between the two angles:
\begin{equation}
\frac{d^2 N_{0}^{ab}}{d\phi^{a} \; d\phi^{b}} =
f(\phi^{a}-\phi^{b}), \;\; \frac{d N_{0}^{a}}{d\phi^{a}} = {\rm
const}, \;\; \frac{d N_{0}^{b}}{d\phi^{b}} = {\rm const}
\label{eq:true_distr_isotropic}
\end{equation}

To study the pair distribution as a function of $\Delta\phi$, we
define the difference and average of $\phi^{a}$ and $\phi^{b}$ as
new orthogonal variables:
\begin{equation}
\Delta\phi \equiv \phi^{a}-\phi^{b}, \;\;\;\;\;\;\; \bar{\Phi}
\equiv (\phi^{a}+\phi^{b})/2
\end{equation}
We integrate the true pairs over $\bar{\Phi}$ to obtain the
projection onto $\Delta\phi$. This is equivalent to binning the
data in $\Delta\phi$.

\begin{eqnarray}
\frac{dN_{0}^{ab}}{d(\Delta\phi)} & = & \int \frac{d^2
N_{0}^{ab}}{d\phi^{a} \; d\phi^{b}} \; d \bar{\Phi}
 =  \int f(\Delta\phi) \; d \bar{\Phi}
\nonumber \\
& =&
 f(\Delta\phi) \; \int d \bar{\Phi}
 \propto f(\Delta\phi)
\label{eq:true_pairs}
\end{eqnarray}

The measured distributions (without superscript) are related to the
true distributions through the experimental acceptance/efficiency
(here just ``acceptance'' for short):
\begin{eqnarray}
\nonumber &&\frac{d^2 N^{ab}}{d\phi^{a} \; d\phi^{b}} =
\varepsilon^{ab}(\phi^{a},\phi^{b}) \; \frac{d^2
N_{0}^{ab}}{d\phi^{a} \; d\phi^{b}} , \\&&\frac{d
N^{a}}{d\phi^{a}} = \varepsilon^{a}(\phi^{a}) \; \frac{d
N_{0}^{a}}{d\phi^{a}} , \; \frac{d N^{b}}{d\phi^{b}} =
\varepsilon^{b}(\phi^{b}) \; \frac{d N_{0}^{b}}{d\phi^{b}}
\label{eq:true_measured_relation}
\end{eqnarray}

\smallskip
\noindent where $\varepsilon^{ab}$, $\varepsilon^{a}$ and
$\varepsilon^{b}$ describe the experimental acceptances for pairs
and singles. The pair acceptance is, to a very good approximation,
equal to the product of the singles acceptances:
\begin{equation}
\varepsilon^{ab}(\phi^{a},\phi^{b}) = \varepsilon^{a}(\phi^{a}) \;
\varepsilon^{b}(\phi^{b}) \label{eq:acceptances}
\end{equation}

\smallskip
\noindent i.e. the experimental acceptance for {\it A} particles is
not influenced by the presence or absence of {\it B} particles in
any particular event, and vice versa~\cite{footnote2}.

The numerator of the correlation function is the measured pair
distribution projected onto $\Delta\phi$
\begin{eqnarray}
N^{\rm same} (\Delta\phi) & = & \int \frac{d^2
N_{0}^{ab}}{d\phi^{a} \; d\phi^{b}}
\; d \bar{\Phi}\nonumber \\
 & = & f(\Delta\phi)
\int \varepsilon^{ab}(\phi^{a},\phi^{b}) \; d \bar{\Phi}
\label{eq:cf_numerator}
\end{eqnarray}

Mixed-event pairs are constructed by combining a particle at
$\phi^{a}$ from one event with particles at $\phi^{b}$ from other,
unrelated events. The mixed-event pairs over $(\phi^{a},\phi^{b})$
factorize and have the form
\begin{equation}
\frac{d^{2} N^{ab}_{\rm mixed} }{d\phi^{a} \; d\phi^{b}} \propto
\frac{d N^{a} }{d\phi^{a}} \frac{d N^{b} }{d\phi^{b}}
\label{eq:measured_factorize}
\end{equation}
The denominator of the correlation function is the projection of
the measured mixed-event pairs:
\begin{eqnarray}
N^{\rm mixed}(\Delta\phi) & = & \int \frac{d^2 N^{ab}_{\rm
mixed}}{d\phi^{a} \; d\phi^{b}} \;
d \bar{\Phi} \nonumber \\
 & \propto &
\int \frac{d N^{a} }{d\phi^{a}} \frac{d N^{b} }{d\phi^{b}} \;
d \bar{\Phi} \nonumber \\
 & \propto &
\int \varepsilon^{a}(\phi^{a}) \varepsilon^{b}(\phi^{b}) \; d
\bar{\Phi} \label{eq:cf_denominator}
\end{eqnarray}

Writing the correlation function with
Equations~\ref{eq:cf_numerator} and~\ref{eq:cf_denominator}, and
applying Eq.~\ref{eq:acceptances} and Eq.~\ref{eq:true_pairs}
yields
\begin{eqnarray}
C(\Delta\phi) &\equiv& \frac{N^{\rm same} (\Delta\phi) }{ N^{\rm
mixed}(\Delta\phi) } \nonumber \\
& \propto & \frac{f(\Delta\phi) \; \int
\varepsilon^{ab}(\phi^{a},\phi^{b}) \; d \bar{\Phi} }{ \int
\varepsilon^{a}(\phi^{a}) \varepsilon^{b}(\phi^{b}) \; d
\bar{\Phi} }
\label{eq:cf_topline} \\
& \propto & f(\Delta\phi) \propto
              \frac{dN_{0}^{ab}}{d(\Delta\phi)}
\label{eq:cf_bottomline}
\end{eqnarray}

\smallskip
\noindent This shows that the shape of the true-pair distribution
is recovered in the correlation function.

\section{THE ROLE OF REACTION PLANE}
\label{appendix:B}

No mention, explicit or implicit, was made of
the reaction plane in the preceding proof; this is not surprising,
since its validity holds for {\em any} source of correlation,
whether from flow, jets, or other. We show that
Eq.~\ref{eq:cf_bottomline} holds for limited detector acceptance
with the reaction plane included explicitly.

For events with reaction plane direction $\Psi$ we define the
conditional probabilities of finding an $a$ or a $b$ particle,
including the effects of acceptance:
\begin{eqnarray}
\nonumber P^{a}(\phi^{a} | \Psi) = \varepsilon^{a}(\phi^{a}) \;
\frac{dN^{a}}{d(\phi^{a} - \Psi)},\\ P^{b}(\phi^{b} | \Psi) =
\varepsilon^{b}(\phi^{b}) \; \frac{dN^{b}}{d(\phi^{b} - \Psi)}
\label{eq:cond_probs}
\end{eqnarray}
We can write the acceptances, and the {\it true} singles
distributions with respect to $\Psi$, into their Fourier
expansions:

\begin{equation}
\varepsilon^{a}(\phi) =  \sum_{p=-\infty}^{p=+\infty} a_{p}
e^{ip\phi} \;\;\;\;\;\; \varepsilon^{b}(\phi) =
\sum_{q=-\infty}^{q=+\infty} b_{q} e^{iq\phi}
\label{eq:acceptances_transforms}
\end{equation}

\noindent where $a_{-p} = a^*_{p}$, $b_{-q} = b^*_{q}$ and

\begin{eqnarray}
\frac{dN^{a}}{d(\phi^{a} -\Psi)} & = & \frac{1}{2\pi}
\sum_{n=-\infty}^{n=+\infty} \nu^{a}_{n} \; e^{in(\phi^{a}-\Psi)}
\nonumber \\& =& \frac{1}{2\pi} \left(1 + \sum_{n=1}^{+\infty} 2
\nu^{a}_{n} \cos n(\phi^{a}-\Psi) \right)
\nonumber \\
\frac{dN^{b}}{d(\phi^{b} -\Psi)} & = & \frac{1}{2\pi}
\sum_{m=-\infty}^{m=+\infty} \nu^{b}_{m} \; e^{im(\phi^{b}-\Psi)}\nonumber \\
 &= &\frac{1}{2\pi} \left(1 + \sum_{m=1}^{+\infty} 2 \nu^{b}_{m} \cos m(\phi^{b}-\Psi) \right)
\label{eq:singles_transforms}
\end{eqnarray}
In the case that $ab$ correlations are due to particle correlation
with respect to the same reaction plane, as would be true of
background pairs, then the {\it measured} same-event pair
distribution can be written as

\begin{eqnarray}
N^{\rm same}(\Delta\phi) &=& \int \frac{d\phi^{a}}{2\pi}
\frac{d\Psi}{2\pi} P^{a}(\phi^{a} | \Psi) P^{b}((\phi^{a} -
\Delta\phi)|\Psi) \nonumber \\&=& \sum_{p=-\infty}^{p=+\infty}
\sum_{n=-\infty}^{n=+\infty} a_{p} \; b_{p}^{*} \; \nu^{a}_{n}
\nu^{b}_{n} e^{i(p+n)\Delta\phi}
\end{eqnarray}
Similarly, the {\it measured} mixed-event pair distribution is

\begin{eqnarray}
N^{\rm mixed}(\Delta\phi) & = & \int
\frac{d\phi^{a}d\Psi^{a}d\Psi^{b}}{8\pi^3}
 P^{a}(\phi^{a} | \Psi^{a})
 P^{b}((\phi^{a} - \Delta\phi)|\Psi^{b})
\nonumber \\
& = & \sum_{p=-\infty}^{p=+\infty} a_{p} \; b_{p}^{*} \; e^{ip\Delta\phi}
\end{eqnarray}

\smallskip
\noindent Using these to construct the correlation function, we
find

\begin{eqnarray}
C(\Delta\phi) & \equiv & \frac{N^{\rm same}(\Delta\phi)}{N^{\rm
mixed}(\Delta\phi)}  = \sum_{n=-\infty}^{n=+\infty} \; \nu^{a}_{n}
\nu^{b}_{n} e^{in\Delta\phi}
\end{eqnarray}
It is clear that the dependence upon the reaction plane angle is
integrated out when forming the correlation function. It is also
clear that the quadruple modulation strength of the correlation
function $\nu^{a}_{2} \nu^{b}_{2}$ for background pairs is exactly
the product of the {\it true} modulation strengths of the {\it
true} singles distributions $\nu^{a}_{2}$ and $\nu^{b}_{2}$.

\section{SIMULATION STUDY OF THE NON-FLOW EFFECT FROM JETS}
\label{appendix:C}

The elliptic flow of the triggers and partners,
which are used to estimate the background contribution in the
correlation function Eq.~\ref{eq:jetpairyield}, are provided by the
BBC reaction plane method. In this section, we show that the large
rapidity separation of $|\Delta\eta|>2.75$ between PHENIX BBC and
central arm greatly suppresses the bias due to jet and dijets to
the reaction plane determination. More details can be also be found
in~\cite{Jia:2006sb}.

The intra-jet correlation is typically limited by the size of the
jet cone, which is much smaller than the $\Delta\eta$ separation
between BBC and central arm. However, due to their broad
distribution of parton $x$ values, the away-side jet has a very
broad distribution in $\Delta\eta$. Hence the inter-jet
correlations can potentially bias the BBC reaction plane
determination. We study the biases by embedding back-to-back jet
pairs into HIJING events. The HIJING events serve as the underlying
Au+Au events, and were checked to reproduce the charged hadron
multiplicity in $\eta$ from PHOBOS~\cite{Back:2002wb} . Elliptic
flow is implemented by applying a track by track weight in each
HIJING event:
\[
w(\textbf{b},p_T,\eta)=1+2v_2(\textbf{b},p_T,\eta)\cos2(\phi-\Psi)
\]
where the $\Psi$ is the direction of the impact parameter
$\textbf{b}$. The centrality and $p_T$ dependence of the $v_2$ is
tuned according to the PHENIX measurement~\cite{Adare:2006ti}. The
$\eta$ dependence of $v_2$ is obtained from
PHOBOS~\cite{Manly:2005zy} minimum bias events. The $v_2$ shape
versus $\eta$ is assumed to be independent of centrality
selections. This gives an overall b, $p_T$ and $\eta$ dependence by
a single function,
\[
v_{2} (p_T,\textbf{b},\eta) = 0.02834 \;\textbf{b} \;{e^{ -
0.5\left(\eta/3.92\right) ^2}} \left( {1 -
\frac{{2.1}} {{1 + e^{1.357 p_T} }}} \right)
\]
We then generate back-to-back jet pairs from the PYTHIA event
generator, requiring a leading particle above 6 GeV/$c$ at
mid-rapidity ($|\eta|<0.35$). Assuming the fractional momentum of
the leading hadron is $\langle z \rangle \approx
0.7$~\cite{Adler:2006sc}, this corresponds to typical jet energy of
$6/\langle z \rangle\approx 9$ GeV/$c$.

\begin{figure}[thb]
\includegraphics[width=1.0\linewidth]{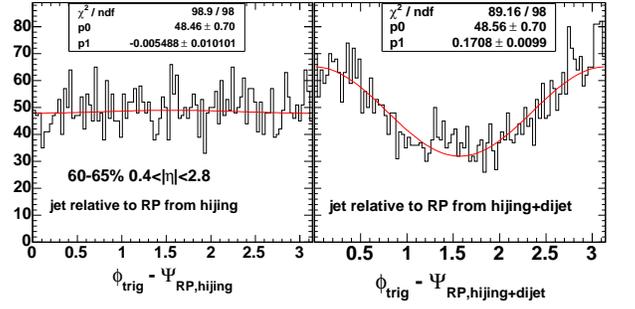}
\caption{\label{fig:rpembed} The distribution of the leading
particle from the dijets relative to the event plane calculated
from HIJING only (left) and event plane from the embedded event
(right).}
\end{figure}

We evaluate the dijet bias by comparing the event plane before and
after the embedding. Dijets tend to bias the event plane towards
the dijet direction, resulting in a false $v_2$ for the jet
particles.  Figure~\ref{fig:rpembed} shows the relative azimuth
distribution between the jet leading hadrons and the event plane
(EP) from the HIJING event (left panels) or the combined event
(right panels). The dijets clearly become correlated with the EP
determined from the combined event, leading to a false $v_2$ for
the leading hadrons. However, since we embed one such dijet pair
for every event, the bias shown in Fig.~\ref{fig:rpembed} should be
interpreted as the bias for those events containing a high-\pt
dijet. Thus it sets an upper limit for the bias effect.

In order to understand the impact of the false $v_2$, we have to
determine their magnitudes in the same way as the real data
analysis, i.e. according to: \[ v_{2} =
\frac{v_{2,\rm{raw}}}{c_{v_2}}
=\frac{\langle\langle\cos2(\phi-\Phi_{\rm{EP}})\rangle\rangle}{\langle\cos2(\Phi_{\rm{EP}}-\Phi_{\rm{RP}})\rangle}
\]
We obtain the raw $v_2$ for by fitting the embedded trigger
distribution (such as Fig.~\ref{fig:rpembed}) for each individual
centrality bin. The raw $v_2$ is then divided by the corresponding
reaction plane resolution, which can be calculated as
$c_{v_2}=\langle\cos2(\Phi_{\rm{EP}}-\Phi_{\rm{RP}})\rangle$.

\begin{figure}[thb]
\includegraphics[width=1.0\linewidth]{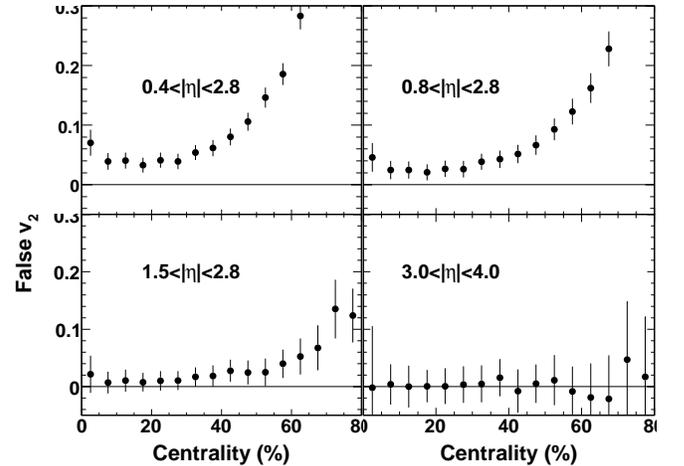}
\caption{\label{fig:v2bias} The false reaction plane $v_2$ of the leading hadron
from the embedded dijet as function of centrality. The $\eta$ range used to determine the event plane
is indicated in each panel. The the embedded dijet is required to have a trigger hadron above 6 GeV/$c$ with in mid-rapidity window of $|\eta|<0.35$.}
\end{figure}

The magnitude of the false $v_2$ depends on the rapidity separation
between the trigger and the sub-event used to determine the EP. Due
to away-side jet swing, this bias could persist to large rapidity
regions.   Figure~\ref{fig:v2bias} shows the centrality dependence of
false trigger $v_2$ for events containing high-\pt dijets for
various rapidity windows used for EP determination. The false $v_2$
decreases as the sub-event used to determine the EP moves towards
large $\eta$. When the sub-event is in the BBC acceptance
($3<|\eta|<4$), the false $v_2$ becomes negligible.

\section{COMPREHENSIVE DATA PLOTS AND DATA TABLES}
\label{appendix:D}

\begin{turnpage}
\begin{figure*}[h]
\includegraphics[width=1.0\linewidth]{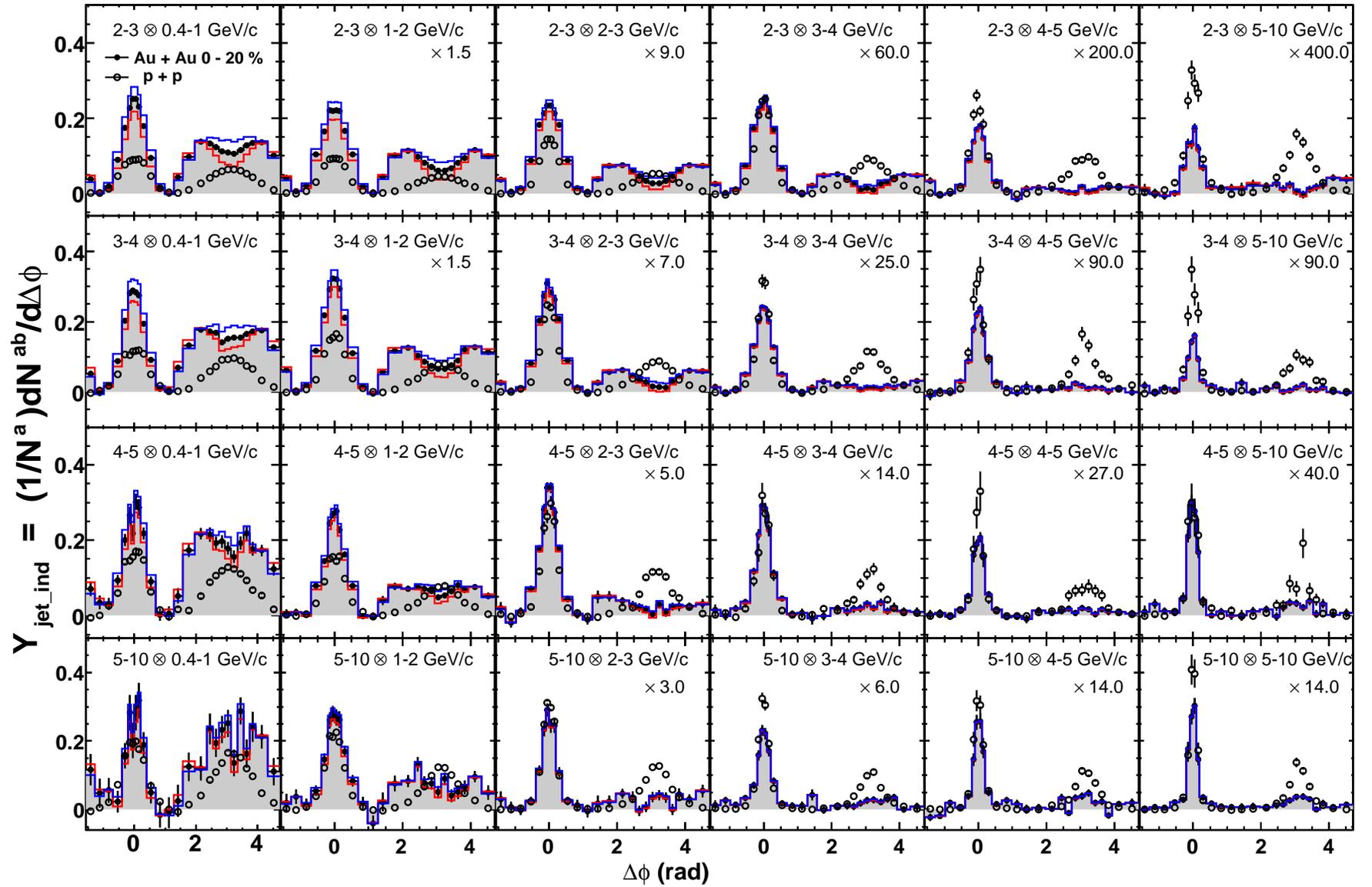}
\caption{\label{fig:shape1} (Color online) Per-trigger yield versus
$\Delta\phi$ for successively increasing trigger and partner \pt
($p_{\rm T}^{a}\otimes p_{T}^{b}$) in $p+p$ (open circles) and 0-20
\% Au+Au (filled circles) collisions. Data are scaled to the
vertical axes of the four left panels. Histograms indicate elliptic
flow uncertainties for Au+Au collisions.}
\end{figure*}
\end{turnpage}

\begin{turnpage}
\begin{figure*}[h]
\includegraphics[width=1.0\linewidth]{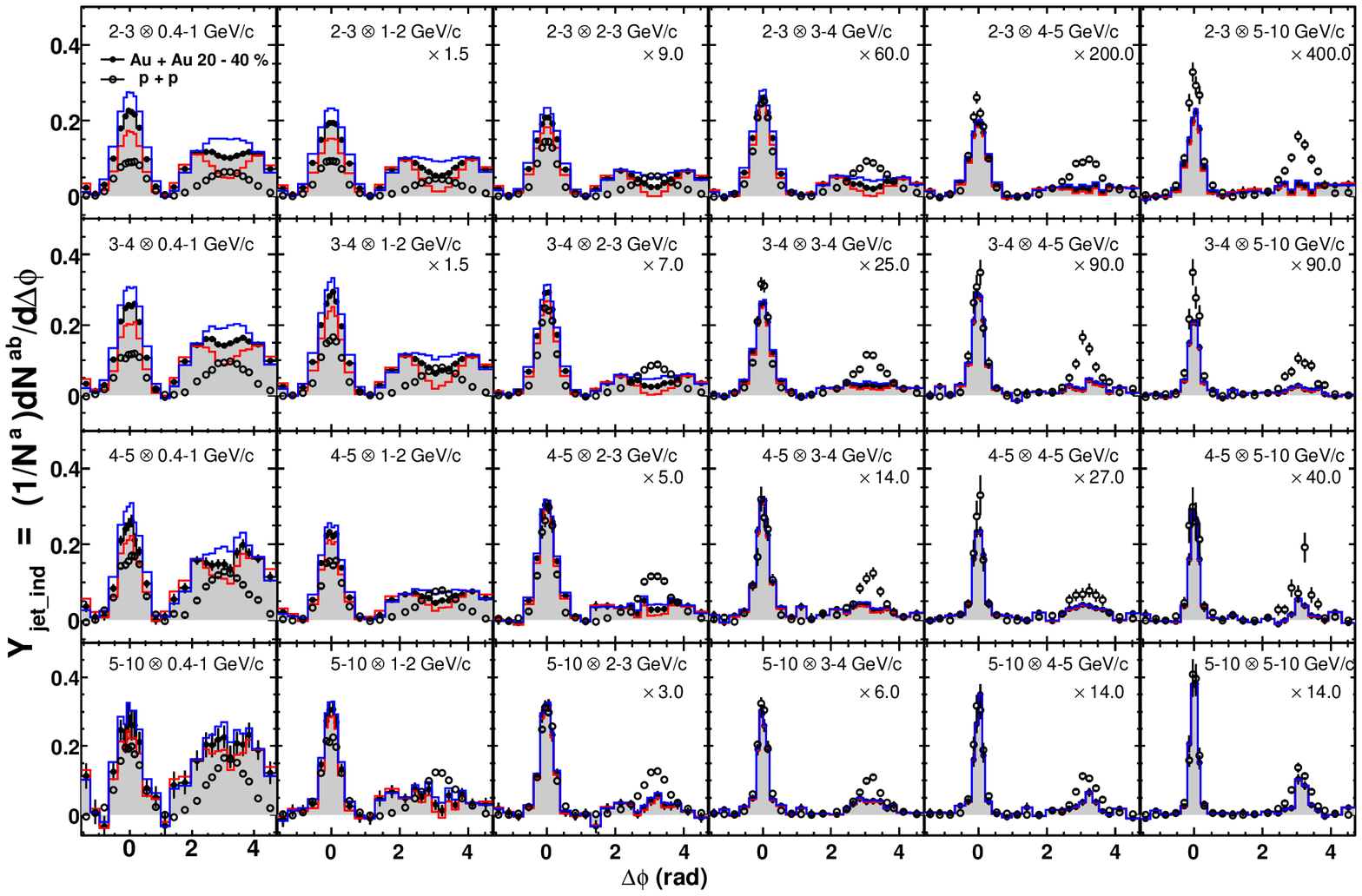}
\caption{\label{fig:shape2} (Color online) Per-trigger yield versus
$\Delta\phi$ for successively increasing trigger and partner \pt
($p_{\rm T}^{a}\otimes p_{T}^{b}$) in $p+p$ (open circles) and
20-40 \% Au+Au (filled circles) collisions. Data are scaled to the
vertical axes of the three left panels. Histograms indicate
elliptic flow uncertainties for Au+Au collisions.}
\end{figure*}

\end{turnpage}

\begin{turnpage}
\begin{figure*}[h]
\includegraphics[width=1.0\linewidth]{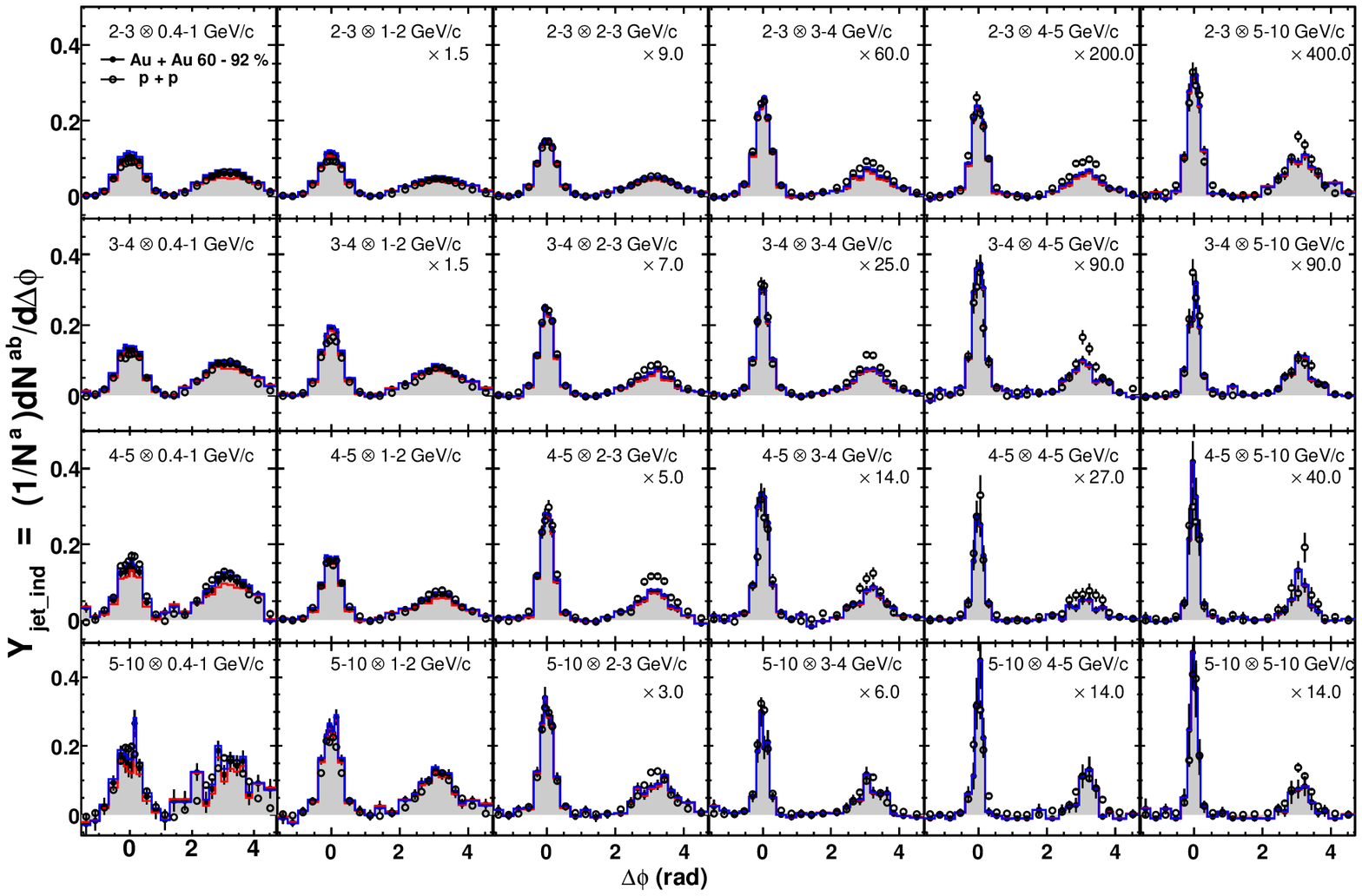}
\caption{\label{fig:shape4} (Color online) Per-trigger yield versus
$\Delta\phi$ for successively increasing trigger and partner \pt
($p_{\rm T}^{a}\otimes p_{T}^{b}$) in $p+p$ (open circles) and
60-92 \% Au+Au (filled circles) collisions. Data are scaled to the
vertical axes of the three left panels. Histograms indicate
elliptic flow uncertainties for Au+Au collisions.}
\end{figure*}
\end{turnpage}

Figures~\ref{fig:shape1}-\ref{fig:shape4} show the comprehensive
array of results, covering the momentum range of 0.4 to 10 GeV/$c$
from which the representative subset shown in Fig.~\ref{fig:shape}
were derived.  These results are described in
Sections~\ref{sec:3} and \ref{sec:4}.

Tables~\ref{tab:compfit4} to \ref{tab:jaa2} show numerical values
of the plotted data.  The corresponding figures are indicated in
the table captions.


\begin{table}[h]
\caption{\label{tab:compfit4} D from FIT1 and FIT2 versus $p_{\rm
T}^{\rm b}$ for three $p_{\rm T}^{\rm a}$ bins
(Fig.~\ref{fig:compfit4}).}
\begin{ruledtabular} 
\begin{tabular}{ccc}
$\langle p_{\rm T}^{\rm b}\rangle$ GeV/$c$  &   D$\pm$Stat+Sys-Sys&
D$\pm$Stat+Sys-Sys\\ & FIT1 &FIT2\\\hline
\multicolumn{3}{c}{$2<p_{\rm T}^{\rm a}<3$ GeV/$c$}\\\hline
0.6  &  0.88$\pm$0.01+0.12-0.12 &1.09$\pm$0.01+0.03-0.05\\
1.2  &  0.94$\pm$0.01+0.09-0.10 &1.14$\pm$0.01+0.04-0.04\\
1.7  &  1.00$\pm$0.01+0.08-0.09 &1.17$\pm$0.01+0.03-0.03\\
2.2  &  1.04$\pm$0.01+0.07-0.08 &1.17$\pm$0.01+0.03-0.03\\
2.7  &  1.10$\pm$0.02+0.06-0.07 &1.19$\pm$0.02+0.03-0.03\\
3.3  &  1.11$\pm$0.05+0.05-0.05 &1.13$\pm$0.05+0.03-0.03\\
4.4  &  1.11$\pm$0.17+0.15-0.12 &1.29$\pm$0.09+0.02-0.03\\
\hline \multicolumn{3}{c}{$3<p_{\rm T}^{\rm a}<4$GeV/$c$}\\\hline
0.6  &  0.90$\pm$0.01+0.11-0.11 &1.12$\pm$0.04+0.03-0.03\\
1.2  &  1.01$\pm$0.01+0.08-0.09 &1.19$\pm$0.02+0.01-0.03\\
1.7  &  1.04$\pm$0.02+0.07-0.08 &1.15$\pm$0.03+0.03-0.03\\
2.2  &  1.10$\pm$0.04+0.05-0.05 &1.12$\pm$0.06+0.03-0.03\\
2.7  &  1.16$\pm$0.12+0.04-0.05 &1.21$\pm$0.07+0.02-0.03\\
3.3  &  1.13$\pm$0.13+0.10-0.05 &1.30$\pm$0.06+0.01-0.01\\
\hline \multicolumn{3}{c}{$4<p_{\rm T}^{\rm a}<5$ GeV/$c$}\\\hline
0.6  &  0.89$\pm$0.05+0.08-0.10 &1.07$\pm$0.13+0.03-0.03\\
1.2  &  0.85$\pm$0.06+0.08-0.07 &0.85$\pm$0.09+0.08-0.05\\
1.7  &  1.13$\pm$0.13+0.08-0.10 &1.20$\pm$0.13+0.03-0.04\\
2.2  &  1.09$\pm$0.16+0.08-0.08 &1.21$\pm$0.16+0.03-0.04\\
2.7  &  1.11$\pm$0.20+0.11-0.10 &1.19$\pm$0.17+0.03-0.04\\
\end{tabular}  \end{ruledtabular}
\end{table}

\begin{table}[h]
\caption{\label{tab:compfit3a}Near-side jet width for $p+p$ and
0-20\% Au+Au for various $p_{\rm T}^{\rm a}$ and $p_{\rm T}^{\rm
b}$ bins (Fig.~\ref{fig:compfit3b}). The width is unchanged by
interchanging $p_{\rm T}^{\rm a}$ and $p_{\rm T}^{\rm b}$.}
\begin{ruledtabular} \begin{tabular}{cccccc}
& \multicolumn{3}{c}{Au+Au 0-20\%}
&\multicolumn{2}{c}{p+p}\\\hline &\multicolumn{5}{c}{$2<p_{\rm
T}^{\rm a}<3$ GeV/$c$}\\\hline $\langle p_{\rm T}^{\rm b}\rangle$
GeV/$c$ & Width & Stat & Sys & Width &Sys\\\hline
0.6  &  0.415  &  0.002   & +0.022-0.041 &0.472  &  0.006\\
1.4  &  0.432  &  0.001   & +0.024-0.028 &0.364  &  0.003\\
2.3  &  0.410  &  0.002   & +0.043-0.019 &0.281  &  0.003\\
\hline&\multicolumn{5}{c}{$3<p_{\rm T}^{\rm a}<4$ GeV/$c$}\\\hline
0.6  &  0.383  &  0.015   & +0.006-0.033 &0.430  &  0.012\\
1.4  &  0.409  &  0.004   & +0.028-0.039 &0.314  &  0.005\\
2.3  &  0.365  &  0.012   & +0.022-0.016 &0.244  &  0.005\\
3.3  &  0.286  &  0.010   & +0.005-0.035 &0.199  &  0.007\\
\hline&\multicolumn{5}{c}{$4<p_{\rm T}^{\rm b}<5$ GeV/$c$}\\\hline
0.6  &  0.394  &  0.037   & +0.036-0.027 &0.402  &  0.018\\
1.4  &  0.340  &  0.017   & +0.036-0.040 &0.299  &  0.008\\
2.3  &  0.274  &  0.012   & +0.011-0.022 &0.225  &  0.007\\
3.3  &  0.223  &  0.010   & +0.012-0.027 &0.201  &  0.010\\
4.4  &  0.179  &  0.009   & +0.002-0.013 &0.148  &  0.011\\
\hline&\multicolumn{5}{c}{$5<p_{\rm T}^{\rm a}<10$ GeV/$c$}\\\hline
0.6  &  0.308  &  0.059   & +0.016-0.036 &0.385  &  0.008\\
1.4  &  0.303  &  0.029   & +0.031-0.040 &0.281  &  0.003\\
2.3  &  0.224  &  0.015   & +0.007-0.028 &0.208  &  0.003\\
3.3  &  0.198  &  0.012   & +0.008-0.023 &0.165  &  0.003\\
4.4  &  0.155  &  0.009   & +0.007-0.014 &0.155  &  0.005\\
6.6  &  0.135  &  0.008   & +0.001-0.001 &0.130  &  0.004\\
\end{tabular}  \end{ruledtabular}
\end{table}
\begin{table}
\caption{\label{tab:compfit3b} Centrality dependence of near-side
jet widths for various $p_{\rm T}^{\rm a}\otimes p_{\rm T}^{\rm b}$
bins (Fig.~\ref{fig:compfit3}).}
\begin{ruledtabular} \begin{tabular}{cccc}
$\langle N_{\rm part} \rangle$ & Width & Stat.Err. &Sys.
Err.\\\hline \multicolumn{4}{c}{$2-3 \otimes 0.4-1$
GeV/$c$}\\\hline
351.4& 0.367 & 0.012 &+0.032-0.043\\
299.0& 0.415 & 0.007 &+0.033-0.044\\
234.6& 0.424 & 0.002 &+0.024-0.027\\
166.6& 0.431 & 0.002 &+0.048-0.041\\
114.2& 0.438 & 0.002 &+0.048-0.041\\
74.4& 0.438 & 0.003 &+0.057-0.043\\
\end{tabular}  \end{ruledtabular}
\end{table}

\begin{table*}[h]
\caption{\label{tab:rhs1} $R_{\rm HS}$ in $p+p$ and 0-20\% Au+Au
for various $p_{\rm T}^{\rm a}$ and $p_{\rm T}^{\rm b}$ bins
(Fig.~\ref{fig:ratio}).}
\begin{ruledtabular} \begin{tabular}{ccccccc}
&\multicolumn{4}{c}{Au + Au 0-20\%} &
\multicolumn{2}{c}{$p+p$}\\\hline $\langle p_{\rm T}^{\rm b}
\rangle$ GeV/$c$ & $R_{\rm HS}$   &    Stat.  &  $v_2$ Err. &
ZYAM Err. & $R_{\rm HS}$   & Total Err. \\\hline
&\multicolumn{6}{c}{$2.0<p_{\rm T}^{\rm a}<3.0$ GeV/$c$}\\\hline
0.66 &   0.965   &   0.008  &  +0.211-0.236  &  +0.007-0.000 &2.428   &   +0.007-0.000\\
1.22 &   0.722   &   0.008  &  +0.180-0.201  &  +0.038-0.002 &2.419   &   +0.038-0.002\\
1.72 &   0.628   &   0.010  &  +0.176-0.196  &  +0.040-0.005 &2.752   &   +0.040-0.005\\
2.23 &   0.531   &   0.016  &  +0.178-0.198  &  +0.051-0.009 &2.673   &   +0.051-0.009\\
2.73 &   0.395   &   0.027  &  +0.198-0.208  &  +0.113-0.021 &3.234   &   +0.113-0.021\\
\end{tabular}  \end{ruledtabular}
\end{table*}
\begin{table*}[h]
\caption{\label{tab:compfit1}Centrality dependence of D from FIT1
and FIT2 (Fig.~\ref{fig:compfit1a}).}
\begin{ruledtabular} \begin{tabular}{cccccccccc}
&\multicolumn{3}{c}{D from FIT1} & \multicolumn{3}{c}{D
from FIT2} &\multicolumn{3}{c}{Fraction of shoulder
yield(FIT2)}\\\hline $\langle N_{\rm part} \rangle$& D &  Stat. Err
& Sys.Err & D &Stat. Err & Sys.Err      &  Frac  &Stat.Err &
Sys.Err\\\hline
351.4 &  1.068 &  0.016 &  +0.032-0.032   &  1.192 &  0.024 &  +0.028-0.028 &  0.800 &  0.053  &   +0.027-0.028\\
299.0 &  1.090 &  0.013 &  +0.055-0.062	  &  1.208 &  0.017 &  +0.028-0.030 &  0.800 &  0.044  &   +0.053-0.059\\
234.6 &  1.030 &  0.010 &  +0.100-0.120	  &  1.126 &  0.022 &  +0.045-0.054 &  0.842 &  0.052  &   +0.108-0.128\\
166.6 &  0.976 &  0.016 &  +0.110-0.153	  &  0.983 &  0.035 &  +0.103-0.036 &  0.981 &  0.111  &   +0.019-0.291\\
114.2 &  0.865 &  0.018 &  +0.141-0.123	  &  0.930 &  0.064 &  +0.091-0.050 &  0.851 &  0.246  &   +0.149-0.253\\
74.4 &  0.786 &  0.030 &  +0.129-0.116	  &  0.916 &  0.045 &  +0.085-0.085 &  0.750 &  0.043  &   +0.250-0.195\\
45.5 &  0.652 &  0.021 &  +0.074-0.066	  &  0.996 &  0.029 &  +0.030-0.030 &  0.293 &  0.040  &   +0.091-0.064\\
14.5 &  0.000 &  0.989 &  +0.010-0.000	  &  1.178 &  0.047 &  +0.050-0.047 &  0.120 &  0.029  &   +0.049-0.032\\
\end{tabular}  \end{ruledtabular}
\end{table*}

\begin{table*}[h]
\caption{\label{tab:slope1} Truncated mean $p_{\rm T}$, $\langle
p_T^\prime\rangle$, for various $\Delta\phi$ ranges and centrality
(Fig.~\ref{fig:slope}).}
\begin{ruledtabular} \begin{tabular}{cccccccccc}
\multicolumn{10}{c}{$2.0<p_{\rm T}^{\rm a}<3.0$ GeV/$c$} \\\hline
&\multicolumn{3}{c}{Near-side(GeV/$c$)} &
\multicolumn{3}{c}{Away-side Shoulder(GeV/$c$)}
&\multicolumn{3}{c}{Away-side Head(GeV/$c$)}\\\hline $\langle
N_{\rm part} \rangle$ &  $\langle p_T^\prime\rangle$    &    Stat
& Sys & $\langle p_T^\prime\rangle$     &    Stat  &    Sys   &
$\langle p_T^\prime\rangle$     &    Stat  &    Sys \\\hline
351.4  &  0.554  &  0.008 &   +0.009-0.019  &  0.445  &  0.011 &   +0.001-0.010  &  0.388  &  0.017 &   +0.006-0.017\\
299.0  &  0.538  &  0.007 &   +0.015-0.022  &  0.453  &  0.009 &   +0.001-0.011  &  0.379  &  0.014 &   +0.014-0.026\\
234.6  &  0.530  &  0.004 &   +0.025-0.012  &  0.439  &  0.006 &   +0.020-0.004  &  0.388  &  0.008 &   +0.042-0.032\\
166.6  &  0.531  &  0.004 &   +0.028-0.015  &  0.449  &  0.007 &   +0.018-0.005  &  0.407  &  0.008 &   +0.033-0.042\\
114.2  &  0.529  &  0.005 &   +0.028-0.020  &  0.441  &  0.008 &   +0.002-0.007  &  0.426  &  0.009 &   +0.005-0.020\\
\end{tabular}  \end{ruledtabular}
\end{table*}

\begin{table*}[h]
\caption{\label{tab:slope2}  Truncated mean $p_{\rm T}$, $\langle
p_T^\prime\rangle$, in HR for various  $p_{\rm T}^{\rm a} \otimes
p_{\rm T}^{\rm b}$ bins (Fig.~\ref{fig:slope1}).}
\begin{ruledtabular} \begin{tabular}{ccccccc}
\multicolumn{7}{c}{$1.0<p_{\rm T}^{\rm b}<7.0$ GeV/$c$}\\\hline
&\multicolumn{3}{c}{$3.0<p_{\rm T}^{\rm a}<4.0$
GeV/$c$}&\multicolumn{3}{c}{$4.0<p_{\rm T}^{\rm a}<5.0$
GeV/$c$}\\\hline $\langle N_{\rm part}\rangle$  &$\langle
p_T^\prime\rangle$& Stat & Sys &$\langle p_T^\prime\rangle$& Stat &
Sys\\\hline
279.9  & 0.383 & 0.020 & +0.017-0.039  & 0.448 & 0.049 & +0.034-0.046\\
140.4  & 0.469 & 0.017 & +0.010-0.014  & 0.602 & 0.042 & +0.047-0.044\\
60.0   & 0.560 & 0.018 & +0.037-0.020  & 0.637 & 0.032 & +0.039-0.030\\
14.5   & 0.643 & 0.018 & +0.014-0.024  & 0.794 & 0.041 & +0.030-0.024\\
2.0    & 0.732 & 0.017 & +0.000-0.002  & 0.838 & 0.032 &
+0.002-0.002\\
\end{tabular}  \end{ruledtabular}
\end{table*}

\begin{table*}[h]
\caption{\label{tab:yield1} Per-trigger yield in NR, HR and SR and
various $p_{\rm T}^{\rm b}$ ranges for $2.0<p_{\rm T}^{\rm a}<3.0$
GeV/$c$. They are bases for Fig.~\ref{fig:spec1}-\ref{fig:iaaaway4}
and  Fig.~\ref{fig:spec2}-\ref{fig:iaanear4}. The systematic
uncertainties due to the single particle efficiency are not
included ($\sim$10\%).}
\begin{ruledtabular} \begin{tabular}{ccccc}
\multicolumn{5}{c}{$2.0<p_{\rm T}^{\rm a}<3.0$ GeV/$c$}\\\\\hline
\\&$p_{\rm T}^{\rm b}$ (GeV/$c$)  & $\int_{|\Delta\phi|<\pi/3,\rm NR}d\Delta\phi\left(\frac{1}{N^{\rm a}}\frac{dN^{\rm ab}}{d\Delta\phi}\right) $&
$\int_{|\Delta\phi-\pi|<\pi/6, \rm
HR}d\Delta\phi\left(\frac{1}{N^{\rm a}}\frac{dN^{\rm
ab}}{d\Delta\phi}\right)$ &
$\int_{\pi/6<|\Delta\phi-\pi|<\pi/3,\rm SR}d\Delta\phi\left(\frac{1}{N^{\rm a}}\frac{dN^{\rm ab}}{d\Delta\phi}\right)$\\
&& Yield$\pm$ Stat. +Sys. -Sys. & Yield$\pm$ Stat. +Sys. -Sys. &
Yield$\pm$ Stat. +Sys. -Sys. \\\\\hline \multirow{9}{*}{0-20\%}
&0.4-1.0   & $(2.24\pm0.01+0.80-0.42) \times 10^{-1}$ & $(1.20\pm0.01+0.45-0.31) \times 10^{-1}$ &$(2.48\pm0.01+0.68-0.05) \times 10^{-1}$\\
&1.0-1.5   & $(9.14\pm0.05+2.03-1.43) \times 10^{-2}$ & $(3.47\pm0.03+1.23-1.02) \times 10^{-2}$ &$(9.60\pm0.06+1.54-0.24) \times 10^{-2}$\\
\hline

\end{tabular}  \end{ruledtabular}
\end{table*}

\begin{table*}[h]
\caption{\label{tab:yield2} Per-trigger yield in various NR, HR and
SR for $3.0<p_{\rm T}^{\rm a}<4.0$ GeV/$c$. They are bases for
Fig.~\ref{fig:spec1}-\ref{fig:iaaaway4} and
Fig.~\ref{fig:spec2}-\ref{fig:iaanear4}. The systematic
uncertainties due to the single particle efficiency are not
included ($\sim$10\%).}
\begin{ruledtabular} \begin{tabular}{ccccc}
\multicolumn{5}{c}{$3.0<p_{\rm T}^{\rm a}<4.0$ GeV/$c$}\\\\\hline
\\&$p_{\rm T}^{\rm b}$ (GeV/$c$)  & $\int_{|\Delta\phi|<\pi/3,\rm NR}d\Delta\phi\left(\frac{1}{N^{\rm a}}\frac{dN^{\rm ab}}{d\Delta\phi}\right)$&
$\int_{|\Delta\phi-\pi|<\pi/6, \rm
HR}d\Delta\phi\left(\frac{1}{N^{\rm a}}\frac{dN^{\rm
ab}}{d\Delta\phi}\right)$ &
$\int_{\pi/6<|\Delta\phi-\pi|<\pi/3,\rm SR}d\Delta\phi\left(\frac{1}{N^{\rm a}}\frac{dN^{\rm ab}}{d\Delta\phi}\right)$\\
&& Yield$\pm$ Stat. +Sys. -Sys. & Yield$\pm$ Stat. +Sys. -Sys. &
Yield$\pm$ Stat. +Sys. -Sys. \\\\\hline \multirow{9}{*}{0-20\%}
&0.4-1.0   & $(2.45\pm0.04+0.98-0.31) \times 10^{-1}$ & $(1.62\pm0.02+0.54-0.26) \times 10^{-1}$ &$(3.23\pm0.05+0.89-0.08) \times 10^{-1}$\\
&1.0-1.5   & $(1.09\pm0.02+0.26-0.14) \times 10^{-1}$ & $(3.90\pm0.10+1.51-1.02) \times 10^{-2}$ &$(1.10\pm0.02+0.22-0.03) \times 10^{-1}$\\
\end{tabular}  \end{ruledtabular}
\end{table*}

\begin{table*}[h]
\caption{\label{tab:yield3} Per-trigger yield in various NR, HR and
SR for $4.0<p_{\rm T}^{\rm a}<5.0$ GeV/$c$. They are bases for
Fig.~\ref{fig:spec1}-\ref{fig:iaaaway4} and
Fig.~\ref{fig:spec2}-\ref{fig:iaanear4}. The systematic
uncertainties due to the single particle efficiency are not
included ($\sim$10\%).}
\begin{ruledtabular} \begin{tabular}{ccccc}
\multicolumn{5}{c}{$4.0<p_{\rm T}^{\rm a}<5.0$ GeV/$c$}\\\\\hline
\\&$p_{\rm T}^{\rm b}$ (GeV/$c$)  & $\int_{|\Delta\phi|<\pi/3,\rm NR}d\Delta\phi\left(\frac{1}{N^{\rm a}}\frac{dN^{\rm ab}}{d\Delta\phi}\right)$&
$\int_{|\Delta\phi-\pi|<\pi/6, \rm
HR}d\Delta\phi\left(\frac{1}{N^{\rm a}}\frac{dN^{\rm
ab}}{d\Delta\phi}\right)$ &
$\int_{\pi/6<|\Delta\phi-\pi|<\pi/3,\rm SR}d\Delta\phi\left(\frac{1}{N^{\rm a}}\frac{dN^{\rm ab}}{d\Delta\phi}\right)$\\
&& Yield$\pm$ Stat. +Sys. -Sys. & Yield$\pm$ Stat. +Sys. -Sys. &
Yield$\pm$ Stat. +Sys. -Sys. \\\\\hline \multirow{9}{*}{0-20\%}
&0.4-1.0   & $(2.50\pm0.11+0.48-0.36) \times 10^{-1}$ & $(1.96\pm0.07+0.32-0.27) \times 10^{-1}$ &$(3.66\pm0.14+0.33-0.19) \times 10^{-1}$\\
&1.0-1.5   & $(1.19\pm0.05+0.17-0.17) \times 10^{-1}$ & $(4.84\pm0.31+1.14-1.17) \times 10^{-2}$ &$(9.97\pm0.63+1.03-0.87) \times 10^{-2}$\\
\end{tabular}  \end{ruledtabular}
\end{table*}

\begin{table*}[h]
\caption{\label{tab:yield4} Per-trigger yield in various NR, HR and
SR for $5.0<p_{\rm T}^{\rm a}<10.0$ GeV/$c$. They are bases for
Fig.~\ref{fig:spec1}-\ref{fig:iaaaway4} and
Fig.~\ref{fig:spec2}-\ref{fig:iaanear4}. The systematic
uncertainties due to the single particle efficiency are not
included ($\sim$10\%).}
\begin{ruledtabular} \begin{tabular}{ccccc}
\multicolumn{5}{c}{$5.0<p_{\rm T}^{\rm a}<10.0$ GeV/$c$}\\\\\hline
\\&$p_{\rm T}^{\rm b}$ (GeV/$c$)  & $\int_{|\Delta\phi|<\pi/3,\rm NR}d\Delta\phi\left(\frac{1}{N^{\rm a}}\frac{dN^{\rm ab}}{d\Delta\phi}\right)$&
$\int_{|\Delta\phi-\pi|<\pi/6, \rm
HR}d\Delta\phi\left(\frac{1}{N^{\rm a}}\frac{dN^{\rm
ab}}{d\Delta\phi}\right)$ &
$\int_{\pi/6<|\Delta\phi-\pi|<\pi/3,\rm SR}d\Delta\phi\left(\frac{1}{N^{\rm a}}\frac{dN^{\rm ab}}{d\Delta\phi}\right)$\\
\multirow{9}{*}{0-20\%}
&0.4-1.0   & $(2.04\pm0.25+0.51-0.52) \times 10^{-1}$ & $(2.25\pm0.17+0.31-0.32) \times 10^{-1}$ &$(3.36\pm0.32+0.45-0.45) \times 10^{-1}$\\
&1.0-1.5   & $(1.17\pm0.11+0.35-0.21) \times 10^{-1}$ & $(5.59\pm0.72+1.85-1.24) \times 10^{-2}$ &$(1.22\pm0.15+0.34-0.20) \times 10^{-1}$\\
&1.5-2.0   & $(8.45\pm0.57+1.22-1.23) \times 10^{-2}$ & $(1.34\pm0.38+0.68-0.69) \times 10^{-2}$ &$(4.14\pm0.78+1.05-1.05) \times 10^{-2}$\\
&2.0-2.5   & $(3.05\pm0.29+0.56-0.57) \times 10^{-2}$ & $(5.22\pm1.93+2.97-3.02) \times 10^{-3}$ &$(1.61\pm0.40+0.53-0.53) \times 10^{-2}$\\
\end{tabular}  \end{ruledtabular}
\end{table*}


\begin{table*}[h]
\caption{\label{tab:jaa1} Pair suppression factor $J_{\rm AA}$ in
NR and HR for various $p_{\rm T}^{\rm a}\otimes p_{\rm T}^{\rm b}$
bins. The systematic uncertainties due to the single particle
efficiency are not included ($\sim$17\%).}
\begin{ruledtabular} \begin{tabular}{ccccccc}
&\multicolumn{3}{c}{$2.0<p_{\rm T}^{\rm a}<3.0$
GeV/$c$}&\multicolumn{3}{c}{$3.0<p_{\rm T}^{\rm a}<4.0$
GeV/$c$}\\\\\hline
\\\multicolumn{2}{c}{$\langle p_{\rm T}^{\rm b}\rangle$ GeV/$c$}  & $J_{\rm AA}(|\Delta\phi|<\pi/3)$&$J_{\rm AA}(|\Delta\phi-\pi|<\pi/6)$ & $\langle p_{\rm T}^{\rm b}\rangle$ GeV/$c$  & $J_{\rm AA}(|\Delta\phi|<\pi/3)$ &$J_{\rm AA}(|\Delta\phi-\pi|<\pi/6)$\\\\\hline
\multirow{9}{*}{\begin{sideways}0-20\%\end{sideways}}
& 0.63  & $1.34\pm0.01+0.52-0.25$ & $1.17\pm0.01+0.46-0.30$& 0.63  & $0.88\pm0.02+0.36-0.11$ & $0.80\pm0.02+0.27-0.13$\\
& 1.20  & $1.39\pm0.01+0.32-0.22$ & $1.03\pm0.01+0.37-0.31$& 1.20  & $0.87\pm0.02+0.21-0.11$ & $0.51\pm0.02+0.20-0.13$\\
& 1.70  & $1.53\pm0.02+0.24-0.19$ & $0.83\pm0.02+0.30-0.27$& 1.70  & $0.94\pm0.02+0.12-0.10$ & $0.37\pm0.02+0.14-0.13$\\
\end{tabular}  \end{ruledtabular}
\end{table*}

\begin{table*}[h]
\caption{\label{tab:jaa2}  Pair suppression factor $J_{\rm AA}$ in
NR and HR for various $p_{\rm T}^{\rm a}\otimes p_{\rm T}^{\rm b}$
bins. The systematic uncertainties due to the single particle
efficiency are not included ($\sim$17\%).}
\begin{ruledtabular} \begin{tabular}{ccccccc}
&\multicolumn{3}{c}{$4.0<p_{\rm T}^{\rm a}<5.0$
GeV/$c$}&\multicolumn{3}{c}{$5.0<p_{\rm T}^{\rm a}<10.0$
GeV/$c$}\\\\\hline
\\\multicolumn{2}{c}{$\langle p_{\rm T}^{\rm b} \rangle$ GeV/$c$} & $J_{\rm AA}(|\Delta\phi|<\pi/3)$& $J_{\rm AA}(|\Delta\phi-\pi|<\pi/6)$ & $\langle p_{\rm T}^{\rm b}\rangle$ GeV/$c$  & $J_{\rm AA}(|\Delta\phi|<\pi/3)$& $J_{\rm AA}(|\Delta\phi-\pi|<\pi/6)$\\\\\hline
\multirow{9}{*}{\begin{sideways}0-20\%\end{sideways}}
& 0.63  & $0.48\pm0.02+0.10-0.07$ & $0.52\pm0.02+0.09-0.07$& 0.63  & $0.28\pm0.04+0.07-0.07$ & $0.39\pm0.03+0.05-0.06$\\
& 1.20  & $0.47\pm0.02+0.07-0.07$ & $0.32\pm0.02+0.08-0.08$& 1.20  & $0.31\pm0.03+0.09-0.06$ & $0.21\pm0.03+0.07-0.05$\\
& 1.70  & $0.50\pm0.02+0.05-0.05$ & $0.21\pm0.02+0.06-0.05$& 1.70  & $0.37\pm0.03+0.05-0.05$ & $0.09\pm0.03+0.05-0.05$\\
\end{tabular}  \end{ruledtabular}
\end{table*}


\begin{references}
\bibitem{footnote1}Note that the pseudo-rapidity coverage of PHENIX
    is smaller than that used in STAR, thus the amount of ridge yield
    seen in PHENIX is also smaller. The projected near-side jet
    shape in $\Delta\phi$ may also depends on the pseudo-rapidity
    coverage.
\bibitem{footnote2}The condition of Eq.~\ref{eq:acceptances} can
    be violated by some experimental effects, such as ghost-pair
    tracking artifacts at small angles. We recover the result of
    Eq.~\ref{eq:acceptances} by applying pair cuts to remove such
    artifacts identically to same-event and mixed-event pairs. Pair
    cuts affect the integrands in the numerator and denominator of
    Eq.~\ref{eq:cf_topline} equally and so leave the end result of
    Eq.~\ref{eq:cf_bottomline} intact.
\end{references}
\end{document}